\documentclass[superscriptaddress,
amsmath,amssymb,aps,prc,
showkeys,
twocolumn,
]{revtex4-1}
\usepackage{amssymb}

\usepackage{amsmath,mathrsfs,amssymb}
\usepackage{eurosym}
\usepackage{graphicx,subfigure}
\usepackage{color}
\usepackage{indentfirst}
\usepackage{extarrows}
\usepackage{hyperref}
\hypersetup{colorlinks=true, citecolor=blue, urlcolor=blue, linkcolor=blue}
\usepackage{ulem}
\usepackage{bm}
\usepackage[latin1]{inputenc}
\newcommand*{\blb}{{\big<}}
\newcommand*{\blk}{{\big>}}
\newcommand*{\bl}{{\big |}}
\usepackage[toc,page]{appendix} 
\usepackage{xcolor}

\begin{document}

\title{ Trojan horse method as surrogate indirect 
approach to study resonant reactions in nuclear astrophysics}
\author{A. M. Mukhamedzhanov}
\email{akram@comp.tamu.edu}
\affiliation{Cyclotron Institute, Texas A$\&$M University, College Station, TX 77843, USA}
\author{A.S. Kadyrov}
\email{a.kadyrov@curtin.edu.au}
\affiliation{Curtin Institute for Computation and Department of Physics and Astronomy, Curtin University, GPO Box U1987, Perth, WA 6845, Australia}
\author{D. Y. Pang}
\email{dypang@buaa.edu.cn}
\affiliation{School of Physics and Beijing Key Laboratory of Advanced Nuclear Materials and Physics, Beihang University, Beijing, 100191, People's Republic of China}

\begin{abstract}
The primary goal of the Trojan horse method (THM) is to analyze resonant rearrangement reactions when the density of the resonance levels is low and statistical models cannot be applied. The main  difficulty of the analysis is related  with the facts that in the final state the THM reaction  involves three particles and that 
the intermediate particle, which is transferred from the Trojan horse particle to the target nucleus  to form a resonance state, is virtual. Another difficulty is associated with the Coulomb interaction between the particles, especially, taking into account that  the goal of the THM is to study resonant rearrangement reactions at very low energies important for nuclear astrophysics. The exact theory of such reactions with three charged particles is very complicated and is not  available. This is why different approximations are used to analyze THM reactions.  
In this review paper  we describe a new approach based on a few-body formalism that provides a solid basis for deriving the THM reaction amplitude taking into account rescattering of the particles in the initial, intermediate and final states of the THM reaction.  Since the THM uses a two-step reaction in which the first step is the transfer reaction populating a resonance state, we address the theory of the transfer reactions. The theory is based on the surface-integral approach and $R$-matrix formalism. We also discuss application of the THM to resonant reactions populating both resonances located on the second energy sheet and subthreshold resonances, which are subthreshold bound states located at negative energies close to thresholds. We consider the application of  the THM  to determine the astrophysical factors of  resonant radiative-capture reactions at energies so low that direct measurements  can hardly be performed due to the negligibly small penetrability factor in the entry channel of the reaction.  We elucidated the main ideas of the THM and outline necessary conditions to perform the THM experiments.

\end{abstract}

\maketitle

\tableofcontents
\section{Introduction}  
  
Often reactions of astrophysical interest are measured in the laboratory at energies
much higher than those relevant to stellar processes. After that, an extrapolation down to stellar energies is performed. This allows one to obtain the astrophysical factors and reaction rates in the so-called Gamow window 
\cite{RolfsRodney} at which the convolution of the Maxwell-Boltzman energy distribution and the cross section of a reaction has a maximum.  However, such extrapolations may lead to significant uncertainties.

There are a few underground laboratories operating or under construction around the world where the astrophysical reactions can be measured. The most famous is the underground laboratory in Gran Sasso (LUNA), Italy. This facility uses a low-energy accelerator to measure cross sections for reactions involving stable beams and
more than 75 targets at significantly lower energies than those previously achieved \cite{Cavanno}. But still extrapolations to astrophysical energies is usually required. 
As an example of the important astrophysical reaction, which requires extrapolation of the direct measurements, is the ${}^{12}{\rm C}+ {}^{12}{\rm C}$ fusion. 
There is another problem which plagues measurements of charged particle reaction rates at low energies. It is electron screening which increases the cross section measured in the laboratory compared to the actual nuclear reaction rates in stellar plasma. Similarly, neutron-induced reactions on unstable short-lived nuclei also cannot be presently measured directly in the laboratory.  

Resonant reactions at very low energies  are important part of the nuclear astrophysical processes occurring in stellar environments. The energies  at which these reactions proceed are much lower than the Coulomb barrier. It  makes their cross sections so small that  it is very difficult or often impossible to measure them directly in the laboratory.  This is owing to the very small barrier penetrability factor from the Coulomb + centrifugal forces producing an exponential fall off of the cross section as  energy decreases.

Indirect techniques, also known as surrogate methods, have been developed over the past several decades to provide ways to determine reaction rates that cannot be measured directly in the laboratory \cite{reviewpaper}.  Important information that is needed to determine the rates for reactions that are dominated by a nuclear resonance is the energy of the resonance and its decay width in the appropriate initial and final channels. Indirect techniques have been applied using both stable and radioactive beams. 
In this review we address the theory and applications of the surrogate reactions proceeding through a resonance in the subsystem. This method allows one to treat a wide variety of astrophysical resonant reactions including resonant rearrangement reactions, reactions proceeding through subthreshold resonances and resonant radiative-capture reactions.  

One of the powerful indirect methods to obtain information about the resonant astrophysical reactions is the Trojan horse method (THM).  The THM is a unique indirect technique that allows one  to determine the astrophysical factors and reaction rates for rearrangement and radiative capture resonant reactions by obtaining the cross section for a binary process through the use of a surrogate ``Trojan horse'' (TH) particle, where direct methods are not able to obtain data due to very small cross sections. The method was originally suggested by Baur \cite{Baur1986} but became a powerful and well known indirect method  under  the  leadership of Prof. Claudio Spitaleri \cite{spitaleri2019,spitaleri2016,reviewpaper,muk2008}.
The  THM is used to analyze resonant reactions when the level density is low and the statistical theory is not applicable. Such reactions play an important role in nuclear and atomic physics. Here we present a theory of the THM based on the few-body formalism.

\begin{figure}[htbp]
\includegraphics[width=0.5\columnwidth]{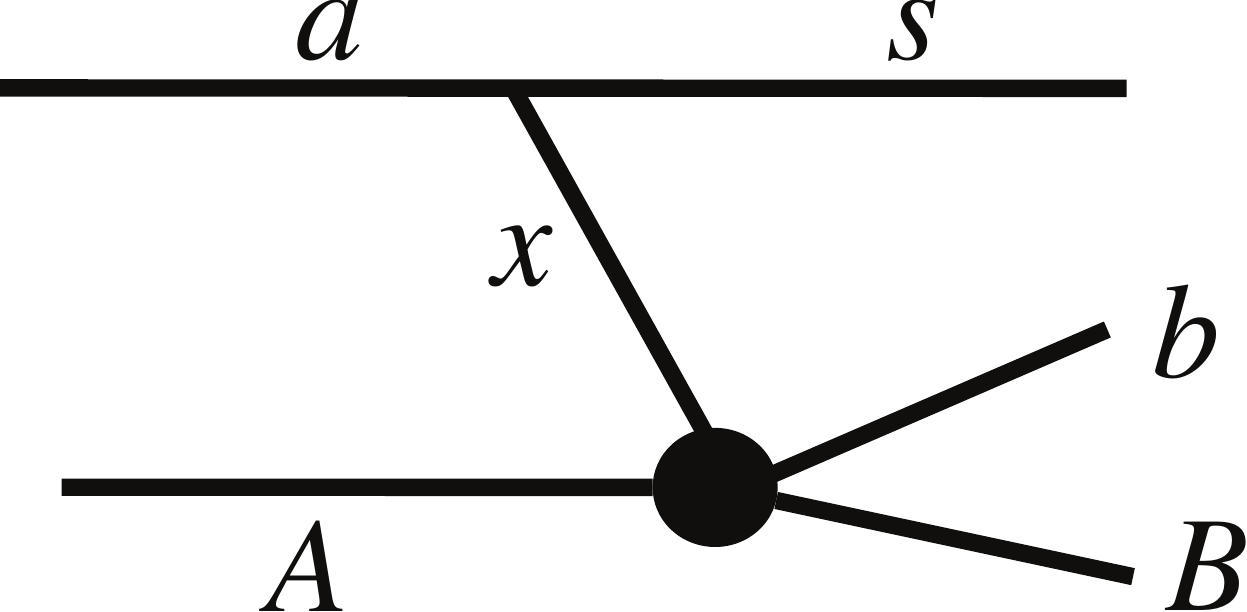}
\caption{The  diagram describing the  knockout reaction in the PWA.  }
\label{fig_polediagram1}
\end{figure}

The THM reactions are a subclass of reactions representing transfer to the continuum followed by rearrangement reactions $x+A \to b+B$.  The THM reactions  leading to  3 particles in the final state  are described by the pole diagram depicted in Fig.  \ref{fig_polediagram1}. 
 In this diagram the three-ray vertex corresponds to the virtual decay $a \to s+x$ and the four-ray vertex describes the subreaction amplitude  $x+A \to b+B$.   In the  plane-wave approach,
 the fully differential cross section (DCS),  ${ {\rm d}^{3}\sigma }/{ {\rm d}\Omega_{bB}\,{\rm d}\Omega_{sF}\,{\rm d}E_{bB} }$,       of the knockout reaction with three particles in the final state corresponding to the pole diagram is
 expressed in terms of the doubly DCS,  $({\sigma_{xA \to bB}}/{ {\rm d}\Omega_{bB}} )^{\rm HOES}$, as
 \begin{align}
 \frac{ {\rm d}^{3}\sigma }{ {\rm d}\Omega_{bB}\,{\rm d}\Omega_{sF}\,{\rm d}E_{bB} }= KF\,|\varphi_{sx}({\rm {\bf p}}_{sx})|^{2}\,\Big(\frac{ {\rm d}\sigma_{xA \to bB} }{ {\rm d}\Omega_{bB} } \Big)^{\rm HOES}.
 \label{DCSPWIA1}
 \end{align}
All the variables are Galilean invariant.  $E_{bB}$ is the $b-B$ relative kinetic energy,  $\,{\rm {\bf p}}_{sx}\,$  is the $s-x$ relative momentum,
 $\,\Omega_{bB}\,$ is  the $b-B$ solid angle and $\,\Omega_{sF}\,$ is the relative solid angle of  $s$ and the center of mass (c.m.) of the $b+B$ system.
 One can also choose a different set of the Galilean  variables characterizing the fully DCS.  KF is a kinematical factor and $\varphi_{sx}({\rm {\bf p}}_{sx})$ is the $s-x$ bound-state wave function in the momentum space. 
 The doubly DCS  is half-off-the-energy-shell  (HOES)  because in the entry channel of the subreaction $x + A \to b+B$ 
particle $x$ transferred from $a$  is off-the-energy-shell (OFES), that is, $E_{x} \neq p_{x}^{2}/(2\,m_{x})$. 
In what follows the momentum of the virtual particle $x$ is denoted by ${\rm {\bf p}}_{x}$ to distinguish it from on-the-energy-shell (OES) momentum ${\rm {\bf  k}}_{x}$.  Since the particle $x$ is virtual, the relative $s-x$ momentum is denoted as
$\,{\rm {\bf p}}_{sx}\,$ to distinguish it from the OES momentum $\,{\rm {\bf k}}_{sx}\,$ for which $\,E_{sx}= k_{sx}^{2}/(2\mu_{sx})$;
$\,\mu_{sx}$ is the reduced mass of particles $s$ and $x$.  

Let us consider in more details the pole diagram. The amplitude of this diagram depends on five independent 
Galilean-invariant variables,  but one of them, the rotation angle around the momentum of the incident particle $A$
is not significant  and can be excluded leaving only four independent invariants  in the amplitude of  diagram \ref{fig_polediagram1}.     Hereafter, we refer to the diagram depicted in Fig.  \ref{fig_polediagram1} as diagram \ref{fig_polediagram1}.   These four invariants determine the complete kinematics of the THM reaction  $A + a \to b + B + s$. 
Usually, in THM experiments the angles and momenta of the final state particles $b$ and $B$ are measured.

From the energy-momentum  conservation law  we get
\begin{align}
\sigma_{x}  &=E_{x}   - \frac{ p_{x}^{2}  }{2\,m_{x} } 
\nonumber\\
&= E_{a} - E_{s}  - \varepsilon_{sx}  - \frac{  ({\rm {\bf k}}_{a}  -  {\rm {\bf k}}_{s})^{2}   }{2\,m_{x}  }
\nonumber\\
&
= E_{sx}  -  \frac{ p_{sx}^{2} }{2\,\mu_{sx} }   = -\frac{ p_{sx}^{2} +  {\kappa}_{sx}^{2}    }{2\,\mu_{sx}   },  
\label{sigmaxs1}
\end{align}
where $\sigma_{x}$   is the measure of the deviation of particle $x$ from OES (for the OES particle $\sigma_{x}=0$), $\,E_{sx}= - \varepsilon_{sx}$,  $\,\kappa_{sx}= \sqrt{2\,\mu_{sx}\,\varepsilon_{sx}  }$  is the $a=(s\,x)$ bound-state wave number,  $\varepsilon_{sx}= 
m_{s} + m_{x}- m_{a}$ is the binding energy for the virtual decay $a \to s+x$  and $m_{i}$  is the mass of particle $i$,  $\,\mu_{sx}$ is the reduced mass of particles $s$ and $x$.

Similarly, from the energy and momentum conservation law at the $x+A \to b+B$ four-ray vertex we obtain 
\begin{align}
\sigma_{x}= E_{xA} -  \frac{p_{xA}^{2}}{2\,\mu_{xA}}.
\label{sigmaxA1}
\end{align}

Then it is evident that, due to the virtual character of particle  $x$, $\,E_{sx}  \neq  p_{sx}^{2}/(2\,\mu_{sx} )$  and 
$\,E_{xA}  \neq p_{xA}^{2}/(2\,\mu_{xA})$. The real particles are $a$ and $A$ in the entry channel and particles $s, \,b$ and $B$ in the
exit channel, see Fig. \ref{fig_polediagram1}. The relative momenta  ${\rm {\bf p}}_{sx}$ and ${\rm {\bf p}}_{xA}$
are given by
\begin{align} 
{\rm {\bf p}}_{sx} = \frac{m_{x}\,{\rm {\bf k}}_{s} - m_{s}\, {\rm {\bf p}}_{x}}{m_{sx}},
\label{psx1}
\end{align}
and
\begin{align}
{\rm {\bf p}}_{xA} = \frac{m_{A}\,{\rm {\bf p}}_{x} - m_{x}\, {\rm {\bf k}}_{A}}{m_{xA}}.
\label{pxA1}
\end{align}
Here $m_{ij}= m_{i}+m_{j}$.
The relative momentum of real (OES) particles is
\begin{align}
{\rm {\bf k}}_{ij} = \frac{m_{j}\,{\rm {\bf k}}_{i} - m_{i}\, {\rm {\bf k}}_{j} }{m_{ij}}.
\label{kij1}
\end{align}
From Eqs. (\ref{sigmaxs1}) and (\ref{sigmaxA1})   we get one of the important energy-momentum relationships of the THM:
\begin{align}
E_{xA} = \frac{p_{xA}^{2}}{2\mu_{xA}} - \frac{p_{sx}^{2}}{2\mu_{sx}}-  \varepsilon_{sx} . 
\label{ExATHM1}
\end{align}
Thus in the THM, due to the virtual character of particle $x$,  we always have $ {p_{xA}^{2}}/{2\mu_{xA}}> E_{xA}$. In the quasi-free (QF) kinematics,
$p_{sx} = 0$ and the $x - A$ relative kinetic energy is
\begin{align}
E_{xA} = \frac{p_{xA(0)}^{2}}{2\mu_{xA}} - \varepsilon_{sx} . 
\label{ExATHMQF1}
\end{align}
where $p_{xA(0)}$ is the $x - A$ relative momentum in the QF kinematics.  
It is convenient to rewrite Eq. (\ref{ExATHM1})  in the reference frame where $k_{a}=0$:
\begin{align}
E_{xA}= \frac{m_{x}}{m_{xA}}\,E_{A}  + \frac{ {\rm {\bf k}}_{s} \cdot  {\rm {\bf k}}_{A} }{m_{xA} }
- \frac{k_{s}^{2}}{2\,\mu_{sx}} - \varepsilon_{sx}.
 \label{ExAkazero1}
 \end{align}

We introduce  the form factor 
\begin{align}
W_{a}({\rm {\bf p}}_{sx}) 
&= \int\,{\rm d}{\rm {\bf r}}_{sx}\,e^{i{\rm {\bf p}}_{sx} \cdot {\rm {\bf r}}_{sx}  }\,
V(r_{sx})\,\varphi_{a}({\rm {\bf r}}_{sx})                 \nonumber\\
&= -\frac{ p_{sx}^{2} +  {\kappa}_{sx}^{2}     }{2\,\mu_{sx}   }\,\varphi_{a}({\rm {\bf p}}_{sx}),
\label{Wa1}
\end{align}
where  $\varphi_{a}({\rm {\bf p}}_{sx})$  is the Fourier  transform  of the 
wave function  $\varphi_{a}({\rm {\bf r}}_{sx})$
of the bound state $a=(sx)$,   $\,{\rm {\bf r}}_{sx}$  is the radius-vector connecting the centers of mass 
of nuclei $s$ and $x$,
$\,V_{sx}(r_{sx})$  is their interaction potential.

If one of the particles $s$  or $x$ is a neutron, the form factor is regular  at  $p_{sx}^{2} +  {\kappa}_{sx}^{2} =0$, that is, 
\begin{align}
\varphi_{a}({\rm {\bf p}}_{sx})  =  - 2\,\mu_{sx}\,\frac{W_{a}({\rm {\bf p}}_{sx})}{p_{sx}^{2} +  {\kappa}_{sx}^{2}  }
\label{varphia1}
\end{align}
has  a  pole  at  $p_{sx}^{2} +  {\kappa}_{sx}^{2}  =0$.  However,   if both particles $s$ and $x$  are charged,  the potential $V_{sx}(r_{sx})$
includes both nuclear and Coulomb parts.  The latter modifies the behavior of  the form factor  at $p_{sx}^{2} +  {\kappa}_{sx}^{2}  =0$ to
\begin{align}
W_{a}({\rm {\bf p}}_{sx})  \stackrel{p_{sx}^{2} +  {\kappa}_{sx}^{2}  \to 0}{=}  [p_{sx}^{2} +  {\kappa}_{sx}^{2} ]^{\eta_{sx}}\,{\tilde W}_{a}({\rm {\bf p}}_{sx}),
\label{Wach1}
\end{align}
where ${\tilde W}_{a}({\rm {\bf p}}_{sx})$ is a function regular  at the singular point,  
\begin{align}
\eta_{ij}= \frac{  Z_{i}\,Z_{j}  \,e^{2}\,\mu_{ij}  }{ \kappa_{ij}    }
\label{etaij1}
\end{align}
is the Coulomb parameter of the bound state $\,(i\,j)$, $\,Z_{i}\,e\,$  is the charge of particle $i$.
The system of units in which $\hbar=c=1$ is used throughout the paper.
Hence the Fourier component of the bound-state wave  function $\,\varphi_{a}({\rm {\bf r}}_{sx})\,$  and, consequently,
the amplitude of diagram  {\ref{fig_polediagram1},   have the branching point singularity  rather than the pole one:
\begin{align}
\varphi_{a}({\rm {\bf p}}_{sx})  \stackrel{  p_{sx}^{2} +  (\kappa_{sx} )^{2}  \to 0  }{=} -2\mu_{sx}\,\frac{ {\tilde W}_{a}({\rm {\bf p}}_{sx})  }{[p_{sx}^{2} +  {\kappa}_{sx}^{2}]^{ 1- \eta_{sx}  }}.
\label{varphipsx1}
\end{align}                   
Therefore, the amplitude of  diagram \ref{fig_polediagram1}  is not  a pole in the presence of the Coulomb
interaction at the vertex  $a \to s+x$. Only if one of the particles $s$ or $x$ is a neutron, that is  $\eta_{sx}=0$,
the singularity at $p_{sx}^{2} +  {\kappa}_{sx}^{2} = 0$ turns into a pole. Strictly speaking, only in this case
diagram \ref{fig_polediagram1}  can be called a pole diagram. But historically, in the literature diagrams 
in which a single particle is transferred are called the pole ones. That is why we also will use the same name for the amplitude
of diagram \ref{fig_polediagram1} but keeping in mind that for charged particles $s$  and $x$ this amplitude has the branching 
point singularity rather than the pole one.

In all the performed THM experiments  the orbital angular momentum $l_{sx}$ of the bound state $(s\,x)$  is $0$.
Although the singularity at $p_{sx}^{2} +  {\kappa}_{sx}^{2}  =0$  is located in the unphysical region,
the modulus of the amplitude of diagram \ref{fig_polediagram1}  has a peak  at $p_{sx}=0$, which is called the quasi-free 
peak. However,  the Coulomb $s-x$ interaction  decreases the QF peak, see Eq. (\ref{varphipsx1}).  Thus the QF kinematics 
provides the best condition for the dominance of the diagram in Fig. \ref{fig_polediagram1}.

The second THM condition that ensures the validity of the so-called plane-wave impulse approximation  (PWIA) follows from Eq. (\ref{ExATHMQF1}):  $E_{xA} >> \varepsilon_{sx}$.
Then $p_{xA} \approx k_{xA}$  and the amplitude for the subreaction $x+A \to b+B$  reaches  its OES limit.  
Then 
in Eq. (\ref{DCSPWIA1}) the HOES doubly DCS
can be replaced by the OES doubly DCS.
Note that while the HOES DCS is actually  $({ {\rm d}\sigma_{xA \to bB}({\rm {\bf k}}_{bB},\,{\rm {\bf p}}_{xA}) }/{ {\rm d}\Omega_{bB} } )^{\rm HOES}$, the OES DCS used in the PWIA is $({ {\rm d}\sigma_{xA \to bB}({\rm {\bf k}}_{bB},\,{\rm {\bf k}}_{xA}) }/{ {\rm d}\Omega_{bB} } )^{\rm OES}$.

Note that  in the THM the TH  particle $a=(sx)$  is usually loosely bound. From the uncertainty principle it follows that for small binding energies the condition  $\,p_{sx} \lesssim \kappa_{sx}$    probes distances $r_{sx}  \gtrsim 1/\kappa_{sx}$
 where the nuclear interaction between $s$ and $x$ is depleted and particle $s$ in the final state can be treated as a spectator. 
 
 In practical applications of the THM, experiments are performed at some fixed value of $E_{aA}$. To cover a broader range of $E_{xA}$, events that deviate from the QF condition of $p_{sx}=0$ are selected within the interval  $p_{sx} \lesssim \kappa_{sx}$. 
 From  Eq.  (\ref{ExAkazero1}) it follows that this can be achieved  by varying both direction and magnitude of ${\rm {\bf k}}_{s}$. 
 
However, there are serious shortcomings of the PWIA. First, it neglects the Coulomb-nuclear interaction  of the particles in the initial and final states of the THM reaction, which often becomes very important, especially for heavier particles. 
The second serious limitation of the PWIA is that the main idea of the THM  is to apply it for astrophysically 
relevant $E_{xA}$ energies, which may be significantly smaller than $\varepsilon_{sx}$.    At such low $E_{xA}$  the OES PWIA approximation
for the doubly DCS (quasi-free approximation) for the subreaction $x+A \to b+B$ is not valid.

The indirect TH method is used for obtaining the astrophysical factors $S(E_{xA})$  for the resonant reactions 
\begin{align}
x+A \to F^{*} \to b+B. 
\label{THMressubreact1}
\end{align} 
To determine the resonant astrophysical factors for reaction (\ref{THMressubreact1}) a two-step THM  resonant 
\begin{align}
a+ A \to s+F^{*} \to s+b+B
\label{THMreaction1}
\end{align}
is used. Here $a=(sx)$ is the Trojan horse particle, 
$F^{*}$ is the intermediate resonance formed by the $x+A$ system. The THM resonant reactions are described by the diagram depicted 
 in Fig. \ref{fig_THMPWAdiagram}.

\begin{figure}[htbp]
  \includegraphics[width=0.6\columnwidth]{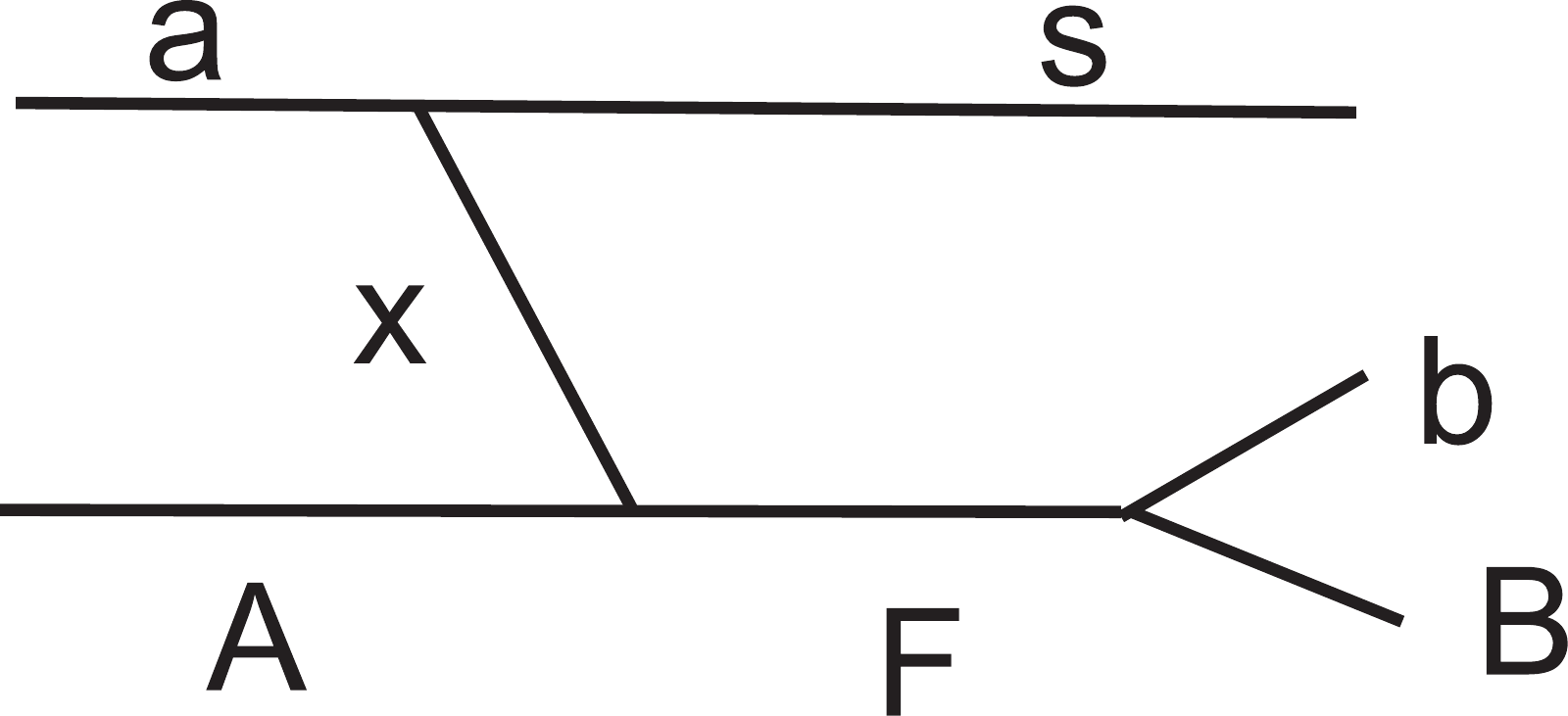}
\caption{The diagram describing the two-step THM resonant reaction in the PWA  }
\label{fig_THMPWAdiagram}
\end{figure}

In this review we address another approach based on the surface-integral formalism \cite{annals, muk2011} which does not require 
the assumption about the quasi-free sub-reaction $x+A \to b +B$ and explicitly takes into account the off-shell character of the transferred particle $x$ within the generalized $R$-matrix approach for three-body reactions. We present theoretical foundations of this approach
focusing on the THM resonant reactions and provide some calculations demonstrating our approach for different resonant reactions.

This review is organized  as follows.  Before addressing the theory of the two-step THM  reactions leading to three charged particles in the final state,  in Section \ref{Transferresonance1}  we present the theory for the first step of the THM  reaction, which is the transfer reaction,  based on the surface-integral formalism.  
The theoretical part is accompanied by calculations of the  ${}^{16}{\rm O}(d,\,p){}^{17}{\rm O}(1d_{3/2})$    
 transfer reaction populating a resonance state where we analyse contributions of  the internal and external terms in prior- and post-form adiabatic DWBA (ADWBA). These calculations show why the prior-form  ADWBA is more preferable than the post one when analysing transfer reactions populating resonance states.
 In  Section \ref{Gentheorsurrreact1}   we address a general theory of the two-step THM  resonant reactions based on a few-body approach. Section \ref{critanalalysisTHM1} presents a critical analysis of  the application of the THM method to obtain the astrophysical factor  for carbon-carbon fusion, which is
of one of the key astrophysical reactions.  In  Section \ref{subthreshres}  we review the theory of the THM for obtaining the astrophysical factors for the resonant reactions proceeding through subtheshold resonances. In particular, we discuss application of this theory to another key astrophysical reaction of ${}^{13}{\rm C}(\alpha,n){}^{16}{\rm O}$, believed to be 
the neutron generator in low mass AGB stars. In  Section \ref{THMradcapt} we focus on the application of  the THM  to measure resonant radiative-capture reactions at energies so low that direct measurements  can hardly be performed due to the negligibly small penetrability factor in the entry channel of the reaction. The developed theory is accompanied in Section \ref{Radcaptapplication1}   by the practical application for the astrophysical reaction  ${}^{12}{\rm C}(\alpha,\,\gamma){}^{16}{\rm O}$ via  indirect reaction ${}^{12}{\rm C}({}^{6}{\rm Li},d\,\gamma){}^{16}{\rm O}.$  Finally, Section  \ref{Summary1}   concludes the review with a brief summary.   

\section{Transfer reactions populating resonance state in the surface-integral approach}
\label{Transferresonance1}

\subsection{Theory of transfer reactions populating resonance states in the surface-integral formalism}
\label{Theorytransfer1}
Assume that one needs to obtain information about the binary resonant reaction  (\ref{THMressubreact1}),
where $F^{*}=(xA)$  is an isolated  resonance state of the charged particles $x$ and $A$, which decays into the channel $b+B$. 
The THM can provide this information  using the surrogate reaction (\ref{THMreaction1}).  The first step of the surrogate reaction is the transfer reaction  populating the resonance state $F^{*}$  and the second step is decay of the resonance $F^{*} \to b+B$.

Since the first step of the reaction (\ref{THMreaction1})  is the transfer of the particle $x$ populating the resonance state $F^{*}$, the amplitude of this transfer reaction is divergent.  To calculate it one needs to know how to handle the divergent amplitude. Two main problems complicate the  practical theory of stripping into resonance states:  (i) the numerical problem of the convergence of the matrix element in the distorted-wave Born approximation (DWBA) when the full transition operator is included and (ii) the ambiguity over what spectroscopic information can be extracted from the analysis of transfer reactions populating the resonance states. The purpose of this section is to address these two problems.  Our approach is based on the surface-integral formalism developed in 
\cite{annals,muk2011}. The reaction amplitude is parametrized in terms of the reduced width amplitudes related with the resonance widths, inverse level matrix, boundary condition and channel radius. These are the same parameters that are used in the conventional $R$-matrix method \cite{lanethomas}. For stripping to resonance states, many-level, and one- and two-channel cases are considered. The theory provides a consistent tool for analyzing binary resonant reactions and stripping to resonance states in terms of the same parameters.
 
Next we apply the surface-integral formalism to describe the transfer reaction populating resonance states. 
Using the DWBA prior form  we  obtain the generalized DWBA $R$-matrix amplitude for the transfer reaction  to resonance states assuming two-channel, multi-level resonance in the subsystem $x+A$. 
The prior-form of the DWBA amplitude for the first step of the THM reaction 
\begin{align}
a+ A \to s + F^{*}
\label{traAsF1}
\end{align}
populating the resonance state $F^{*}$ is
\begin{widetext}
\begin{align}
M_{{M_F}{M_s};{M_A}{M_a}}^{DW(prior)}({\rm {\bf k}}_{sF},{{\bf{k}}_{aA}}) = &  \sum\limits_{m_{s_{xA}} m_{l_{xA}}M_{x} }\,\,\big<s_{xA}m_{s_{xA}} \,l_{xA}m_{l_{xA}} \big| J_{F}M_{F}\big>\big<J_{x}M_{x}\,\,J_{A}M_{A} \big| s_{xA}m_{s_{xA}}\big>    \nonumber\\
& \times \big<s_{sx}m_{s_{sx}}\,l_{sx}m_{l_{sx}}\,\, \big |\,J_{a}M_{a}\big>\big<J_{s}M_{s}\,\,J_{x}M_{x} \big| s_{sx}m_{s_{sx}}\big>
{\cal M}^{DW(prior)},
\label{MDWBACoul1}
\end{align}
\end{widetext}
with
\begin{align}
{\cal M}^{DW(prior)}({\rm {\bf k}}_{sF},\,{\rm {\bf k}}_{aA})              
 = \big<\chi_{sF}^{(-)}\,\Upsilon_{xA}^{(-)}|
 \,\Delta\,{\overline V}_{aA}|\varphi_{a}\,\chi_{aA}^{(+)}\big>,
\label{dwpriorres1}
\end{align}
where  
$\chi_{aA}^{(+)}$ is the distorted wave describing the relative motion in the initial channel with the outgoing boundary condition, $\chi_{sF}^{(-)}$ is the distorted wave describing the relative motion in the final channel with the incoming boundary condition,   $\,{\rm {\bf k}}_{ij}$ is the $i-j$ relative 
momentum.  Also, $s_{ij}$ $(m_{s_{ij}})$ is the channel spin (its projection) in the channel $i-j$, $l_{ij}$ $ (m_{l_{ij}})$ is the relative orbital angular momentum of particles $i$ and $j$, $J_{i}$ $(M_{i})$ is the spin (its projection) of particle $i$.

The transition operator in the prior form is
\begin{align}
\Delta\,{\overline V}_{aA} = U_{sA} + {\overline V}_{xA} - U_{aA}.
\label{DeltaVpr1}
\end{align}
Here, $U_{ij}$ is the $i-j$ optical potential and ${\overline V}_{xA} = \big<\varphi_{A}|V_{xA}|\varphi_{A}\big> $ is the mean-field real potential supporting the resonance state in the system  $F=(x\,A)$,
$\,\varphi_{i}$ is the bound-state wave function of nucleus $i$. 
$\Upsilon_{xA}^{(-)}=\big<\varphi_{A}|\Psi _{xA}^{(-)} \big>$ 
is the overlap of the  resonant wave function $\Psi_{xA}^{(-)}$ in the channel $x+A$ and the bound-state wave function of nucleus  $\,\varphi_{A}$. 
We neglect the internal degrees of freedom of nucleus $x$.

One can also use the post-form of the DWBA amplitude
\begin{align}
{\cal  M}^{DW(post)}({\rm {\bf k}}_{sF},\,{\rm {\bf k}}_{aA})               
= \big<\chi_{sF}^{(-)}\,\Upsilon_{xA}^{(-)}|
 \, \Delta\,{ V}_{sF}  |\varphi_{a}
 \,\chi_{aA}^{(+)}\big>,
\label{dwpostres1}
\end{align}
with the post-form transition operator given as
\begin{align}
\Delta\,{
V}_{sF} = U_{sA} + V_{sx} -U_{sF} ,
\end{align}
where $\,V_{sx}$ is the $s-x$ potential supporting the $a=(sx)$ bound state. However, we prefer to use the prior-form due to the faster  convergence over variable ${\rm {\bf r}}_{xA}$, which is the coordinate vector conjugate of the relative momentum ${\rm {\bf k}}_{xA}$ (see section \ref{Calculations1}).

The matrix element in Eq. (\ref{dwpriorres1}) involves integration over variable $r_{xA}$. 
Following Refs. \cite{muk2011} and \cite{muk2014} we split the matrix element \eqref{dwpriorres1} 
into an internal part with $r_{xA} \leq R_{ch}$ and the external part with $r_{xA} > R_{ch}$:
\begin{align}
{\cal M}^{DW(prior)}({\rm {\bf k}}_{sF},\,{\rm {\bf k}}_{aA}) 
=& M_{int}^{DW(prior)}({\rm {\bf k}}_{sF},\,{\rm {\bf k}}_{aA})   \nonumber\\   
& + M_{ext}^{DW(prior)}({\rm {\bf k}}_{sF},\,{\rm {\bf k}}_{aA}),
\label{dwpriorintextres1} 
\end{align}
with 
\begin{align}
{\cal M}_{int}^{DW(prior)}&({\rm {\bf k}}_{sF},\,{\rm {\bf k}}_{aA})           \nonumber\\                        
&=  \big<\chi_{sF}^{(-)}\,\Upsilon_{xA}^{(-)}\,|\,\Delta\,{\overline V}_{aA}|\varphi_{a}\,\chi_{aA}^{(+)}\big>\Big |_{r_{xA} \leq R_{ch}}   
\label{dwpriorintres1}
\end{align}
and
\begin{align}
{\cal M}_{ext}^{DW(prior)}&({\rm {\bf k}}_{sF},\,{\rm {\bf k}}_{aA})         \nonumber\\   
&= \big<\chi_{sF}^{(-)}\,\Upsilon_{xA}^{(-)}\,|\,\Delta\,{\overline V}_{aA}|\varphi_{a}\,\chi_{aA}^{(+)}\big>\Big |_{r_{xA} > R_{ch}}.
\label{dwextpriorres1}
\end{align}
Here $R_{ch}$ is the channel radius, which is chosen so that at $r_{xA} > R_{ch}$ the nuclear $x-A$ interaction can be neglected and the $V_{xA}$ potential can be replaced by the Coulomb term $V_{xA}^{C}$.  Since $s$ and $x$ are close to each other due to the presence of the bound-state wave function  $\varphi_{sx}$  the $V_{sA}$ potential can also be approximated by the Coulomb term.  Then in the external region over the variable ${\rm {\bf r}}_{xA}$ we have
 \begin{align}                                 
&{\cal M}_{ext}^{DW(prior)}                \nonumber\\
& =\big<\Psi_{{\rm {\bf k}}_{sF}}^{C(-)}\,\Upsilon_{xA}^{(-)}\big| V_{sA}^{C} + V_{xA}^{C} - U_{aA}^{C} \big| \varphi_{sx}\,\Psi_{{\rm {\bf k}}_{aA}}^{C(+)}\big>\Big|_{r_{xA} > R_{ch}}.
\label{LDWext1}
\end{align}
In this amplitude $\Upsilon_{xA}^{(-)}$ can be replaced by the resonant wave function in the external region 
given by Eq. (36) from \cite{mukfewbody2019}:
\begin{align}
{\tilde \phi}_{R(xA)}(r_{xA})  =&  e^{  -i\,\delta^{p}_{s_{xA}l_{xA}J_{F}}(k_{0(xA)}) }\, \sqrt{  \frac{ \mu_{xA}   }{ k_{0(xA)}  }\,\Gamma_{(xA)}     }               \nonumber\\
& \times \frac{ O^{*}(k_{0(xA)}, r_{xA})}{ r_{xA}  },
\label{tildephi1}
\end{align}
where $\delta^{p}_{s_{xA}l_{xA}J_{F}}(k_{0(xA)})$ is the $x-A$ potential (non-resonant)  scattering phase shift 
with the quantum numbers of the resonance $F^{*}$,  $k_{0(xA)}$ is the real part of the $x-A$ resonance momentum,  $O(k_{0(xA)}, r_{xA})$ is the outgoing Coulomb scattering wave.

The external matrix element ${\cal M}_{ext}^{DW(prior)}$ in the prior form is small and in some cases with a reasonable choice of the channel radius $R_{ch(xA)}$ can even be neglected \cite{muk2011}. However, it is important for analysis of the stripping to resonance states because the external part in the post form does not converge. In this sense the usage of the prior form in the external part has a clear benefit. 

The splitting of the amplitude into the internal and external parts in the subspace over the $x-A$ coordinate ${\rm {\bf r}}_{xA}$ is necessary to rewrite the prior DWBA amplitude in the generalized $R$-matrix approach for stripping to resonance states. 
The main contribution to the prior form amplitude ${\cal M}^{DW(prior)}$ comes from the internal part ${\cal M}_{int}^{DW(prior)}$. Since the internal part is given by a volume integral, its calculation requires the knowledge of $\Upsilon_{xA}^{(-)}$ in the internal region. The model dependence of this function in the nuclear interior ($r_{xA} \leq R_{ch}$), where different coupled channels contribute, brings one of the main difficulties  and leads to the main uncertainty in the calculation of the internal matrix element. 
Using a surface-integral representation we can rewrite the volume integral of the internal matrix element in terms of the volume integral in the post form and the dominant surface integral taken over the sphere at $r_{xA}=R_{ch}$. With a reasonable choice of the channel radius $R_{ch}$ the contribution from the internal volume integral in the post form can be minimized to make it significantly smaller than the surface matrix element. The latter can be expressed in terms of the $R$-matrix parameters such as the observable reduced width amplitude (observable resonance width), boundary condition and channel radius.

For the purpose of transforming ${\cal M}_{int}^{DW(prior)}$ into a surface integral in the subspace over variable ${\rm {\bf r}}_{xA}$ we rewrite the transition operator in the internal region as
\begin{align}
\Delta & {\overline V}_{aA} = U_{sA} + {\overline V}_{xA} - U_{aA}                 \nonumber\\
& = [{\overline V}_{xA} + U_{sF}] + (U_{sA} + V_{sx} - U_{sF}) - [V_{sx} + U_{aA}]. 
\label{DeltaVdAint1}
\end{align}
The bracketed transition operators are the potential operators in the Schr\"odinger equations for the initial and final channel wave functions. Hence, for the internal prior form of the DWBA we obtain 
\begin{align}
{\cal M}_{int}^{DW(prior)}({\rm {\bf k}}_{sF},\,{\rm {\bf k}}_{aA})           
=& {\cal M}_{int}^{DW(post)}({\rm {\bf k}}_{sF},\,{\rm {\bf k}}_{aA})              \nonumber\\
&+ {\cal  M}_{S}^{DW}({\rm {\bf k}}_{sF},\,{\rm {\bf k}}_{aA}),
\label{priorDWBAres}
\end{align}
where 
\begin{align}
{\cal M}_{int}^{DW(post)}&({\rm {\bf k}}_{sF},\,{\rm {\bf k}}_{aA})  \nonumber\\
&= \big<\chi_{sF}^{(-)}\,\Upsilon_{xA}^{(-)}
 \,|\,\Delta\,{V}_{sF}| 
\varphi_{a}\,\chi_{aA}^{(+)}\big>\Big |_{r_{xA} \leq R_{ch}}
\label{intdwpostres1}
\end{align}
is the internal post-form DWBA amplitude and
\begin{align}
{\cal M}_{S}^{DW}({\rm {\bf k}}_{sF},\,{\rm {\bf k}}_{aA}) &= - \big<\chi_{sF}^{(-)}\,\Upsilon_{xA}^{(-)}|\,{\overleftarrow {\hat T}} - {\overrightarrow {\hat T}}|\,\varphi_{a}\,\chi_{aA}^{(+)}\big>    \nonumber\\
&= - \big<\chi_{sF}^{(-)}\,\Upsilon_{xA}^{(-)}|\,{\overleftarrow {\hat T}}_{xA} - {\overrightarrow {\hat T}}_{xA}|\,\varphi_{a}\,\chi_{aA}^{(+)}\big>
\label{Sintdwprior1}
\end{align}
is the surface term. Here ${\hat T}$ is the total kinetic energy operator in the c.m. of the reaction and  ${\hat T}_{xA} $ is the kinetic energy operator of the relative motion of particles $x$ and $A$. Arrows indicate the direction in which the differential kinetics energy operator acts. It is important to note that with a proper choice of the optical potential $U_{sF}$ the matrix element ${\cal M}_{int}^{DW(post)}$ can be minimized so that its model dependence would not have impact on the total matrix element ${\cal M}^{DW(prior)}$.  That is why in what follows we disregard 
 ${\cal M}_{int}^{DW(post)}$.

We show now how to simplify $M_{S}^{DW}({\rm {\bf k}}_{sF},\,{\rm {\bf k}}_{aA})$ by 
reducing it to a surface-integral form. In the surface-integral form the resonance overlap function  $\Upsilon_{xA}^{(-)}$  is replaced with the $(xA)$ real  resonance wave function ${\tilde \phi}_{{ R}(xA)}$
given by Eq. (\ref{tildephi1}).

We use a three-body model of the constituents $s,\,x$ and $A$, all assumed to be structureless particles. In a more general approach we need to introduce the projection operators to ensure that particles $x$ and $A$ are in the ground states in the intermediate states of the transfer reaction. In this case the bound-state wave function $\varphi_{a}$ and the resonance wave function ${\tilde \phi}_{{R}(xA)}$ should be replaced by the overlap functions. These overlap functions can be approximated by the product of the two-body wave functions and the square roots of the corresponding spectroscopic factors. Since in the THM only the energy dependence of the DCS are measured, these spectroscopic factors can be dropped. Here we also use the two-body wave functions rather than the overlap functions.
Then
\begin{widetext}
\begin{align}
{\cal M}_{S}^{DW(prior)} =& \big<\Psi_{{\rm {\bf k}}_{sF}}^{C(-)}\,{\tilde \phi}_{{R}(xA)}\big| { {\overleftarrow {\hat T} }}_{xA}
- { {\overrightarrow {\hat T} }}_{xA} \big| \varphi_{a}\,\Psi_{{\rm {\bf k}}_{aA}}^{C(+)}\big>\Big|_{r_{xA}=R_{ch}}
\nonumber\\
=&\frac{{R_{ch}^2}}{{2{\mu _{xA}}}}\int {d{{\bf{r}}_{sF}}} \Psi _{ - {{\bf{k}}_{sF}}}^{C( + )}({{\bf{r}}_{sF}})\int d {\Omega _{{{\bf{r}}_{xA}}}}
\nonumber \\ & \times
{\Big[{\varphi _{a}}({{\bf{r}}_{sx}})\,\Psi _{{{\bf{k}}_{{\bf{aA}}}}}^{C( + )}({{\bf{r}}_{aA}})\frac{{\partial {\phi _{R(xA)}}({{\bf{r}}_{xA}})}}{{\partial {r_{xA}}}} - {\phi _{R(xA)}}({{\bf{r}}_{xA}})\frac{{\partial {\varphi _{a}}({{\bf{r}}_{sx}})\Psi _{{{\bf{k}}_{{\bf{aA}}}}}^{C( + )}({{\bf{r}}_{aA}})}}{{\partial {r_{xA}}}}\Big]\Big|_{{r_{xA}} = {R_{ch}}}}.
\end{align}

Then the external part of the
amplitude can be expressed in terms of the resonance width $\Gamma _{xA}$   and reduces to
 \begin{align}                                  \nonumber\\
{\cal M}_{ext}^{DW(prior)}=&e^{i\,\delta^{p}_{s_{xA}l_{xA}J_{F}}(k_{0(xA)})}\, {\sqrt{ \frac{\mu_{xA}}{k_{0(xA)}}\Gamma_{(xA)} }}\, \big<\Psi_{{\rm {\bf k}}_{sF}}^{C(-)}\,\frac{O^{*}(k_{0(xA)}, r_{xA})}{r_{xA}}\,\big| V_{sA}^{C} + V_{xA}^{C} - U_{aA}^{C} \big| \varphi_{a}\,\Psi_{{\rm {\bf k}}_{aA}}^{C(+)}\big>\Big|_{r_{xA} > R_{ch}}.
\label{LDWext2}
\end{align} 

Thus, the internal matrix element consists of two terms, the internal post-form Coulomb DWBA amplitude ${\cal M}_{int}^{DW(post)}$ and
the surface term ${\cal M}_{S}^{DW(prior)}$.  The internal Coulomb or Coulomb+nuclear DWBA in the post form
should be small due to the highly oscillatory behavior of the binned resonance wave functions (this will be demonstrated in  part  \ref{DWBA1}). Note also that the smaller the resonance energy, the smaller is the contribution of the internal region.
Then the dominant contribution to the matrix element ${\cal M}^{DW(prior)}$ comes from the surface term ${\cal M}_{S}^{DW(prior)}$   and   
${\cal M}_{ext}^{DW(prior)}$.

We transform now the surface matrix element into zero-range DWBA amplitude. To this end we use
\begin{align}
{\rm {\bf r}}_{aA}= {\rm {\bf r}}_{xA} + \frac{m_{s}}{m_{sx}}\,{\rm {\bf r}}_{sx},  \qquad
{\rm {\bf r}}_{sF}=  \frac{m_{A}}{m_{xA}}\,{\rm {\bf r}}_{xA} +{\rm {\bf r}}_{sx}.
\label{raArsF1}
\end{align}
Rewriting the wave functions $\Psi _{{{\rm {\bf k}}_{aA}}}^{C( + )}({{\rm {\bf r_{aA}}}})\,$ and $\Psi _{ - {{\rm {\bf k}}_{sF}}}^{C( + )}({{\rm {\bf r}}_{sF}})$  in the momentum space we get
\begin{align}
 {\cal M}_{S}^{DW(prior)} =& \frac{{R_{ch}^2}}{{2{\mu _{xA}}}} \int {d{{\rm {\bf r}}_{sF}}}\, \int\,\frac{ {\rm d}{\rm {\bf p}}_{sF} }{(2\,\pi)^{3} }\,
\int\,\frac{{\rm d}{\rm {\bf p}}_{aA}}{(2\,\pi)^{3}}\,
\Psi_{-{{\rm {\bf k}}_{sF}}}^{C( + )}({{\rm {\bf p}}_{sF}})\,\Psi _{{{\rm {\bf k}}_{aA}}}^{C( + )}({{\rm {\bf p_{aA}}}})\,{\varphi _{a}}({{\rm {\bf r}}_{sx}}) {e^{-i{\rm {\bf p}}_{sx} \cdot {{\rm {\bf r}}_{sx}}}}
\nonumber\\
& \times \int d{\Omega _{{\rm {\bf r}}_{xA}}}\,                                                                                      \Big[{e^{i{{\bf{p}}_{xA}} \cdot {{\bf{r}}_{xA}}}}\,\frac{{\partial {{\phi}_{{R}(xA)}}({{\bf{r}}_{xA}})}}{{\partial {r_{xA}}}} - {{\phi}_{{R}(xA)}}({{\bf{r}}_{xA}})\,\frac{{\partial \,{e^{i{{\bf{p}}_{xA}} \cdot {{\bf{r}}_{xA}}}}}}{{\partial {r_{xA}}}}{\Big]\Big|_{{r_{xA}} = {R_{ch}}}},
\label{LSprior1}
\end{align}
where
\begin{align}
{\rm {\bf p}}_{xA}=  {{\rm {\bf p}}_{aA}} -  \frac{{{m_A}}}{{{m_F}}}{{\rm {\bf p}}_{sF}} , \qquad  {\rm {\bf p}}_{sx}={{\rm {\bf p}}_{sF}} - \frac{{{m_s}}}{{{m_a}}}\,{{\rm {\bf p}}_{aA}}.
\label{pxApsx1}
\end{align}
Taking into account the fact that $r_{xA} = R_{ch}$ is larger than the nuclear interaction radius we replace the relative momentum  ${\rm {\bf p}}_{xA}$ with the momentum ${\rm {\bf k}}_{xA}= {{\rm {\bf k}}_{aA}} - \frac{{{m_A}}}{{{m_F}}}{{\rm {\bf k}}_{sF}}$, which is expressed in terms of the OES momenta ${{\rm {\bf k}}_{aA}}$  and ${{\rm {\bf k}}_{sF}}$. However, the $x-A$ relative momentum ${\rm {\bf k}}_{xA}$ is OFES because the transferred particle $x$ is OFES.

 We consider ${\cal M}_{S}^{DW(prior)}$ at the real part of the $(xA)$ resonance energy, i.e., $k_{xA}= k_{0(xA)}$ and $k_{sF} = k_{0(sF)}$.
Then returning  to the coordinate-space representation for ${\cal M}_{S}^{DW(prior)}$ we get
\begin{align}
& {\cal M}_{S}^{DW(prior)} = \frac{{R_{ch}^2}}{{2{\mu _{xA}}}}\,{\cal M}^{DWZR(prior)}                                                                                     
\int d\Omega_{{\rm {\bf r}} _{xA}}\,\Big[{e^{ - i{{\bf{k}}_{xA}} \cdot {{\bf{r}}_{xA}}}}\,\frac{{\partial {{\phi} _{{R}(xA)}}({{\bf{r}}_{xA}})}}{{\partial {r_{xA}}}} - {\phi _{{R}(xA)}}({{\bf{r}}_{xA}})\,\frac{{{e^{ - i{{\bf{k}}_{xA}} \cdot {{\bf{r}}_{xA}}}}}}{{\partial {r_{xA}}}}{\Big]\Big|_{{r_{xA}} = {R_{ch}}}}.
\label{LSprior2}
\end{align}
\end{widetext}
Here,
\begin{align}
 {\cal M}^{DWZR(prior)}=& \int {d{{\rm {\bf r}}_{sx}}}\, \Psi_{ - {{\rm {\bf k}}_{0(sF)}}}^{C( + )}({{\rm {\bf r}}_{sx}})\,{\varphi _{a}}({{\rm {\bf r}}_{sx}})
\nonumber\\&\times  
\Psi_{{\rm {\bf k_{aA}}}}^{C( + )}\left(\frac{{{m_s}}}{{{m_a}}}{\rm {\bf r}}_{sx}\right)
\label{MDZRpr1}
\end{align}
is the DWBA amplitude, which does not depend on the resonant wave function ${\tilde \phi}_{{R}(xA)}$  and $V_{xA}$ potential. This equation looks like the zero-range DWBA (ZRDWBA). However, in contrast to the standard zero-range approximation, Eq. (\ref{MDZRpr1}) can be used for arbitrary value of the orbital momentum of the resonance state $(xA)$.
Note that replacing in Eq. (\ref{MDZRpr1}) the distorted waves by the plane waves leads to the PWA introduced in \cite{reviewpaper} and used in \cite{Nature}.

Integrating in Eq. (\ref{LSprior2}) over $\,\Omega_{{\rm {\bf r}}_{xA}}\,$ and using again Eq. (\ref{tildephi1})  for the external resonant wave function we arrive at the surface term of the DWBA reaction amplitude singled out using  the surface-integral formalism:
\begin{widetext}
\begin{align}
 {\cal M}_{S}^{DW(prior)} =& {i^{ - {l_{xA}}}} {e^{ - i\,{\delta^{p}_{s_{xA}l_{xA}J_{F}}}({k_{0(xA)}})}}\,\sqrt {\frac{1}{{{\mu _{xA}}\,{k_{0(xA)}}}}\,{\Gamma _{xA}}} \, \frac{1}{2}\,O_{{l_{xA}}}({k_{0(xA)}}{R_{ch}})\,j_{{l_{xA}}}({k_{0(xA)}}{R_{ch}})               \nonumber\\
& \times {\cal W}_{l_{xA}}\,{Y_{{l_{xA}},{m_{{l_{xA}}}}}}({\widehat {\bf{k}}_{0(xA)}})\,{\cal M}^{DWZR(prior)}.
\label{MDWBACoul2}
\end{align}
\end{widetext}

Note that from the energy conservation  in the transfer reaction
\begin{align}
E_{aA}- \varepsilon_{sx}= E_{sF}+ E_{xA}
\label{encons1}
\end{align}
it follows that when $E_{xA}$ approaches the complex resonance energy $E_{R(xA)}$ while the energy $E_{sF}$ 
reaches the complex energy $E_{R(sF)}$, which corresponds to the resonance in subsystem $F=(xA)$.    
Hence $k_{0(sF)}$ is the real part of the complex $s-F$ relative momentum $k_{R(sF)}$ corresponding to the $k_{0(xA)}$, which is the real part of the complex resonance $x-A$ momentum  $k_{R(xA)}$.
The off-shell factor ${\cal W}_{l_{xA}}$ is written as
\begin{align}
{\cal W}_{l_{xA}} =& \Big[
 j_{{l_{xA}}}({k_{0(xA)}}{r_{xA}})\Big[R_{ch}\,\frac{ {\partial {\ln[O_{{l_{xA}}}}({k_{0(xA)}}{r_{xA}})]}}{{\partial {r_{xA}}}}
-1\Big] 
\nonumber\\& - R_{ch}\,\frac{\partial{{\ln\,[j_{{l_{xA}}}}({k_{0(xA)}}{r_{xA}})]}}{{\partial {r_{xA}}}}\Big]{\Big|_{{r_{xA}} = {R_{ch}}}},
\label{calWr1}
\end{align}
where  $O_{l_{xA}}$ is the outgoing spherical wave and $j_{l_{xA}}$ is the spherical Bessel function.

Equation (\ref{MDWBACoul2})  is a very important result. It contains the off-shell factor  ${\cal W}_{l_{xA}}$  reflecting  the virtual character of the transferred particle $x$. It also contains the boundary conditions expressed in terms of the logarithmic derivatives and generalizes the $R$ matrix for method for binary reactions.  Hence in the surface-integral formalism the OSE limit of the transferred particle  is not required.   

Thus in the surface-integral formalism the dominant contribution to the DWBA amplitude of reaction (\ref{traAsF1}) in the prior  form  is 
given by the sum of the dominant surface and external terms and the final expression for the prior-form DWBA amplitude is
\begin{widetext}
\begin{align}
M_{{M_F}{M_s};{M_A}{M_a}}^{(prior)}({k_{0(sF)}}{{\bf{\hat k}}_{sF}},{{\bf{k}}_{aA}}) \approx & \sum\limits_{m_{s_{xA}} m_{l_{xA}}M_{x} }\,\,\big<s_{xA}m_{s_{x A}} \,l_{xA}m_{l_{xA}} \big| J_{F}M_{F}\big>\big<J_{x}M_{x}\,\,J_{A}M_{A} \big| s_{xA}m_{s_{xA}}\big>                       \nonumber\\
& \times \big<s_{sx}m_{s_{sx}}\,l_{sx}m_{l_{sx}}\,\, \big |\,J_{a}M_{a}\big>\big<J_{s}M_{s}\,\,J_{x}M_{x} \big| s_{sx}m_{s_{sx}}\big>\,\Big[ {\cal M}_{S}^{DW(prior)}  +  {\cal M}_{ext}^{DW(prior)} \Big].
\label{MDSextpr1}
\end{align}
\end{widetext}
This expression can be used for the analysis of the transfer reaction  (\ref{traAsF1}). To analyze the THM reaction (\ref{THMreaction1})  one can use Eq. (\ref{MDSextpr1}) with the added part describing the second step of the THM reaction.

An important feature of  Eq. (\ref{MDSextpr1})  is that  it is expressed in terms of the resonance width  in the decay channel $F^{*} \to x+A$.  The amplitude $\, M_{{M_F}{M_s};{M_A}{M_a}}^{(prior)}({k_{0(sF)}}{{\bf{\hat k}}_{sF}},{{\bf{k}}_{aA}})\,$ is the amplitude of the transfer reaction (\ref{traAsF1}) in the surface-integral approach, which is the  first step of the THM reaction.  
If we add the second step then we will be able to express the
THM reaction amplitude in terms of the  OSE astrophysical factor despite   the presence of the off-shell factor.
There is one more important feature of Eq. (\ref{MDSextpr1})  to be noted.  The amplitude $M_{{M_F}{M_s};{M_A}{M_a}}^{(prior)}({k_{0(sF)}}{{\bf{\hat k}}_{sF}},{{\bf{k}}_{aA}})$, according to Eqs. (\ref{MDWBACoul2})  and  (\ref{LDWext2}), is expressed in terms of  $\sqrt{\Gamma _{xA}}$. As we will see below in Section \ref{DoublDCS1}, this allows one to single out the astrophysical $S$ factor from the THM doubly DCS.

\subsection{Numerical results for reaction $\mathbf{{}^{16}{\rm O}(d,\,p){}^{17}{\rm O}(1d_{3/2})}$}
\label{Calculations1}

In this section we present DWBA DCS calculations for a particular transfer reaction populating the resonance state, which corroborate our theoretical findings although the code for the surface-integral formalism is not yet available.  We select  the reaction
${}^{16}{\rm O}(d,\,p){}^{17}{\rm O}(1d_{3/2})$  at $E_{d}=36$ MeV populating a resonance state of energy
$E_{x}=5.085$ MeV, which corresponds to the resonance level at $0.94$ MeV. 
In all the calculations shown below we use the single-particle approach for the $n-A$ resonant scattering wave function calculated  in the Woods-Saxon potential with the radial parameter $r_{0}=1.25$ fm and diffuseness $a=0.65$ fm.

We  compare the post- and prior-form  calculations  following the procedure developed in Section \ref{Transferresonance1}.  The post and prior  adiabatic distorted-wave approach (ADWA) and prior  coupled-channel  Born approximation (CCBA) are used for comparison. 
For the prior ADWA amplitude we use Eq. (\ref{dwpriorres1}). The prior ADWA is the standard  prior DWBA  in which the initial
deuteron potential  is given by the sum of the  optical  $U_{pA}$ and $U_{nA}$ potentials calculated at half of the deuteron energy using the zero-range Johnson-Sopper prescription \cite{johnson70}. 
The resonance scattering wave function $\Psi_{nA}^{(-)}$ is taken in the form of the binned function for the resonant partial wave \cite{FRESCO}  which has asymptotically both incident and outgoing waves. 
    Hence, the resonant wave function  used in FRESCO code is different from the one considered in Eq. (\ref{Gamowwfi1}). The latter has the outgoing wave and used in subsection \ref{DWBApr1}.

In the CCBA the final $n+A$ resonant wave function is  given by the binned resonant wave function $\chi_{ {\rm {\bf k}}_{pF}({\bf k}_{nA})  }^{(res)(-)}({\rm {\bm r}}_{pF})$ which  is coupled with two bound states in ${}^{17}{\rm O}$:  the ground state $1d_{5/2}$ and the first excited state $2s_{1/2}$.  Schematically we can write the final-state wave function in CCBA as
\begin{widetext}
\begin{align}
\Psi_{f}^{CDCC(-)}({\rm {\bm r}}_{pF},\,{\rm {\bf r}}_{nA}) =& \,\varphi_{nA}^{(0)}({\rm {\bf r}}_{nA})\,\chi_{{\rm {\bf k}}_{pF}}^{(0)(-)}({\rm {\bm r}}_{pF})  +  \varphi_{nA}^{(1)}({\rm {\bf r}}_{nA})\,\chi_{{\rm {\bf k}}_{pF}}^{(1)(-)}({\rm {\bm r}}_{pF})                            
+ {\overline \psi}_{ {\rm {\bf  k}}_{nA},\,l_{nA}=3 }^{(res)(-)}({\rm {\bf r}}_{nA})\,\chi_{ {\rm {\bf k}}_{pF}({\bf k}_{nA})  }^{(res)(-)}({\rm {\bm r}}_{pF}).
\label{cdccwf12}
\end{align}
\end{widetext}
Here, for simplicity, we omitted spins. The radial and momentum spherical harmonics are absorbed into 
${\overline \psi}_{ {\rm {\bf  k}}_{nA} }^{(res)(-)}({\rm {\bf r}}_{nA})$. 

In Fig.  \ref{fig_Rx17O36MeVadwaccba} we present the ratio $R_{x}$  of the ADWA and CCBA DCSs for the deuteron stripping to resonance  $\,{}^{16}{\rm O}(d,\,p){}^{17}{\rm O}(1d_{3/2})\, $  at $\,E_{d}=36\,$ MeV calculated at  zero proton scattering angle with cutoff over $r_{nA}$  to the total   ADWA and CCBA DCSs  calculated at  zero proton scattering angle  without cut-off. This figure  provides a very important  justification for the surface integral  formalism used in subsection   \ref{Theorytransfer1}.   
The dark blue short dashed line and light-blue long dashed-dotted line show the ratios $\,R_{x}\,$ of the zero-angle prior ADWA and CCBA DCSs, respectively, to the full DCS, in which the  radial integral over $r_{nA}$  is  calculated  for $r_{nA} \geq  r_{nA}^{min}$ while $r_{nA}^{max}$  is extended to infinity.  
Note that $r_{nA}^{min}$  ($r_{nA}^{max}$)  determines the low-limit (upper limit) of the matrix element  radial integral.   
As we see,  when $r_{nA}^{min}$   varies from $0$ until $2$ fm the  prior DCS does not change. Hence we can start the integration in the radial matrix element from $r_{nA}^{min}=2$  fm. The fact that for $r_{nA}^{min} \geq  5 $ fm the prior DCS becomes negligible means that the upper limit of the integration in the radial matrix element is $r_{nA}^{max} =5$  fm  and we can use the radial
integral of the matrix element in the interval $2  \leq r_{nA} \leq 5$ fm.  Thus the prior form  is  quickly converging
justifying the use the prior form when treating  transfer reactions populating resonant states.  The same conclusion 
we can drawn from the analysis  of the magenta dotted and green dashed lines.  To obtain these lines 
the matrix element  is calculated from $r_{nA}^{min}=0$ to $r_{nA}^{max}$, where $r_{nA}^{max}$ varies.  From Fig. 
\ref{fig_Rx17O36MeVadwaccba}  it follows that we can use $r_{nA}^{max}= 5$ fm. We also see that  at $r_{nA}^{max}= 2$ fm  the prior DCS vanishes.  Again, as concluded for the dark-blue short-dashed and light-blue long dashed-dotted lines,
we can calculate for the magenta dotted and green dashed lines that the radial matrix element can be calculated from $r_{nA}^{min}=2$ fm to  $r_{nA}^{max}= 5$ fm.
The surface term (\ref{MDWBACoul2}) obtained in subsection \ref{Theorytransfer1} allows one to  replace the volume matrix element  calculated from $r_{nA}^{min}=2$ fm to  $r_{nA}^{max}= 5$ fm by the surface term calculated at 
$R_{ch}=5$ fm.  The surface term approximation is based on the assumption that the internal post form at $r_{nA} \leq 5$ fm becomes negligible. It follows from the observation of the solid red line  shown in Fig. \ref{fig_Rx17O36MeVadwaccba}, which represents the post form. Small oscillations at $r_{nA} \geq 5$ fm, which are seen in the magenta dotted and green dashed lines, are presented by the external term  (\ref{LDWext2}).
  Another important  justification  for using of  the prior formalism  is  a poor convergence of the post form, which follows from the observation of the solid red line. 

\begin{figure}
\includegraphics[width=\columnwidth]{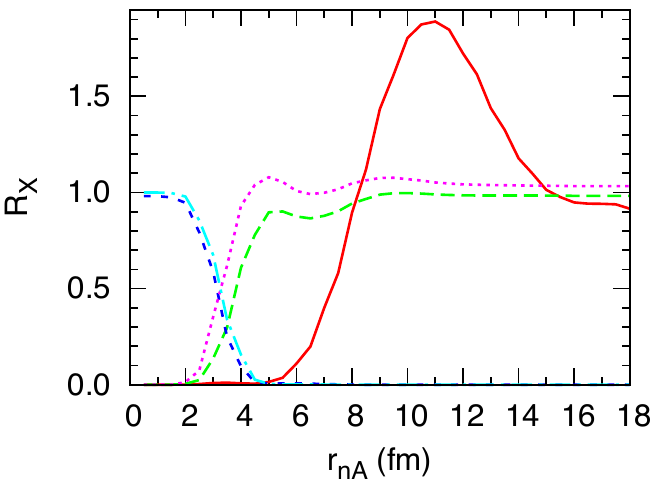}
\caption{(Color online) Dependence of the normalized  ADWA and CCBA  DCS ratios $\,R_{x}\,$  on  $\,r_{nA}\,$   for the deuteron stripping to resonance  $\,{}^{16}{\rm O}(d,\,p){}^{17}{\rm O}(1d_{3/2})\, $  at $\,E_{d}=36\,$ MeV.   Dark blue short dashed line and light-blue long dashed-dotted line - the ratios $\,R_{x}\,$ of the zero-angle prior ADWA and CCBA DCSs, respectively, in which the  radial integral over $r_{nA}$  is  calculated  for  
    $r_{nA} \geq  r_{nA}^{min}$, to the full DCS. 
Similarly, magenta dotted and green dashed lines are the ratios $\,R_{x}\,$ of the  zero-angle prior ADWA and CCBA DCSs, respectively, in which the  radial integral over $r_{nA}$  is  calculated  in the interval  $0 \leq  r_{nA} \leq  r_{nA}^{max}$,  to the full DCS.
 The red solid  line is the  $R_{x}$ dependence on 
 $r_{nA}^{max}$ calculated for the post ADWA form.  Hence $r_{nA}$  on the abscissa  is   $\,r_{nA}^{min}$ for the blue short and long dashed lines  and $r_{nA}^{max}$ for the dotted magenta, dashed green  and solid red  lines. 
 { First published in \cite{muk2014}}. } 
\label{fig_Rx17O36MeVadwaccba}
\end{figure}

\section{General theory of surrogate reactions 
in a few-body approach}
\label{Gentheorsurrreact1}

In the previous section we developed the DWBA theory of the transfer reactions populating a resonance state. 
In this section within a genuine three-body approach we consider theory of transfer reactions proceeding through a resonance state in the intermediate subsystem and leading to three particles in the final state \cite{mukfewbody2019}. All the initial, intermediate and three-body final-state
interactions are taken into account. We derive the fully differential cross section (DCS) in terms of the transfer reaction amplitude obtained in the previous section. 

We begin by considering the surrogate reaction \eqref{THMreaction1}.
In nuclear physics, such surrogate reactions are used in the THM allowing one to obtain a vital astrophysical information about a binary resonant subreaction \cite{Baur1986,Spitaleri1999,reviewpaper,spitaleri2019}.
In the THM the initial channel is $a + A$, where $a = (sx)$ is the Trojan Horse (TH) particle, rather than just $x + A$. 
Collision $a+A$ is followed by the transfer reaction $a+ A \to s+ F^{*}$ populating a resonance state $F^{*}$ in the subsystem 
$F=x+A$, which decays to another channel $b + B$. In the THM reaction (\ref{THMreaction1}), this results in three particles $b + B + s$ 
in the final state, rather than the two-particle final state $b +B$ in the binary reaction. 

The indirect THM allows one to bypass the Coulomb barrier issue in the system $x+A$ by using reaction (\ref{THMreaction1}) 
rather than the binary reaction (\ref{THMressubreact1}). However, the employing of the surrogate reaction leads to complications of 
both experimental and theoretical nature. Experimentally, coincidence experiments are needed. From the theoretical point of view,
the presence of the third particle $s$ can affect the binary reaction  (\ref{THMressubreact1}), especially if the particle $s$ 
is charged. In this case, it interacts with the intermediate resonance $F^{*}$ and with the final products $b$ and $B$ via the Coulomb forces, which are very important at low energies and larger charges of the participating nuclei. 

The THM reactions induced by collision of the  ${}^{14}{\rm N}$ and ${}^{12}{\rm C}$ nuclei lead to three charged particles in the final state. 
Recently, the THM reaction 
\begin{align}
{}^{14}{\rm N} + {}^{12}{\rm C}  \to  d+ {}^{24}{\rm Mg}^{*} \to d + b +B,
\label{14N12Creact1}
\end{align}
where $b= p$ and $\alpha$, $B={}^{23}{\rm Na}$ and ${}^{20}{\rm Ne}$, respectively, was used to obtain an information about the ${}^{12}{\rm C} + {}^{12}{\rm C}$ fusion \cite{Nature}, which is a very important reaction in nuclear astrophysics  \cite{RolfsRodney}.
In this reaction the Coulomb effects should have dramatic effect on the cross section. In particular, the post-collision Coulomb interaction may have a crucial impact on the DCSs  \cite{Muk1991}. 

We use a few-body approach to derive the amplitude and the fully DCS of the surrogate reaction
(\ref{THMreaction1}) taking into account the Coulomb interaction of the particle $s$ with the resonance $F^{*}$ in the intermediate state 
and the Coulomb interactions of $s$ with $b$ and $B$ in the final state. Our final results reveal a universal effect of the Coulomb interaction  important for both nuclear and atomic surrogate reactions.
For nuclear surrogate reactions the short-range nuclear interactions should be taken into account alongside the long-range Coulomb interactions. However, the consideration of the nuclear surrogate reaction is simplified due to the fact that the nuclear rescattering effects in the intermediate  $s- F^{*}$ state and within the $s-b$ and $s-B$ pairs in the final state give rise to diagrams that can be treated as a background and disregarded for narrow resonances. 

The main goal of the THM is to investigate an energy behavior of  the fully DCS which is needed to determine the astrophysical factors \cite{reviewpaper}. The obtained fully DCS allows one to investigate the Coulomb effects on the resonance line shape
both in atomic and nuclear collisions. To calculate the energy dependence of the THM fully DCS one needs to calculate the DCS of the transfer reaction. To treat the final-state three-body Coulomb effects we use the paper \cite{muk1985}.
In this work the formalism of the three-body Coulomb asymptotic states (CAS) was used to calculate the reaction amplitudes with three charged particles in the final state. 

\subsection{The amplitude of the breakup reaction in a few-body approach proceeding through a resonance in the intermediate subsystem}
\label{Breakupampl1}

Let us consider the  surrogate reaction (\ref{THMreaction1}), which is the two-step THM reaction.
The difficulty with the analysis of such a reaction stems from the fact that the resonance decays into  the channel 
$b+ B$, which is different from the entry channel $x+A$ of the resonant subreaction. 
As we mentioned earlier, the goal of the THM  is to extract from the TH  reaction the  astrophysical factor
for the resonant  rearrangement reaction \eqref{THMressubreact1}.

To simplify considerations we neglect the spins of the particles with the relative orbital angular momenta $l$ in the pairs $x+A$ and $b+B$  set $\,l=0$.  For brevity of the notation we omit the angular momenta keeping in mind that their values are zero.  
The starting expression for the  breakup reaction amplitude in the c.m.  of the few-body system can be written as 
\begin{align}
{ M}=  
\blb {\psi_{ {\rm {\bf k}}_{B},{\rm {\bf k}}_{b}  }^{(0)}}
\bl\blb X_{f}\bl{ U}_{0A}\bl\varphi_{a}\,\varphi_{A}\blk \bl
{\psi_{{\rm {\bf k}}_{aA}}^{(0)}}\big>,
\label{Reacampl1}
\end{align}
where  $\,X_{f}= \varphi_{B}\,\varphi_{b}$. The particle $s$  in the THM is a spectator and we can disregard its internal structure and treat it as a  structureless point-like particle.
Wave function  $\,\psi_{ {\rm {\bf k}}_{aA}}^{(0)}$ represents the plane-wave describing the relative motion of the noninteracting  particles $a$ and $A$ in the initial state of the THM reaction,  $\,\psi_{{\rm{\bf k}}_{B},\, {\rm {\bf k}}_{b}}^{(0)}\,$ is the  three-body plane wave of  particles $s,$ $\,b$ and $B$ in the final state, ${\rm {\bf k}}_{i}$ is the momentum of particle $i$. 
Note that for the moment we use for the charged particles screened Coulomb potentials. 
The transition operator $\, U_{0A}\,$ corresponds to the breakup reaction from the initial channel $\,a+A\,$ to the final three-body channel $\,s+ b+B$.  The two-fragment partition $\,\alpha + (\beta\,\gamma)\,$ with free particle $\,\alpha\,$ and the bound state $\,(\beta\,\gamma)\,$ is denoted by the free particle index $\,\alpha\,$.

The transition operator $\,U_{0A}\,$ satisfies the equation
\begin{align}
{ U}_{0A}=  V_{f}  +   V_{f}\,G\,{\overline V}_{sx}.
\label{U0A1}
\end{align} 
Here and in what follows we use the following notations:  $V_{i\,j} = V_{i\,j}^{N} +V_{i\,j}^{C}$  is the interaction potential between particles $i$ and 
$j$ given by the sum of the nuclear and Coulomb potentials,  ${\overline V}_{1\,2}  = V_{1\,3} + V_{2\,3}$, and  $\;V_{f}= V_{bB} + V_{sB} + V_{sb}$.

As we mentioned above, the difficulty of the problem is due to the fact that in the TH process the final three-body system $\,s+b+B$, which is formed after the resonance  decay $ F^{*} \to b+B$,  is different from the initial three-body system $s+x+A$  before the resonance $F^{*}$  was formed. This  change  is caused by the rearrangement resonant reaction  (\ref{THMressubreact1}). 
The full Green's function resolvent  in Eq. \eqref{U0A1} is
\begin{align}
G= \frac{1}{z - {\hat T}_{sF} -V_{sF} - {\hat H}_{F}},
\label{Greenfunction1}
\end{align}
where  $\,V_{sF}$  is the $s-F$ interaction potential,  $\,{\hat H}_{F}$ is the internal Hamiltonian of the system $\,F= x+A=b+B$.

Our final  goal is to single out the resonance term  in the subsystem $F=x+A=b+B$,  which generates a peak in the fully DCS of the breakup reaction (\ref{THMreaction1}), to obtain the reaction amplitude corresponding to the diagrams shown in Fig. \ref{fig_resdiagram1}. 
\begin{figure}[htbp]
  \includegraphics[width=0.8\columnwidth]{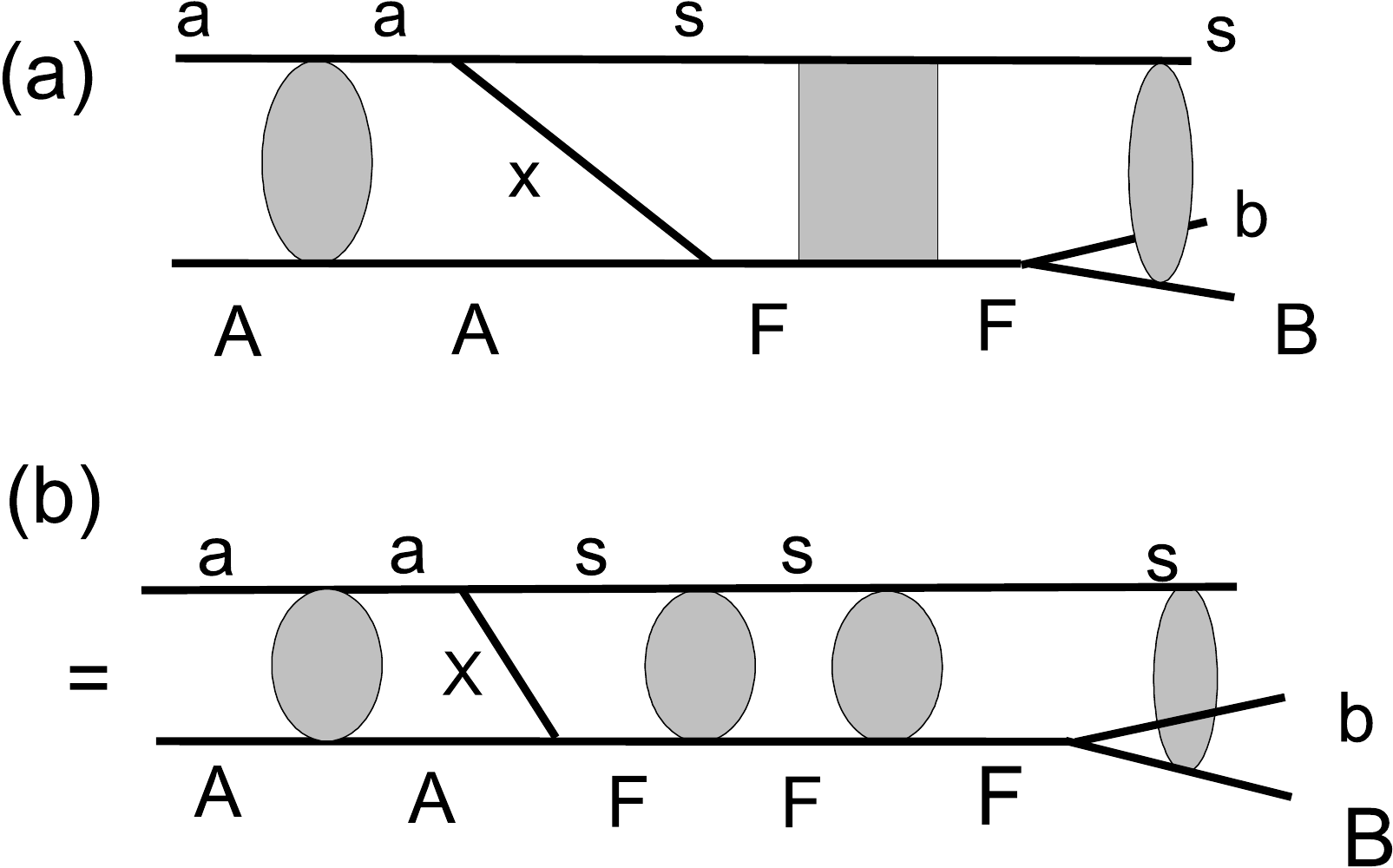}
\caption{The diagrams describing the TH mechanism including the Coulomb interactions in the initial, final and intermediate states. The grey bulb on the left side of diagram (a) is the Coulomb $a+A$ scattering in the initial channel described by the Coulomb scattering wave function. The grey rectangle is the Green's function in the final state describing the propagation of the system $s+F$, where $F$ is the resonance. The grey bulb on the right side describes the intermediate state three-body Coulomb interaction given by the three-body Coulomb wave function. In diagram (b), the Green's function is replaced by its spectral decomposition, which includes the Coulomb scattering wave functions (grey bulbs) describing the Coulomb rescattering of the spectator $s$ in the intermediate state.  }
\label{fig_resdiagram1}
\end{figure}

First we single out  the resonance term  with all the Coulomb rescatterings in the initial, intermediate and final states.
To this end one can rewrite  ${ U}_{0A}\,$ as
\begin{align}
{ U}_{0A}=&  V_{f}  +   V_{f}\,G\,{\overline V}_{sx}
\nonumber\\
=& V_{f} + (V_{bB}+  {\overline V}_{bB}^{C})\,G\,{\overline V}_{sx}                            
+ {\overline  V}_{bB}^{N}\,G\,{\overline V}_{sx},
\label{U0A2}
\end{align}
where ${\overline V}_{bB}^{N,C}= V_{sb}^{N,C}   +  V_{sB}^{N,C}$.
The resonance term in the subsystem $F$, which can be singled out from  the full Green's function $G,\,$  will be smeared out by nuclear interactions 
$\,{\overline V}_{bB}^{N}= V_{sb}^{N} + V_{sB}^{N}.\,$   Hence the term  $ {\overline  V}_{bB}^{N}\,G\,{\overline V}_{sx}$  does not produce the resonance peak in the TH reaction amplitude generated by the resonance in the subsystem $F\,$  and can be treated as a background. That is why the  only term  in Eq. (\ref{U0A2}), which is responsible for  a resonance behavior of the TH reaction amplitude caused  by the resonance in the subsystem $F,\,$  is
\begin{align}
 { U}_{0A}^{{\rm R}} =  (V_{bB}+ {\overline V}_{bB}^{C} )\,G\,{\overline V}_{sx} .
 \label{U0AR3}
 \end{align}
 Note that the Coulomb potentials $\, V_{sB}^{C}$ and $\,V_{sb}^{C}$ do not smear out the resonance in the subsystem $\,F,\,$ which can be singled out from $\,G$ \cite{blokh84}.
  
 Substituting  ${ U}_{0A}^{{\rm R}}$  into the matrix element (\ref{Reacampl1})  instead of $\,{ U}_{0A}\,$  leads to
\begin{align}
{ M}' = 
\blb{{\overline \Phi}_{bB}^{C(-)}}
\bl\blb{X_f}\bl{V_{bB}\,G\,{\overline V}_{sx}}\bl\varphi_{a}\,\varphi_{A}\blk\bl
{ \psi_{{\rm {\bf k}}_{aA}}^{(0)}}\blk,
\label{Min2}    
\end{align}
where $
{\overline V}_{sx}= V_{sB}^{C}  + V_{sb}^{C}$.
To obtain Eq. (\ref{Min2})  the two-potential formula is applied. The wave function $\,{\overline \Phi}_{bB}^{C(-)}$  in the bra state is a solution of the Schr\"odinger equation
\begin{align}
{\overline \Phi}_{bB}^{C(-)*}\big( E_{f} - {\overleftarrow {\hat T}}_{sbB} -  V_{sB}^{C} - V_{sb}^{C} \big) = 0,
\label{3BCSheq1}
\end{align}
with the incoming-wave boundary condition. Here the operator ${\overleftarrow {\hat T}}_{sbB}\,$ acting to the left is the total kinetic energy operator  of  the three-body system $s+b+B$ and   $\,E_{f}$ is the total kinetic energy of the system $\,s+b+B$  in the  c.m.  system. 

Now rewriting the potential $\,{\overline V}_{sx}\,$ as $\,{\overline V}_{sx} = {\overline V}_{sx}^{N} + {\overline V}_{sx}^{C} $   and applying again the two-potential formula, we reduce Eq. (\ref{Min2}) to
 \begin{align}                                     
{ M}'=  
\blb{{\overline \Phi}_{bB}^{C(-)}}\bl\blb{X_f}\bl{V_{bB}\,G\,{\overline V}_{sx}^{N}}\big|\varphi_{a}\,\varphi_{A}\blk\bl
{\Psi_{{\rm {\bf k}}_{aA}}^{C(+)} }\blk.
\label{Mf1}
\end{align} 
Here, $\,\Psi_{{\rm {\bf k}}_{aA}}^{C(+)}\,$ is the $\,a+A\,$  Coulomb scattering wave function, in the Coulomb potential  ${\overline V}_{sx}^{C}= V_{sA}^{C} + V_{xA}^{C}$ with the outgoing-wave boundary condition.

It is important to explain again why we are taking into account only the Coulomb distortions in the initial and final states rather than the Coulomb plus nuclear distortions. 
Firstly, as it will be shown below the Coulomb rescatterings  transform the resonance pole into the branching point without smearing out the resonance peak.
Secondly, in the THM, as used currently \cite{reviewpaper}, only the energy dependence of the  extracted astrophysical factor  is measured  while its absolute value 
is determined by the normalization to the available direct data.
Thirdly, the Coulomb rescatterings in the initial, intermediate and final states, in contrast to the nuclear distortions, can significantly modify  the energy dependence of the THM DCS for the reactions at sub-Coulomb and near the Coulomb barrier energies and, therefore, must  be taken into account. 

In the first step of the THM reaction the particle $x$, being in the ground state, is transferred from the bound state $a=(s\,x)$ to the  nucleus $A$ forming a resonance $\,F^{*}= x+A$.  
The THM reaction amplitude, in which the particle $x$ is transferred in the ground state,
 can be obtained from Eq. (\ref{Mf1}):
\begin{align}                                     
{ M}'=  
\blb{{\overline \Phi}_{bB}^{C(-)}}\bl\blb{X_f}\bl{V_{bB}\,G\,{\overline V}_{sx}^{N}}\bl X_{i}\blk \bl I_{x}^{a}\,
{\Psi_{{\rm {\bf k}}_{aA}}^{C(+)} }\blk,
\label{Mf11}
\end{align} 
where $X_{i}= \varphi_{x}\,\varphi_{A}$ and $\varphi_{x}$ is the bound-state wave function in the ground state. We also introduce the overlap function of the bound-state wave functions of nuclei $a$ and $x$:
\begin{align}
I_{x}^{a}(r_{sx}) = \blb \varphi_{x}(\xi_{x}) \bl \varphi_{a}(\xi_x; r_{sx}) \blk,
\label{ovfIax1}
\end{align}
which is the projection of the bound-state wave function $\,\varphi_{a}\,$ on the two-body channel $\,s+x$.
The integration in the matrix element is carried over the internal coordinates $\,\xi_{x}\,$ of nucleus $x$. 
Since we assume that the relative  $\protect{s-x}$ orbital angular  momentum  in the bound  state $\,a=(sx)\,$ equals zero the overlap function depends on $\,r_{sx}\,$, the radius connecting the c.m. of nuclei $\,s\,$ and $\,x$, rather than on $\,{\rm {\bf r}}_{sx}.\,$  We recall that the internal degrees of freedom of the spectator $\,s$ are neglected.

Singling out the resonance contribution $F^{*}$   in the intermediate state of the two-step THM reaction  (\ref{THMreaction1})
 requires lengthy tedious transformations. These are given in Appendix \ref{spectraldecGrfunct1}.

As described in Appendix \ref{spectraldecGrfunct1},  the amplitude of two-step  THM reaction Eq. (\ref{THMreaction1}) reads as
\begin{align}
{\cal M}_{\rm R} &= -\frac{ i}{4\,\pi}\,\sqrt{\frac{\mu_{bB}}{k_{{\rm R}(bB)}  }}
\frac{e^{i\,\delta^{p}_{s_{xA}l_{xA}J_{F}}(k_{0(bB)})}\Gamma_{bB}^{1/2}\,{\cal M}_{(tr)} }{E_{0(bB)} -E_{bB} - i \frac{\Gamma}{2}}                    \nonumber\\
& \times N_{\rm C}(E_{bB},\,\zeta).
\label{RAL1}
\end{align}
Here ${\cal M}_{(tr)}= M_{{M_F}{M_s};{M_A}{M_a}}^{(prior)}({k_{0(sF)}}{{\bf{\hat k}}_{sF}},{{\bf{k}}_{aA}})$ is the amplitude of the reaction (\ref{traAsF1}) , which is the first step of the THM reaction  (\ref{THMreaction1}). 
$M_{{M_F}{M_s};{M_A}{M_a}}^{(prior)}({k_{0(sF)}}{{\bf{\hat k}}_{sF}},{{\bf{k}}_{aA}})$ is
 given by Eqs.  (\ref{MDSextpr1}), (\ref{LSprior2})  and (\ref{LDWext2}).

Also
\begin{align}
 N_{\rm C}(E_{bB},\,\zeta) =& \frac{ \Gamma(1+ i\,\eta_{bs})\,\Gamma(1 + i\,\eta_{Bs})}
 {\Gamma\Big(1+i[\eta_{bs} + \eta_{Bs}] \Big)   }
 \nonumber \\ &\times
 F\Big( - i{\eta _{Bs}},\,-i{\eta _{bs}},1;\,-1 \Big) (-\gamma(0))^{i\,\eta_{bs}}
   \nonumber\\ 
& \times (-\nu)^{i\,\eta_{Bs}}\,\left(E_{0(bB)} -E_{bB} - i \frac{\Gamma}{2}\right)^{-i\,\zeta}
 \label{N1}
\end{align}
is the Coulomb renormalization factor which is equal to unity when the Coulomb interactions are turned off. 
Note that Eq. (\ref{RAL1})  is the three-body generalization of the resonant DWBA reaction amplitude  obtained in \cite{muk2011}. It explicitly takes into account the Coulomb $s-F$ interaction in the intermediate state and the final-state $s-b-B$ three-body Coulomb interaction not accounted for in Ref. \cite{muk2011}.
Also here
\begin{align}
\zeta=\eta _{bs} + \eta _{sB} - \eta _{\rm R},
\label{zeta1}
\end{align}
$\nu= -k_{0(sF)}{\rm {\bf {\hat k}}}_{sF} \cdot  {\rm {\bf  k}}_{sB}  + \,k_{0(sF)}\, k_{sB},$
$\gamma(0)= -k_{0(sF)}{\rm {\bf {\hat k}}}_{sF} \cdot  {\rm {\bf k}}_{sb} + \,k_{0(sF})\,k_{sb}$     
can be obtained  from  Eqs. (62)-(64)  from \cite{mukfewbody2019} assuming the narrow resonance. 
In addition, $\eta_{{\rm R}}= Z_{s}\,Z_{F}\,\alpha\,\mu_{sF}/k_{{\rm R}}$ is the  Coulomb parameter 
of the interaction of the spectator $s$ and the resonance $F^{*}$ in the intermediate state,  
$\alpha =e^{2}/(\hbar\,c)$  is the fine-structure  constant,
$\eta_{ij}=(Z_{i}\,Z_{j}\,\alpha\,\mu_{ij}/k_{ij}$ is the Coulomb parameter of the particles $i$ and $j$,
$\,E_{sF} - E_{{\rm R}}= E_{{\rm R}(bB)} - E_{bB} = E_{0(bB)} - E_{bB} - i\,\Gamma/2 $,    
$\, E_{{\rm R}}= k_{{\rm R}}^{2}/(2\,\mu_{sF})$, $k_{{\rm R}}$ is given by Eq. (\ref{k0sF1}),
$E_{sF}$ is  relative kinetic energy of the particle $s$ and the c.m. of the system $b+B$ in the final state, and $E_{bB}$ is the $b-B$ relative kinetic energy in the final state. 

Thus, using a few-body approach we derived the expression for the TH reaction proceeding though the resonance in the binary subsystem. The intermediate $s-F^{*}$ and the final-state three-body Coulomb interactions have been taken into account explicitly using the three-body formalism. 
The THM reaction amplitude  proceeding through the intermediate resonance in the binary subsystem $F$ has in the denominator the resonant energy factor  $E_{0(bB)} - E_{bB} -i\,\Gamma/2$. However, a conventional Breit-Wigner resonance pole 
$( E_{0(bB)} - E_{bB} - i\,\Gamma/2)^{-1}$ is converted into the branching point singularity
 $(E_{0(bB)} - E_{bB} -i\,\Gamma/2)^{-1- i\,\zeta}$. This transformation of the resonance behavior of the THM reaction amplitude is caused by the Coulomb interaction of the particle $s$ with the resonance in the intermediate state and with products $b$ and $ B$ of the resonance in the final state.

The final-state Coulomb effects have a universal feature and should be taken into account 
whether one considers nuclear reactions leading to the three-body final states or in atomic processes. If $\,\eta_{bs},\,\eta_{Bs}\,$ and ${\rm Re}(\eta_{{\rm R}})\,$ have the same sign, the Coulomb $\,s-F^{*}\,$ interaction in the intermediate state weakens the impact of the final state Coulomb $s-b$ and $s-B$ interactions. This is  because the intermediate state Coulomb parameter $\eta_{\rm R}$ is subtracted from the final-state Coulomb parameters $\,\eta_{bs}+\eta_{Bs}$, see Eq. (\ref{zeta1}).
For example, if the Coulomb $s-F^{*}$ interaction in the intermediate state is turned off, that is $\eta_{{\rm R}} =0$, the resonance behavior of the TH reaction amplitude coincides with that from Ref. \cite{Senashenko}  where the angular and energy  dependences  of the electrons formed from the autoionization resonances induced by collisions of fast protons with atoms were investigated.

\subsection{Triply  DCS}
\label{TRdoubleDCS1}

Now we present some useful equations for the fully and doubly DCSs which can be used in the analysis of the THM resonant reactions. 
The fully DCS at $k_{bB} \to k_{0(bB)}$ is given by \cite{dol73}
\begin{align}
\frac{{{d^3}\sigma }}{{d{\Omega _{{{bB}}}}d{\Omega _{{{sF}}}}d{E_{sF}}}} = \frac{{{\mu _{aA}}{\mu _{sF}}}}{{{{(2\pi )}^3}}}\frac{{{k_{0(sF)}}}}{{{k_{aA}}}}\frac{k_{0(bB)}}{\mu_{bB}}{\overline {\big|{\cal M}_{ R}\big|^{2}}},
 \label{TripleDCSspin1}
\end{align}
where
\begin{widetext}
\begin{align}
\overline {\big|{\cal M}_R{\big|^2}} =& \frac{1}{ {\widehat J}_{a} {\widehat J}_{A}  }\sum\limits_{ {M_B}{M_b}{M_s}{M_a}{M_A} } \big|  M_{ {M_B}{M_b}{M_s};{M_A}{M_a} }(k_{0}{\rm {\bf {\hat k}}}_{sF},{\rm {\bf k}}_{bB},{\rm {\bf k}}_{aA}) \big|^2            \nonumber\\
 =& \frac{1}{ {\hat J}_{a} {\hat J}_{A} }\sum\limits_{  M_{F} M_{F}' M_{A} M_{a} M_{s} }\,M_{{M_F}{M_s};{M_A}{M_a}}^{(prior)}({k_{0(sF)}}{{\bf{\hat k}}_{sF}},{{\bf{k}}_{aA}})\,[M_{ M_{F}^{'}{M_s};{M_A}{M_a} }^{(prior)}({k_{0(sF)}}{{\bf{\hat k}}_{sF}},{{\bf{k}}_{aA}})]^*             \nonumber\\
& \times \frac{|N_{\rm C}|^2}{  ( E_{0} - E_{sF} )^{2}  + {\Gamma ^2}/{4}  }\sum\limits_{ M_{B}M_{b} } W_{ M_{B} M_{b} }^{M_{F}} ({\rm {\bf k}}_{{0}(bB)})\, \left[W_{ M_{B} M_{b} }^{M_{F}'}({\rm {\bf k}}_{{0}(bB)}) \right]^*.
\label{THMreactampl1}
\end{align}
Here $ M_{ {M_B}{M_b}{M_s};{M_A}{M_a} }(k_{0}{\rm {\bf {\hat k}}}_{sF},{\rm {\bf k}}_{bB},{\rm {\bf k}}_{aA}) $   is the amplitude of the two-step THM reaction  (\ref{THMreaction1})  with three particles in the final state and 
$M_{{M_F}{M_s};{M_A}{M_a}}^{(prior)}({k_{0(sF)}}{{\bf{\hat k}}_{sF}},{{\bf{k}}_{aA}})$ is the amplitude of the transfer reaction (\ref{traAsF1}), which is the first step of the THM reaction, and is defined by Eq. (\ref{MDSextpr1}).
Term
\begin{align}
W_{ M_{B} M_{b}}^{M_{F}}({k_{0(bB)}}) = & \sqrt{4\,\pi}
\sum\limits_{ s_{bB}l_{bB} m_{s_{bB}}m_{l_{bB}} } 
\big<s_{bB}m_{s_{bB}} \,l_{bB}m_{l_{bB}} \big| J_{F}M_{F}\big>\big<J_{b}M_{b}\,\,J_{B}M_{B} \big| s_{bB}m_{s_{bB}}\big>\,Y_{{l_{bB}}{m_{{l_{bB}}}}}({{\rm {\bf k}}_{0(bB)}})                   \nonumber\\
 & \times   e^{i\,\delta^{p}_{s_{bB}l_{bB}J_{F}}        (k_{0(bB)})}   \sqrt{\frac{\mu_{bB}\,\Gamma_{(bB)}}{k_{{0}(bB)}}}.
\label{vertexformfctor1}
\end{align}
\end{widetext}
represents the vertex form factor for the resonance decay $F^{*} \to b+B$.
The notations for spin-angular variables have been introduced in section  \ref{Transferresonance1}.

Taking into account that
\begin{align}
\big|\Gamma[1+ i\,\eta] \big|^{2} = \frac{\pi\,\eta}{\sinh(\pi\,\eta)}
\label{Gammash1}
\end{align}
we get for the Coulomb renormalization factor $N_{\rm C}$ \cite{mukfewbody2019}:
\begin{align}
| N_{\rm C}(E_{bB},\,\zeta)|^{2}= & \frac{{\sinh[\pi ({\eta _{sb}} + {\eta _{sB}})]}}{{\sinh(\pi {\eta _{sb}})\sinh(\pi {\eta _{sB}})}}\,\frac{{\pi {\eta _{sb}}{\eta _{sB}}}}{{({\eta _{sb}} + {\eta _{sB}})}}
\nonumber \\& \times 
\frac{{\pi {\eta _\zeta }}}{{\sinh(\pi {\eta _\zeta })}}
|F( - i{\eta _{sB}}, - i{\eta _{sb}},1; - 1){|^2} 
\nonumber \\& \times 
{\exp{\left[2\zeta \arctan \frac{{2({E_{0(bB)}} - {E_{bB}})}}{\Gamma }\right]}}.
\label{NC21}
\end{align}

\subsection{Doubly DCS of transfer reaction populating resonance state}
\label{DoublDCS1}

It is convenient to integrate the fully DCS over $\,\Omega_{{bB}}\,$ to get the doubly DCS, which is expressed
in terms of the DCS of the reaction (\ref{traAsF1}),
corresponding to the first step of the TH reaction.
However, in the case under consideration, due to the presence of the Coulomb renormalization factor $ N_{\rm C}(E_{bB},\,\zeta)$, the DCS obtained from integrating the fully DCS over $\,\Omega_{{bB}}\,$ cannot be expressed in terms of the DCS of the first step. The reason is that $ N_{\rm C}(E_{bB},\,\zeta)$ depends on the integration variable $\,\Omega_{{bB}}$.  However, in the following cases one can neglect this dependence: 
\begin{enumerate}
\item 
When $\,|\eta_{sb}| \ll  1\,$ and $\,\eta_{sB} \approx \eta_{{0(sF)}},\,$ where $\eta_{0(sF)}= Z_{s}\,Z_{F}\,\alpha\,\mu_{sF}/k_{0(sF)}$,
the imaginary part of $\,\eta_{{ R}}\,$ can be neglected because of the narrow resonance. In this case, $\,|N_{\rm C}| \approx 1\,$ and the integration over $\,\Omega_{{sF}}\,$ can be performed without any complications. \\
\item
 When $|\eta_{sb}| \ll 1$ and $m_{B} \gg m_{s},\,m_{b}$. 
 By choosing the Galilean momenta
$\,{\rm {\bf k}}_{s}= {\rm {\bf k}}_{sF}\,$ and $\,{\rm {\bf k}}_{bB}$ as independent variables, one can write
\begin{align}
{\rm {\bf k}}_{sB}=  \frac{m_{B}\,M}{m_{sB}\,m_{bB}}\,{\rm {\bf k}}_{sF} + \frac{m_{s}}{m_{sB}}\,{\rm {\bf k}}_{bB}
\approx {\rm {\bf k}}_{sF}.
\label{ksB1}
\end{align}
Then $\eta_{sB}= Z_{s}\,Z_{B}\,\alpha\,\mu_{sB}/k_{sF}\,$ and $\, N_{\rm C}(E_{bB},\,\zeta)\,$  does not depend on $\,{\rm {\bf k}}_{bB}\,$ and integration over
$\,\Omega_{{\rm {\bf k}}_{bB}}\,$  can be performed in a straightforward way.
\end{enumerate}

For simplicity, we assume that $| N_{\rm C}(E_{bB},\,\zeta)|=1$.
Integrating the fully DCS over $\,\Omega_{{bB}}\,$ and using the orthogonality of the spherical harmonics one obtains the doubly DCS:
\begin{align}
\frac{{d^{2}\sigma }}{{d{\Omega _{{{sF}}}}d{E_{sF}}}} = \frac{1}{{2\,\pi}}\frac{{{\Gamma _{bB}}}}{{{{({E_{0(bB)}} - {E_{bB}})}^2} + {{{\Gamma ^2}}}/{4}}}\frac{{d\sigma }}{{d{\Omega _{{{\rm {\bf k}}_{sF}}}}}},
\label{doubleDCS1}
\end{align}
where
\begin{align}
\frac{{d\sigma }}{{d{\Omega _{{{sF}}}}}} =& \frac{{{\mu _{aA}}{\mu _{sF}}}}{{4{\pi ^2}}}\frac{{{k_{0(sF)}}}}{{{k_{aA}}}} 
\nonumber\\& \times
\sum\limits_{ M_{F}M_{s}M_{A}M_{a}}
 |  M_{{M_F}{M_s};{M_A}{M_a}}^{(prior)}({k_{0(sF)}}{{\bf{\hat k}}_{sF}},{{\bf{k}}_{aA}}){|^2}
\label{singlyDCS1}
\end{align}
is the singly DCS of  reaction (\ref{traAsF1}).
Note that integrating the doubly DCS over $E_{sF}$ gives
\begin{align}
\int\limits_0^\infty\, d{E_{sF}}\frac{{d\sigma }}{{d{\Omega _{{{sF}}}}d{E_{sF}}}} = \frac{{{\Gamma _{bB}}}}{\Gamma }\frac{{d\sigma }}{{d{\Omega _{{{sF}}}}}},
\label{DoubleDCS1}
\end{align}
where $\Gamma_{bB}$ is the partial resonance width for the decay of the resonance to the channel $b+B$.
The reaction amplitude in Eq. (\ref{singlyDCS1}) describes only transfer reaction into the resonance without consideration of the resonance propagator and decay into the final channel $b+B$.

\subsection{DWBA amplitude of reaction populating resonance state }
\label{DWBApr1}
In the THM it is enough to consider the doubly DCS  ${{d^{2}\sigma }}/({{d{\Omega _{{{\rm {\bf k}}_{sF}}}}d{E_{sF}}}})$ from which one needs to single out the astrophysical $S(E_{xA})$ factor for the coupled two-channel resonant binary subreaction
\begin{align}
x+A \to F^{*} \to b+B
\label{binarysubreaction1}
\end{align}
at $E_{xA} \to E_{R(xA)}$, where $E_{R(xA)} = E_{0(xA)} - i\,\Gamma/2$ is the resonance energy in the channel $x+A$, 
\begin{align}
S(E_{xA})  \stackrel{E_{xA} \to E_{R(xA)}   }{ =}& \frac{5\cdot 10^{-3} \pi}{\mu_{xA}}\, \frac{ {\hat J}_{F} }{ {\hat J}_{x}\,{\hat J}_{A}  }\,\lambda_{N}^{2}\,m_{N}^{2}\,e^{2\,\pi\,\eta_{xA}}      \nonumber\\
& \times \frac{\Gamma_{bB}\,\Gamma_{xA}}{\big(E_{0(xA)}  - E_{xA} \big)^{2}  + {\Gamma^{2}}/{4}},
\label{SxA1}
\end{align}
where $m_{N}=931.5$ MeV is the atomic mass unit,  $\lambda_{N}$  is the nucleon Compton wave length. In this expression the dimension of the $S$ factor is MeV\,b. For weaker transitions we use units of keV\,b. 
Comparing Eqs. (\ref{SxA1}) and (\ref{doubleDCS1})  one can observe that to single out the $S(E_{xA})$ astrophysical factor from the latter it is enough to single out from the DCS ${{d\sigma }}/{{d{\Omega _{{{sF}}}}}} $ the resonance width $\Gamma_{xA}$.   This has been done  in section  \ref{Transferresonance1}.  
The derived transfer  reaction amplitude  $M_{{M_F}{M_s};{M_A}{M_a}}^{(prior)}({k_{0(sF)}}{{\bf{\hat k}}_{sF}},{{\bf{k}}_{aA}})$, according to Eqs. (\ref{MDSextpr1}),  (\ref{MDWBACoul2})  and  (\ref{LDWext2}), 
is expressed in terms of $\sqrt{\Gamma_{xA}}$.  

In many cases the external part in Eq. (\ref{MDSextpr1}) is small compared to the surface term and can be neglected. Then Eq. (\ref{doubleDCS1}) can be rewritten as
\begin{align}
\frac{{d\sigma }^{THM}}{{d{\Omega _{{{sF}}}}d{E_{sF}}}} =& S(E_{xA})\,e^{-2\,\pi\,\eta_{xA}}\,P_{l_{xA}}^{-1}(k_{(0)xA}, R_{ch})\,
\frac{ {\hat J}_{x}\,{\hat J}_{A}  }{ {\hat J}_{F} }            
\nonumber\\
& \times \frac{{\hat l}_{xA}R_{ch} }{80\, \pi^{2}}\,\lambda_{N}^{-2}\,m_{N}^{-1}\,\big|{\cal W}_{l_{xA}}\big|^{2}\;
\frac{{d\sigma }^{DWZR(prior)}}{{d{\Omega _{{{sF}}}}}}.
\label{doubleTHMDCS1}
\end{align}
We assigned to it the superscript ``THM" because this doubly DCS can be used to analyze THM data.  We assume that
$\,{\widehat {\bf{k}}_{0(xA)}}\,$ is directed along the axis $\,z$, that is, 
$\,{Y_{{l_{xA}},{m_{{l_{xA}}}}}}\big({\widehat {\bf{k}}_{0(xA)}}\big)= \sqrt{{(2\,l_{xA}+1)}/{4\,\pi}}\,\delta_{m_{l_{xA}}\,0}$.
With this for the DCS of  reaction (\ref{traAsF1}) populating the resonance state $\,F^{*}$ we get
\begin{align}
\frac{{d\sigma }^{DWZR(prior)}}{{d{\Omega _{{{sF}}}}}} = &\frac{{{\mu _{aA}}{\mu _{sF}}}}{{4{\pi ^2}}}\frac{{{k_{sF}}}}{{{k_{aA}}}}  
\sum\limits_{M_{F}M_{s}M_{A}M_{a}}
\nonumber\\ & \times
\Big|M_{{M_F}{M_s};{M_A}{M_a}}^{DWZR(prior)}({k_{0(sF)}}{{\bf{\hat k}}_{sF}},{{\bf{k}}_{aA}}){\Big|^2}.
\label{singleDCS1}
\end{align}

\subsection{THM amplitude in plane-wave approach}
\label{THMPWA1}

In this part, we will simplify the equations derived in the previous section using the plane-wave approach (PWA). 
The PWA, both in the initial and final states, a priori, cannot be applied for the analysis of the THM reactions. The reason is that the THM involves a collision of charged particles $a$ and $A$. To increase the DCS of the THM reaction,  the $E_{aA}$ relative energy is selected to be above the Coulomb barrier in the system $a+A$. But it cannot be too high because the resonance energies $E_{0(xA)}$  measured using the THM kinematics, depend on $E_{(xA)}$: the higher  $E_{(xA)}$ the higher minimum resonance energy $E_{0(xA)}$ that can be measured. Hence, for nuclei with higher charges like in Ref. \cite{Nature},  the Coulomb distortion in the initial state $a+A$ cannot be neglected. 
    It would be even more accurate to add nuclear distortion in the initial state.  However, taking into account the fact that in the THM, only the energy dependence of the DCS is measured, we assume that even such a drastic approximation as the PWA can be applied in some particular cases provided the PWA and DWBA excitation functions are similar.

 The prior-form of the PWA follows immediately from the prior-form of the DWBA   
\begin{align}
{\cal M}^{PWA(prior)}= \big<\Psi_{{\rm {\bf k}}_{sF}}^{(0)}\,{\tilde \phi}_{{ R}(xA)}\big|V_{xA}+ V_{sA}\big| \varphi_{sx}\,\Psi_{{\rm {\bf k}}_{aA}}^{(0)}\big>.
\label{PWAprior1}
\end{align}
The new notation here is the plane wave 
$\Psi_{{\rm {\bf k}}_{ij}}^{(0)}= e^{{\rm {\bf k}}_{ij} \cdot {\rm {\bf r}}_{ij}}$ describing the wave function of the  relative motion of the noninteracting particles $i$ and $j$. 

\begin{figure}
\includegraphics[width=\columnwidth]{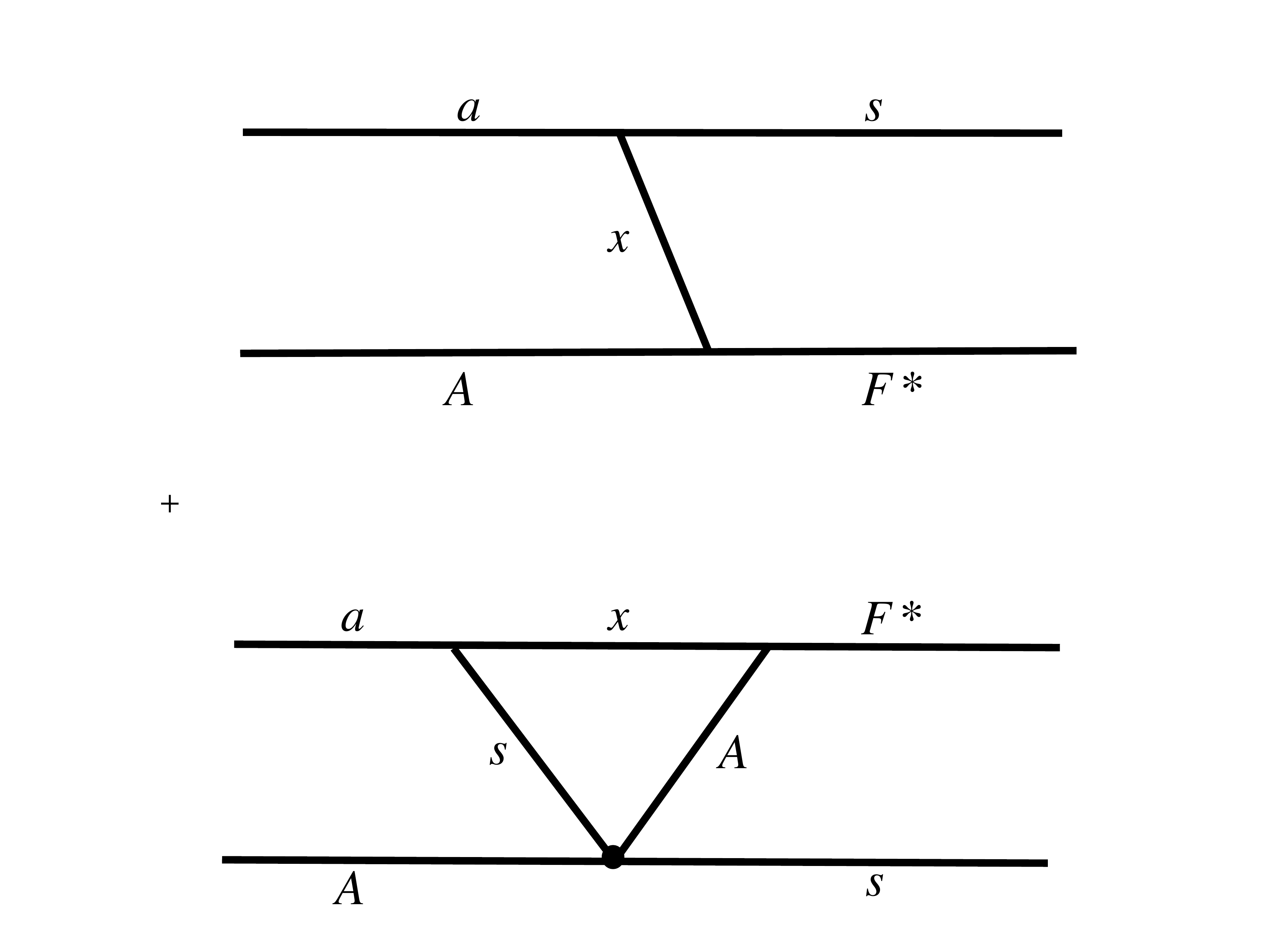}
\caption{The diagrams describing the PWA transfer reaction amplitude $a+A \to s+F^{*}$. The first diagram, which is the pole diagram, describes the simplest transfer reaction mechanism of nucleus $x$. The second diagram is the exchange triangular diagram. The black filled sphere is the OFES $s+A$ Coulomb+nuclear scattering amplitude in the Born approximation generated by the potential $V_{sA}$.} 
\label{fig_poletriangle}
\end{figure}

Figure \ref{fig_poletriangle} shows the Feynman diagram representation 
of the amplitude ${\cal M}^{PWA(prior)}$. The amplitude of the first diagram has a peak at scattering angles of particle $s$ near the forward direction. The second diagram in Fig. \ref{fig_poletriangle}  is the exchange triangular diagram. As a function of the scattering angle of the particle $s$ its amplitude has two singularities: due to the $s+A$ Coulomb potential the triangular diagram has the same singularity as the pole diagram in Fig. \ref{fig_poletriangle}. Including the triangular diagram leads to the renormalization of the first diagram, see Appendix D from \cite{muk2012}. The second singularity of the amplitude corresponding to the triangle diagram in the plane of the scattering angle of the particle $s$ is generated by the skeleton triangle diagram (this singularity is generated by the propagators of the triangle diagram) and leads to the peak in the angular distribution at backward angles. 

As it has been done in the previous subsection, 
we split the PWA matrix element into the internal and external parts to get  
\begin{align}
&{\cal M}^{PWA(prior)}= {\cal M}^{PWA(post)}_{int} + {\cal  M}^{PWA(prior)}_{S}    \nonumber\\
+ &{\cal M}^{PWA(prior)}_{ext}.
\label{intextPWA1}
\end{align}
Here,  
\begin{align}
{\cal M}^{PWA(post)}_{int}= \big<\Psi_{{\rm {\bf k}}_{sF}}^{(0)}\,{\tilde \phi}_{{ R}(xA)}\big|  V_{sx} + V_{sA}\big| \varphi_{sx}\,\Psi_{{\rm {\bf k}}_{aA}}^{(0)}\big>
\label{PWApostint1}
\end{align}
is the  post-form of the internal PWA amplitude, 

\begin{align}
&{\cal M}_S^{PWA(prior)}  = \frac{1}{2}\,e^{  - i\,\delta^{p}_{ s_{xA}l_{xA}J_{F}  }(k_{0(xA)})      }\,\sqrt {\frac{1}{{{\mu _{xA}}\,{k_{0(xA)}}}}\,{\Gamma _{xA}}} \, 
\nonumber\\
& \times\,O_{{l_{xA}}}({k_{0(xA)}}{R_{ch}})\,{\varphi _{sx}}({{\rm {\bf k}}_{sx}})\, {\cal W}_{l_{xA}}
{i^{ - {l_{xA}}}}{Y_{{l_{xA}},{m_{{l_{xA}}}}}}({\widehat {\bf{k}}_{0(xA)}})
\label{LSPWAfinal1}
\end{align}
is the surface term of the prior-form PWA amplitude, ${\rm {\bf k}}_{sx}= {\rm {\bf k}}_{s} -  \frac{m_{s}}{m_{a}}\,{\rm {\bf k}}_{a}$, ${\rm {\bf k}}_{i}$ is the OES momentum of the particle $i$,
 $\,{\varphi _{a}}({{\rm {\bf k}}_{sx}})\,$ is the Fourier transform of the $s$-wave $(sx)$ bound-state wave function $\,{\varphi _{a}}({{\rm {\bf r}}_{sx}})$.
Finally, the  prior-form of the external PWA amplitude is written as 
 \begin{align}                                  \nonumber\\
{\cal M}_{ext}^{PWA(prior)}=& \big<\Psi_{{\rm {\bf k}}_{sF}}^{(0)}\,{\tilde \phi}_{{R}(xA)}\big|V_{xA}^{C} + V_{sA}^{C} \big| \varphi_{a}\,\Psi_{{\rm {\bf k}}_{aA}}^{(0)}\big>\Big|_{r_{xA} > R_{ch}}.
\label{PWAext1}
\end{align}
Taking into account that the distance between the particles $s$ and $x$ is controlled by the bound-state wave function ${\varphi _{a}}({{\rm {\bf r}}_{sx}})$ for $r_{xA} > R_{ch}$, at large enough channel radius $R_{ch}$ we  replaced $V_{sA}$ in Eq. (\ref{PWAext1}) by the Coulomb part $V_{sA}^{C}$.

\section{Application of DWBA formalism for analysis of resonant THM reaction of ${}^{12}{\rm C} + {}^{12}{\rm C}$ fusion}
\label{critanalalysisTHM1}

In this section we present a critical analysis of the application of the indirect Trojan Horse method (THM) to measure the astrophysical $S^\ast$ factor of ${}^{12}{\rm C}+{}^{12}{\rm C}$ fusion \cite{Nature}. In the case under consideration, $a={}^{14}{\rm N}$, $A={}^{12}{\rm C}$, $x={}^{12}{\rm C}$, $s=d$, and $F={}^{24}{\rm Mg}^\ast$. Four different channels in the final state were populated in the THM experiment: $p_{0} + {}^{23}{\rm Na}$, $p_{1} + {}^{23}{\rm Na}$  (0.44 MeV), $\alpha_{0} + {}^{20}{\rm Ne}$, and $\alpha_{1} + {}^{20}{\rm Ne}$ (1.63 MeV) \cite{Nature}.
The THM is a 
unique indirect technique that allows one to measure the astrophysical factors of the resonant reactions at low energies, where direct methods are not able to obtain data due to very small cross sections.  The critical analysis of the THM in this review is not aimed to taint the whole method which demonstrated  its power in more than hundred publications (see \cite{spitaleri2019}  and references therein). We critically review only the analysis of the data in \cite{Nature}. 

The THM resonant reaction (\ref{THMreaction1}) involves two steps.  The first step is transfer reaction (\ref{traAsF1})  populating the resonance state $F^{*}=x+A$, and the second step is decay of the resonance $F^{*} \to b+B$. In other words, the THM reaction is a process leading to three particles in the final state. This makes analysis of such a reaction quite complicated. Special kinematical conditions should be  fulfilled and angular correlation data are needed to make sure that the reaction is dominantly contributed by the resonant THM mechanism. Only then it will allow one to extract from the THM reaction an information about resonant binary reaction (\ref{THMressubreact1}). 
In \cite{Nature}  the ${}^{12}{\rm C}({}^{14}{\rm N},d)(\alpha + {}^{20}
{\rm Ne})$ and  ${}^{12}{\rm C}({}^{14}{\rm N},d)(p + {}^{23}{\rm Na})$  reactions were measured to obtain the $S^{*}$ factors for the carbon-carbon fusion and a sharp rise of the astrophysical $S$ factor for carbon-carbon fusion determined using the indirect THM was reported. 
Here we outline the most important inconsistencies in the THM analysis and data from \cite{Nature}. All the notations are given in \cite{MPK2019}. \\

To analyze the measured data Tumino {\it et al.} \cite{Nature} used a simple PWA developed by one of us (A.M.M.) rather than a generalized $R$-matrix approach based on the surface-integral approach discussed in subsection \ref{DWBApr1} and originally published in Ref.  \cite{muk2011}.  The approach uses distorted waves in both initial and final states.  
The PWA follows from this more general approach when the distorted waves are replaced with the plane waves.  The PWA was successfully applied for analyses of  many THM reactions in which the spectator is a neutron. It was also applied to reactions at energies above the Coulomb barrier in the initial and final states, and when the interacting nuclei have small charges \cite{reviewpaper,spitaleri2019,spitaleri2016}.  In the PWA it is assumed that the angular distribution of the spectator is forward-peaked in the center-of-mass system (quasi-free kinematics) and that the bound-state wave function of the spectator can be factorized out (see subsection \ref{DWBApr1} and Eq. (117) of Ref. \cite{muk2011} and Eq. (2) of Ref. \cite{Nature}).   Usage of the PWA can be justified only if the PWA and the DWBA   give  similar energy dependence for the DCS of the transfer reaction. This is because in the THM  only the energy dependence of the astrophysical factor is measured while its absolute value is determined by normalizing the THM data to available direct data at higher energies.

Tumino {\it et al.} \cite{Nature} reported that the astrophysical $S^\ast(E)$ factors extracted from the THM experiment demonstrate a steep rise when the resonance energy $E$ decreases. This rise would have profound implications on different astrophysical scenarios as the carbon-carbon fusion rate calculated from the astrophysical $S^\ast$ factors deduced in Ref.  \cite{Nature} significantly exceeds all previous estimations of the reaction rate obtained by extrapolating the direct data to the low-energy region. For example, the reaction rate calculated in Ref. \cite{Nature} at temperature  $T \sim 2 \times 10^{8}$ K exceeds the adopted value \cite{Fowler,illiadis} by a factor of $500$.

The authors of Ref. \cite{Nature} were rightly concerned about the Coulomb barrier in the initial state. That is why in the experiment the initial energy was above the Coulomb barrier. However, given the energy of the emitted particles,  neglecting the Coulomb effects in the final channel is unjustified. 
Below we present a detailed analysis of the TH resonant reactions based on the distorted-wave formalism. We take into account the distortions in the initial, intermediate and final states.

\subsection{Kinematics of the THM reaction}
\label{THMkinematics1}

In Ref. \cite{Nature} the normalization of the THM data to the direct data was done in the energy interval $E=2.5-2.63$ MeV, where $E$ is the ${}^{12}{\rm C}- {}^{12}{\rm C}$ relative kinetic energy. Here and in what follows we use $E_{xA} \equiv E$.
To check whether the PWA is justified, we consider the kinematics of the THM in the energy interval covered by the THM experiment \cite{Nature}.
In the experiment the relative ${}^{14}{\rm N}-{}^{12}{\rm C}$   energy in the entrance channel was $E_{aA}=13.845$ MeV. From energy conservation in the THM reaction it follows that $E_{aA}+Q= E_{f}$, where 
 $\,E_{f}=E_{sF}+E_{bB}$ is the total kinetic energy of the final three-body system $s+b+B$  and $\,Q= m_{a} + m_{A} - m_{s} - m_{b} - m_{B}$. From this equation we get that  the total kinetic energy in the final  $\,d+p + {}^{23}{\rm Na}$ channel is $\,E_{f}=5.8$ MeV.

Let us consider the  ${}^{12}{\rm C}-{}^{12}{\rm C}$ relative energy  $E=2.63$ MeV \cite{Nature} which is the highest point of the THM normalization interval. For the binary reaction $\,{}^{12}{\rm C} + {}^{12}{\rm C} \to p+{}^{23}{\rm Na}$, we have $Q_{2}=2.24$ MeV, where $Q_{2} =m_{x} + m_{A} - m_{b}- m_{B}$. Accordingly, the energy in the $\,p+{}^{23}{\rm Na}$ channel corresponding to $E=2.63$ MeV is $E_{p\,{}^{23}{\rm Na}}=4.87$ MeV. Hence, the relative kinetic energy of the deuteron and the c.m. of the $\,p + {}^{23}{\rm Na}$ system corresponding to this energy is $E_{d\,{}^{24}{\rm Mg}}=0.93$ MeV. This energy is well below the Coulomb barrier in the $d-{}^{24}{\rm Mg}$ system, which is about 3 MeV. Even on the lower end of the normalization interval corresponding to $E=2.5$ MeV, the relative energy is $E_{d\,{}^{24}{\rm Mg}}=1.06$ MeV.

At the  energy of $\,E=1.5$ MeV in the $\,{}^{12}{\rm C}-{}^{12}{\rm C}$ channel, which corresponds to the  energy $\,E_{p\,{}^{23}{\rm Na}}= 3.74$ MeV in the exit channel, the relative energy $\,E_{d\,{}^{24}{\rm Mg}}=2.06$ MeV. This is still below the Coulomb barrier. Note that the resonance energies that can be observed in the THM experiment are below $3.56$ MeV. This is due to the fact that above $3.56$ MeV, the resonance energy in the $p+{}^{23}{\rm Na}$ channel is $3.56+Q_{2}>5.8$ MeV. In other words, the $d-{}^{24}{\rm Mg}$ relative energy is $E_{d\,{}^{24}{\rm Mg}}<0$.
Even for the  energy of $E=0.805$ MeV, which corresponds to $E_{p{}\,^{23}{\rm Na}}= 3.05$  MeV, the $d-{}^{24}{\rm Mg}$ relative energy is $E_{d\,{}^{24}{\rm Mg}}=2.75$ MeV. The latter is close to but still below the Coulomb barrier.
Thus we may conclude that the Coulomb interaction plays a very important role in the energy interval exploited in
\cite{Nature} and, therefore, cannot be neglected.

\subsection{DWBA DCS}
\label{DWBA1}

The presence of the strong Coulomb interaction for such deep sub-Coulomb processes in the final state of the transfer reaction significantly increases the DCS in the backward hemisphere, shifting the peak of the angular distribution of the deuterons to the backward angles.  It completely contradicts to the PWA DCS in the c.m. system, which has a pronounced peak at forward angles. Even at the lowest observed resonances at $0.8-0.9$ MeV in the THM experiment \cite{Nature} the angular distribution of the deuterons noticeably deviates from the PWA one if the Coulomb (or Coulomb plus nuclear) rescattering effects in the initial and final states of the ${}^{12}{\rm C}$-transfer reaction are included.

But what is even more important is the fact that the presence of the strong Coulomb interaction significantly changes the absolute values of the DCSs of the ${}^{12}{\rm C}$ transfer reaction and their energy dependence. The absolute values of the DCSs in the THM normalization interval become smaller than the corresponding PWA ones by more than three orders of magnitude and they increase rapidly when the resonance energy decreases. That is one of the main reasons for the drop of the THM astrophysical factors found in this work compared to those extracted in  \cite{Nature} using the PWA.

Below  we compare  the PWA and DWBA DCSs.  To ensure the convergence of the matrix elements for the transfer reaction involving resonant states of ${}^{24}{\rm Mg}$ we use the binned resonant wave functions. 
In the next three figures we present three different curves corresponding to three different potentials describing the resonances with $l_{xA}=2$ in
${}^{24}{\rm Mg}$ given in Table \ref{table1}.
We use the standard notations for the potential parameters shown in Table \ref{table1}: $V$ is the depth of the Woods-Saxon potential,
$r$ and $a$ are the radial parameter and the diffuseness.

\begin{table}
\caption{ Parameters of the ${}^{12}{\rm C}-{}^{12}{\rm C}$ potentials used to calculate the resonance bin wave functions.}
\begin{center}
\begin{tabular}{ cccccc}
\hline
 $E$ (MeV) & No. & $V$ (MeV) & $r$ (fm) & $a$ (fm)  & width (MeV) \\
 \hline
 $2.7$ & Potential $1$ & $58.87$  &  $1.25$  & $2.40$   &  $3.595 \times 10^{-3}$  \\
 $2.7$ & Potential $2$ & $94.51$  &  $1.05$  & $2.40$   &  $4.924 \times 10^{-3}$    \\
 $2.7$ & Potential $3$ & $221.95$  &  $1.25$  & $1.85$   &  $3.104 \times 10^{-4}$   \\
 $1.5$ & Potential $1$ & $110.57$  &  $2.80$  & $3.05$   &  $2.189 \times 10^{-3}$  \\
 $1.5$ & Potential $2$ & $60.07$  &  $2.60$  & $3.05$   &  $2.253 \times 10^{-4}$    \\
 $1.5$ & Potential $3$ & $150.8$  &  $2.80$  & $2.30$   &  $1.055 \times 10^{-4}$   \\
 $0.8$ & Potential $1$ & $140.47$  &  $4.50$  & $4.50$   &  $2.165 \times 10^{-5}$  \\
 $0.8$ & Potential $2$ & $185.284$  &  $4.20$  & $4.50$   &  $2.022 \times 10^{-5}$    \\
 $0.8$ & Potential $3$ & $198.798$  &  $4.50$  & $4.04$   &  $9.566 \times 10^{-6}$   \\
 \hline
\end{tabular}
\label{table1}
\end{center}
\end{table}

The bin wave functions for the ${}^{12}{\rm C}- {}^{12}{\rm C}$ resonance states are generated using the potentials from Table \ref{table1}  
and are obtained by integrating the ${}^{12}{\rm C}- {}^{12}{\rm C}$ scattering wave functions over the energy $E$  with the interval width of $0.5$ MeV
 centered at the resonant energy. This width corresponds to a typical experimental energy resolution.
The bin wave functions are real and are normalized using a factor of $\sin[\delta(k_{{}^{12 }{\rm C}\,{}^{12}{\rm C}})]\exp{[-i\delta(k_{{}^{12}{\rm C}\,{}^{12}{\rm C}})]}$ \cite{FRESCO}, where $\delta(k_{{}^{12}{\rm C}\,{}^{12}{\rm C}})$ is the 
${}^{12}{\rm C}- {}^{12}{\rm C}$ scattering phase shift. The bin sizes affect the resulting bin wave functions, and, hence, the amplitude of the THM transfer reaction but they do not affect much the shapes of the angular distributions.

\begin{figure}[htbp]
\centering
\includegraphics[width=\columnwidth]{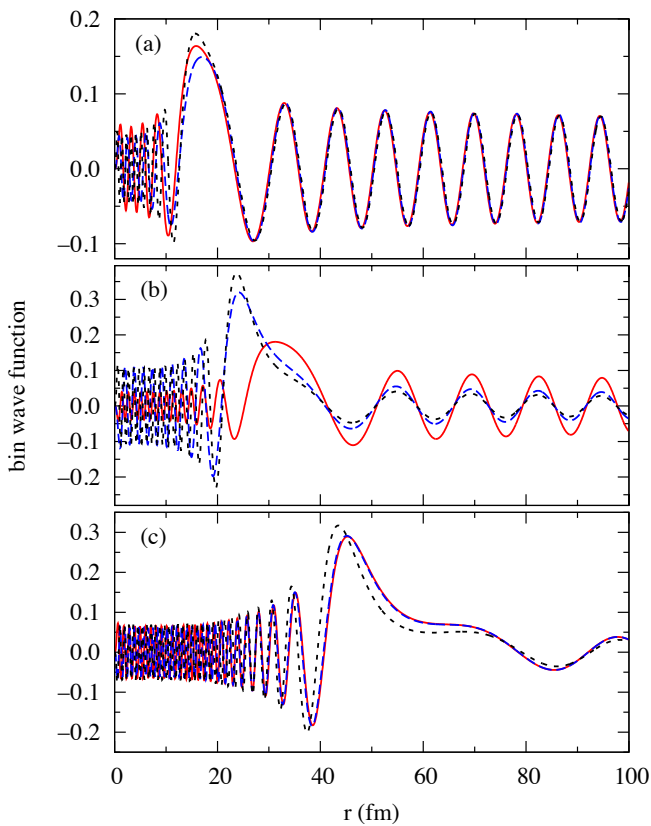}
\caption{(Color online) The bin wave functions calculated for three resonance energies $\,E=2.7,\,1.5\,$ and $\,0.8$ MeV of the ${}^{12}{\rm C}- {}^{12}{\rm C}$ system. Each panel contains three lines corresponding to three different potentials for each resonance energy. The red solid, blue dashed and black dotted curves correspond to the potentials  $1,\,2$ and $3$ from Table \ref{table1}. Panel (a): $E=2.7$ MeV; panel (b): $E=1.5$ MeV; panel (c): $E=0.8$ MeV. 
First published in \cite{MPK2019}.
}
\label{fig_binwfs1}
\end{figure}

The resonance energies given in Table \ref{table1} are selected from the high end, middle and low energy interval measured in \cite{Nature}.
Note that we are not able to reproduce exactly the location of the resonances reported in \cite{Nature} but the obtained resonance energy
are pretty close to the corresponding experimental ones.
The bin wave functions for the three resonance energies constructed using the potentials from Table \ref{table1} are depicted in Fig. \ref{fig_binwfs1}.
The highly oscillatory behavior of the resonance wave functions is a clear evidence that the internal Coulomb or Coulomb+nuclear DWBA in the post form should be small (see Section II B). 

The PWA DCSs for three resonance energies $\,E=2.7,\,1.5\,$ and $\,0.8$ MeV of the  ${}^{12}{\rm C}- {}^{12}{\rm C}$ system are shown in  Fig. \ref{fig_PWA1}. Each panel contains three lines corresponding to three different potentials for each resonance energy, see Table \ref{table1}.
\begin{figure}[htbp]
\centering
\includegraphics[width=\columnwidth]{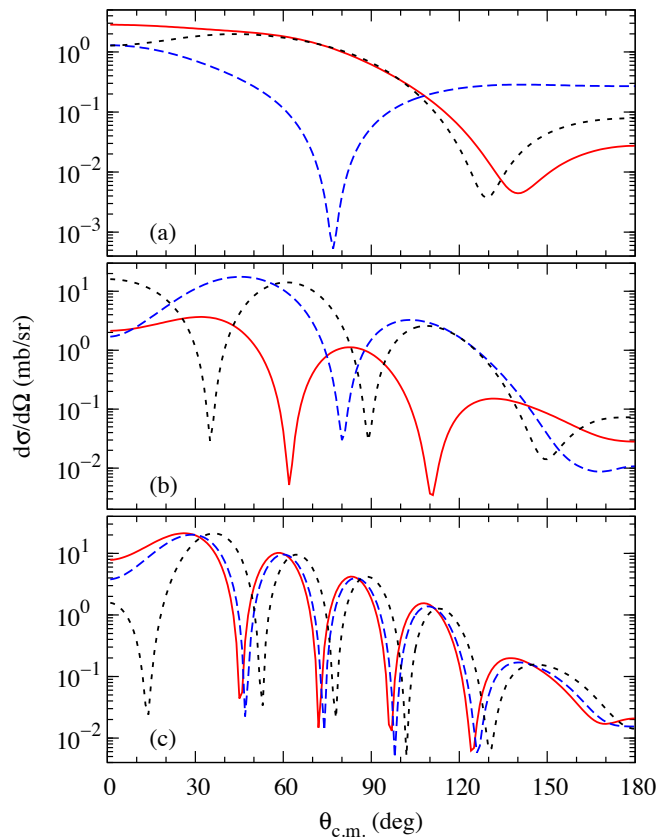}
\caption{(Color online) The PWA DCSs for the ${}^{14}{\rm N} + {}^{12}{\rm C} \to  d+ {}^{24}{\rm Mg}^{*}$ reaction at the relative kinetic energy $E_{{}^{12}{\rm C}\,{}^{14}{\rm N}}=13.85$ MeV populating three resonant states in ${}^{24}{\rm Mg}$: $\,E=2.7,\,1.5\,$ and $\,0.8$ MeV. Each panel contains three lines corresponding to three different potentials for each resonance energy. The red solid, blue dashed and black dotted curves correspond to the potentials  $1,\,2$ and $3$ from Table \ref{table1}. Panel (a): $E=2.7$ MeV; panel (b): $E=1.5$ MeV; panel (c): $E=0.8$ MeV.
First published in \cite{MPK2019}.
}
\label{fig_PWA1}
\end{figure}

Next we show the DWBA DCSs calculated using the bin wave functions shown in Fig. \ref{fig_binwfs1}. Figure \ref{fig_dwbaCandCN} presents the Coulomb DWBA DCSs calculated at the same three resonance energies of the system ${}^{12}{\rm C}- {}^{12}{\rm C}$.  
We performed calculations using the pure Coulomb DWBA (thin lines) and the Coulomb + nuclear DWBA (thick lines). The optical-model potential parameters are taken from the compilation \cite{PereyPerey}, namely, parameters for the ${}^{14}{\rm N}-{} ^{12}{\rm C}$ potential at $27.3$ MeV and the $d-{}^{24}{\rm Mg}$ potential at $3.3$ MeV are used for the entrance and exit channels, respectively. The relative energy between the deuteron and the c.m. of the ${}^{24}{\rm Mg}$ subsystem depends on the excitation energy of the latter. In principle, different optical potentials should be used in the exit channel for each ${}^{24}{\rm Mg}$ excitation energy. However, our calculations suggest that the DCSs  of the  transfer reaction depend weakly on the choice of the exit-channel optical model potentials. This is because the relative  $d+{}^{24}{\rm Mg}$ energies in the exit channel are so low that the Coulomb interaction dominates over the exit-channel distorted waves. For this reason, the same exit-channel optical potential is used for all the cases.

\begin{figure}[htbp]
    \centering
\includegraphics[width=\columnwidth]{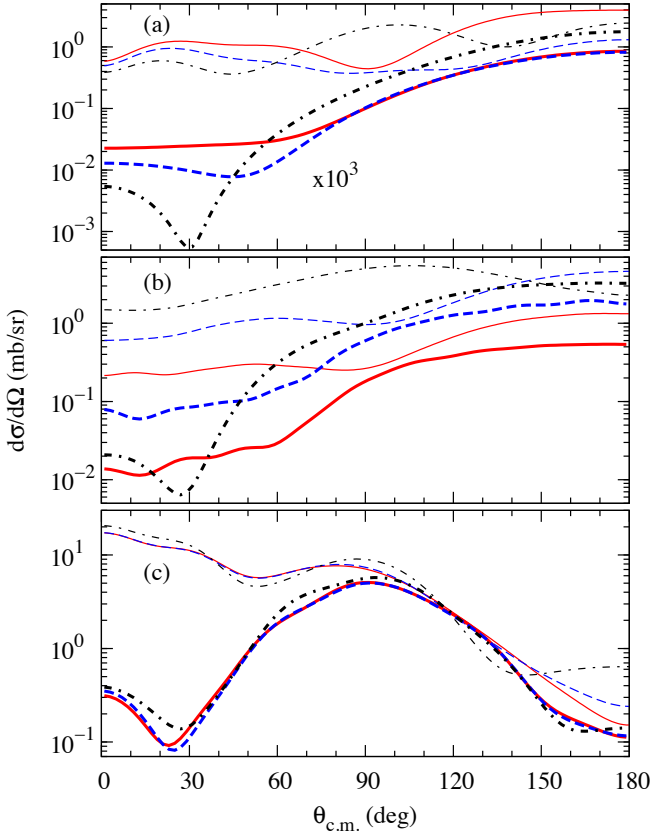}
\caption{(Color online) The DWBA DCSs for the ${}^{14}{\rm N} + {}^{12}{\rm C} \to d+ {}^{24}{\rm Mg}^{*}$ reaction at the relative kinetic energy $E_{{}^{12}{\rm C}\,{}^{14}{\rm N}}=13.85$ MeV populating three resonant states in ${}^{24}{\rm Mg}$. Panel (a): $\,E=2.7$ MeV; panel (b): $E=1.5\,$ MeV and panel (c): $E=0.8$ MeV. Each panel contains six lines. The thin (thick) red solid, blue dashed and black dotted curves correspond to the Coulomb (Coulomb + nuclear) DWBA DCSs  calculated using the  ${}^{12}{\rm C}-{}^{12}{\rm C}$ bin wave functions for the potentials  $1,\,2$ and $3$ from Table \ref{table1}, respectively. Note that the DWBA DCSs in panel (a) are multiplied by $10^{3}$.
First published in \cite{MPK2019}.}
\label{fig_dwbaCandCN}
\end{figure}

From the presented figures we can draw the following conclusions:
\begin{itemize}
\item
The PWA and the DWBA DCSs differ significantly both in the angular distributions and energy dependences.  In particular, the DWBA calculations show that in the interval of the resonance energies $E = 1.5-2.7$ MeV the angular distributions have backward peaks in contrast to the PWA ones. This is a very most important point. As we will see below,  the different energy dependences of the PWA and DWBA DCSs lead to very different energy dependences of the astrophysical factors calculated using the PWA and DWBA.  

\item
The ratio of the DCSs from the PWA and the DWBA at $E=2.7$ MeV and $0.8$ MeV are completely different.  The DWBA DCSs at any angle at $E=2.7$ MeV are significantly smaller than those at $E=0.8$ MeV. This happens only if the Coulomb or the Coulomb plus nuclear distortions are taken into account. This is an additional corroboration of the fact that at the resonance energies of the THM normalization interval $E=2.5$-2.63 MeV considered in \cite{Nature}, the THM reactions are deep sub-Coulomb. This makes their DWBA DCSs extremely small. The absolute value of the DWBA DCS increases when $E$ decreases because the energy of the outgoing deuteron increases as the Coulomb barrier is approached.

As we will see later [see Eq. (\ref{SfactorarbE1}) below for the $S$ factor] the DWBA DCS appears in the denominator. A very small DCS at high $E$ should significantly increase the THM astrophysical factor. As the energy $E$ decreases the DWBA DCS increases and the $S(E)$ factor quickly drops. For comparison we set our renormalization factor $R(E)$ [see Eq. (\ref{RF1}) below] equal to unity at $E=2.664$ MeV, which is on the upper border of the THM normalization interval considered in \cite{Nature}. The significant rise of the  DWBA DCS toward small $E$ is the factor that most contributes to the  drop of the THM $S(E)$.

\end{itemize}

\subsection{Renormalization of THM astrophysical factors}
\label{renSfactor1}

Here we compare the astrophysical factors obtained in the PWA and DWBA  and obtain the renormalization factor of the THM $S$ factor  \cite{Nature}
taking into account the distortion effects in the initial and final states of the transfer reaction.

We start from the THM doubly DWBA DCS given by Eq. (\ref{doubleTHMDCS1}) which can be rewritten as
\begin{align}
&\frac{{{{\rm{d}}^2}{\sigma^\textrm{THM}}}}{{{\rm{d}}E{\mkern 1mu} {\rm{d}}{\Omega _{{{\bf{k}}_{sF}}}}}} = K(E)\,S(E){\mkern 1mu}\,
\big|{\cal W}_{l_{xA}}\big|^{2}\;
\frac{{d\sigma }^{DWZR(prior)}}{{d{\Omega _{{{\rm {\bf k}}_{sF}}}}}}.
\label{doubleTHMcrSfactor1}
\end{align}
Here,
\begin{align}
K(E)= \,e^{-2\,\pi\,\eta_{xA}}\,P_{l_{xA}}^{-1}(k_{(0)aA}, R_{ch})\,
\frac{ {\hat J}_{x}\,{\hat J}_{A}  }{ {\hat J}_{F} }\,\frac{{\hat l}_{xA}R_{ch} }{80\,\pi^{2}}
\frac{1}{
\lambda_{N}^{2} \,m_{N}}
\label{kinemfactor1}
\end{align}
is a trivial kinematical  factor,  ${{d\sigma }^{DWZR(prior)}}/{{d{\Omega _{{{\rm {\bf k}}_{sF}}}}}}$ is the zero-range  DWBA cross section of the  ${}^{14}{\rm N} + {}^{12}{\rm C} \to d+ {}^{24}{\rm Mg}^{*}$ reaction populating the isolated resonance state, $S(E)$ is the astrophysical factor.

Correspondingly, the THM astrophysical factor  determined from  Eq. (\ref{doubleTHMcrSfactor1}) is
\begin{align}
S(E) =& N_{F}\,K(E)\,\frac{1}{\big|{\cal W}_{l_{xA}}\big|^{2}}\frac{{{{\rm d}^2}{\sigma^\textrm{THM}}}}{{{\rm d}E\,{\rm d}{\Omega _{{{\rm {\bf k}}_{sF}}}}}}
\nonumber \\ &\times
\frac{1}{{{{{\rm d}{\sigma^\textrm{DWZR(prior)}}(E,\cos {\theta _s})}}/{{{\rm d}{\Omega _{{{\rm {\bf k}}_{sF}}}}}}}}.
\label{SfactorarbE1}
\end{align}
Here $N_{F}$ is an overall, energy-independent factor for normalization of the THM data to direct data, $\theta_{s}$ is the scattering angle of the spectator $s$ in the c.m. of the THM reaction. We recall that in the THM only the energy dependence of the astrophysical factor is measured. Its absolute value is determined by normalizing  the THM $S(E)$ factor to the direct data available at higher energies.

Equations (\ref{doubleTHMcrSfactor1}) and (\ref{SfactorarbE1}) are pivotal for understanding the challenges in extraction of the $S(E)$ factor from the THM DCS. Since in the normalization interval of $E=2.5-2.66$ MeV the outgoing deuterons are below the Coulomb barrier,  ${{{\rm{d}}{\sigma^\textrm{DWZR(prior)}}(E,\cos {\theta _s})}}/{{{\rm{d}}{\Omega _{{{\bf{k}}_{sF}}}}}}$  is small and rapidly increases when the resonance energy $E$ decreases. This increase of ${{{\rm{d}}{\sigma^\textrm{DWZR(prior)}}(E,\cos {\theta _s})}}/{{{\rm{d}}{\Omega _{{{\bf{k}}_{sF}}}}}}$ should reflect in the behavior of ${{{{\rm{d}}^2}{\sigma^\textrm{THM}}}}/({{{\rm{d}}E{\mkern 1mu} {\rm{d}}{\Omega _{{{\bf{k}}_{sF}}}}}})$ and the THM $S(E)$ factor. As we already mentioned, in Ref. \cite{Nature} a simple PWA was used instead of the distorted waves. The DCS as a function of $E_{dF}$ obtained using the PWA changes very little compared to the change of the DWBA DCS. This is the main reason why the THM $S(E)$ factors show unusually high rise when $E$ decreases.

In \cite{Nature} the normalization interval was chosen to be $E=2.5-2.63$ MeV. However, there are two resonances with negative parities that are questionable because the collision of the two identical bosons ${}^{12}{\rm C} + {}^{12}{\rm C}$ cannot populate resonances with the negative parity. There are two resonances with positive parities cited in \cite{Nature}: at $2.664$ and $2.537$ MeV. It was underscored  in \cite{Nature} that the THM data reproduce the higher-lying resonance. That is why here we use the resonance at $2.664$ MeV for the normalization of the THM data to direct ones. Thus, we assume here that the normalization factor $N_{F}$ is determined by normalizing the THM astrophysical factor to the directly measured resonance at $E=2.664$ MeV. Practically we selected the normalization of the THM data on the edge of the energy interval measured in \cite{Nature}.

To find the renormalization of the THM astrophysical factor presented in \cite{Nature} we recall that in the PWA the THM astrophysical factor for an isolated resonance is given by
\begin{align}
&S^\textrm{(PWA)}(E) = \frac{N_{F}}{K(E)}\,\frac{{{{\rm d}^2}{\sigma^\textrm{THM}}}}{{{\rm d}E\,{\rm d}{\Omega _{{{\rm {\bf k}}_{sF}}}}}}\frac{1}{\varphi_{a}^{2}(E)\,|{\cal W}_{l_{xA}}|^{2}}.
\label{SPWA1}
\end{align}
The factor ${\cal W}_{l_{xA}}$ is given in Eq. (\ref{calWr1}) \cite{muk2011}, $\varphi_{a}(E)$ is the Fourier transform of the  $a=(s\,x)$ bound-state wave function. The Fourier transform, actually, depends on the $s-x$ relative momentum $ k_{sx}$, which in the case under consideration is 
 $k_{d\,{}^{12}{\rm C}}$ and expressed in terms of ${\rm {\bf k}}_{d}$.
From the energy conservation it follows that 
$\,E_{sF} = E_{aA}- \varepsilon_{sx} - E$. Hence the Fourier transform of the bound-state wave function $\,\varphi_{a}(k_{sx})\,$ depends on $\,E$.

By taking the ratio of the $S(E)$ factors given by Eqs. (\ref{SfactorarbE1}) and (\ref{SPWA1}) we get the renormalization factor for the THM astrophysical factor presented in \cite{Nature} due to the Coulomb (Coulomb + nuclear) distortions in the initial and final states of the THM transfer reaction:
\begin{align}
&R(E)= \frac{\varphi_{a}^{2}(E)}{\varphi_{a}^{2}(E_{N})}\, \frac{{{\rm{d}}{\sigma^\textrm{DWZR(prior)}}(E_{N},\cos {\theta _s})/{\rm{d}}{\Omega _{{{\bf{k}}_{sF}}}}}}{{{\rm{d}}{\sigma^\textrm{DWZR(prior)}}(E,\cos {\theta _s})/{\rm{d}}{\Omega _{{{\bf{k}}_{sF}}}}}},
\label{RF1}
\end{align}
where $E_{N}$ is the THM normalization energy.

\subsection{Astrophysical factors for the ${}^{12}{\rm C}-{}^{12}{\rm C}$ fusion from THM reaction}
\label{Sfactors1}

In this subsection we present new ${}^{12}{\rm C}-{}^{12}{\rm C}$ fusion $S^\ast = e^{0.46\,E}\,S(E)$ factors obtained by  renormalizing the THM astrophysical factors presented in \cite{Nature}. For renormalization the factor $R(E)$ given in Eq. \eqref{RF1} is used. The DWBA DCSs are calculated using the FRESCO code \cite{FRESCO}. For comparison we also calculate ${{\rm{d}}{\sigma^\textrm{DWZR(prior)}}(E,\cos {\theta _s})/{\rm{d}}{\Omega _{{{\bf{k}}_{sF}}}}}$ including the nuclear distortions. To calculate the optical-model distorted waves we use the optical potentials from Ref. \cite{PereyPerey} as described above. Following \cite{Nature} we use the normalization energy $E_{N}= 2.664$ MeV.

The energy dependence of the DWBAZR(prior) DCS at the deuteron scattering angle of $\,15$ degree in the c.m. of the reaction is shown in Fig. \ref{fig_ZRexcf1}.
\begin{figure}[htbp]
\includegraphics[width=\columnwidth]{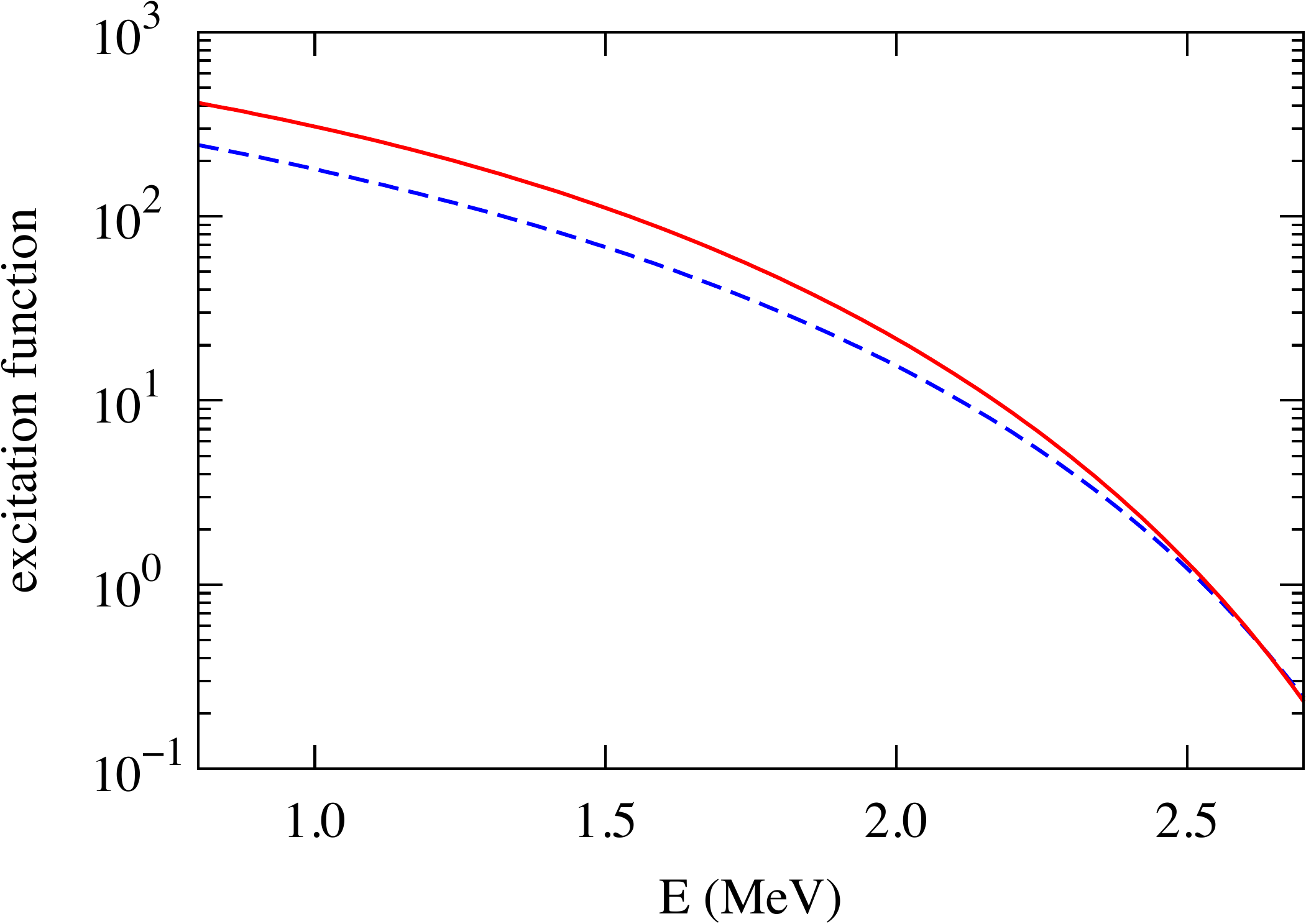}
\caption{(Color online) The excitation function of the THM transfer reaction $\,{}^{14}{\rm N} + {}^{12}{\rm C} \to d +{}^{24}{\rm F}^{*}\,$ calculated using the  DWBAZR(prior) at the scattering angle of the deuteron of $\,15$ degree in the c.m. of the reaction. The solid red line is for the pure Coulomb DWBAZR(prior) and the dashed blue line for the Coulomb + nuclear DWBAZR(prior).
First published in \cite{MPK2019}.
}
\label{fig_ZRexcf1}
\end{figure}

In Fig.  \ref{fig_Stot1} the total renormalized astrophysical factor is compared with original one from \cite{Nature} . The total renormalized astrophysical factor  is $R(E)\,S^\ast(E)$, where $\,S^\ast(E)$ is the total astrophysical factor taken from \cite{Nature}.  One can see that at the resonance energies $E=0.8-0.9$ MeV the renormalization factor $R(E)$ decreases the THM astrophysical factors from \cite{Nature}  by about a factor of $ 10^{3}$.
\begin{figure}[htbp]
\includegraphics[width=\columnwidth]{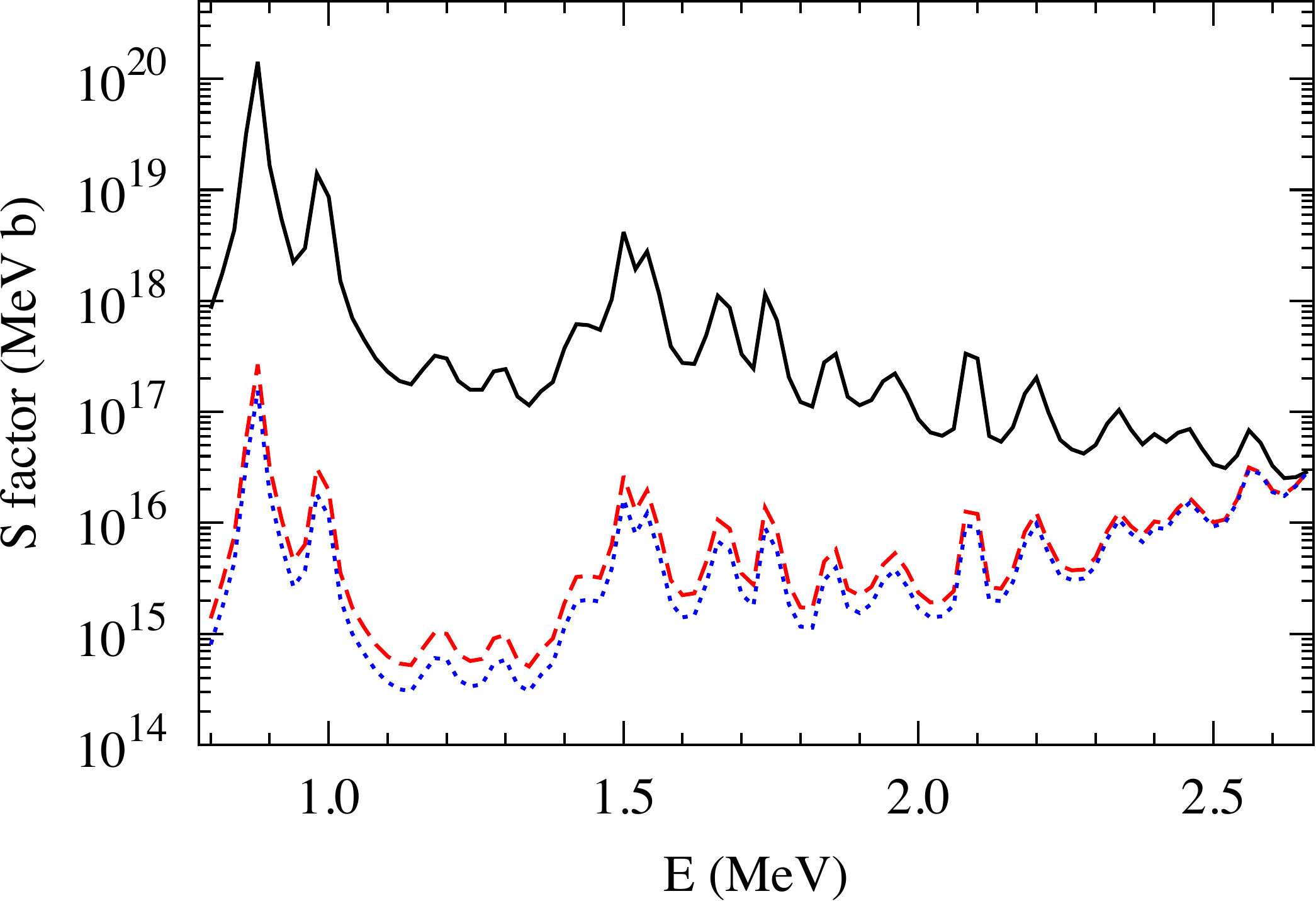}
\caption{(Color online) Total $S^{*}(E)$ factors for ${}^{12}{\rm C}+ {}^{12}{\rm C}$ fusion. Black solid line is the $S^{*}(E)$ factor from \cite{Nature}. The red dashed line is the renormalized $R(E)\,S^{*}(E)$ factor calculated using the pure Coulomb distortions. The blue dotted line is the renormalized $R(E)\,S^{*}(E)$ factor calculated using the Coulomb plus nuclear distortions.
First published in \cite{MPK2019}.
}
\label{fig_Stot1}
\end{figure}

We conclude from  Fig. \ref{fig_Stot1}  that the inclusion of the distorted waves in the initial and final states eliminates the sharp rise of the $S^{*}(E)$ factors extracted in \cite{Nature} using the THM in the PWA.   Our renormalized $S^\ast$ factors do not, and are not supposed to, exhibit new resonances. They just follow the resonance structure of the astrophysical factors obtained in \cite{Nature}.

Our estimations of the DWBAZR(prior) DCSs of the $\,{}^{12}{\rm C}\,$  transfer reaction show  that in the THM normalization interval of $\,2.5-2.664$ MeV, the DWBA DCSs are of the order of $10^{-4}-10^{-5}$ mb/sr. Such small DCSs can hardly be measured in the coincidence experiment. That is why the THM data are not reliable at higher energies. The absence in the THM data of a strong, isolated resonance at $E \sim 2.1$ MeV observed in  experiment \cite{Spillane} confirms the doubts about the quality of the high-energy THM data at $E > 2$ MeV, which is important for the normalization of the THM data.

\subsection{Concluding remarks}

Concluding this section we emphasise once again that the THM is a powerful and unique indirect technique that allows one to measure the astrophysical factors of the resonant reactions at low energies, where direct methods are not able to obtain data due to very small cross
sections. However, we stress the following points.
\begin{enumerate}
\item We question the validity of the results for the astrophysical factors reported in \cite{Nature} using the PWA. Since the THM deals with three-body reactions rather than binary ones, a reliable theoretical analysis of the  THM data becomes critically important. For the THM reactions with the neutron-spectator or for the reactions with the energies above the Coulomb barrier and for interacting nuclei with small charges, the simple PWA works quite well and the THM results are expected to be reliable. However, this is not the case for the THM reaction under consideration, which aims to determine the astrophysical factors of ${}^{12}{\rm C}+ {}^{12}{\rm C}$ fusion. In this process we deal with the strong Coulomb interactions in the initial and final states of the THM  transfer reaction. 
The PWA follows from the more general distorted-wave-Born approximation (DWBA) in which the distorted waves are replaced by the plane waves and can be used only if the PWA calculations provide a reasonable agreement with DWBA ones. It is not the case under consideration \cite{MPK2019}. It has been demonstrated in \cite{MPK2019}  that the rise of the $S^{*}$ factors at low energies seen in the aforementioned work was an artifact of using the PWA, which is not usable for the case under consideration. It was shown that such a rise disappears if the Coulomb (or Coulomb-nuclear) interactions in the initial and  final states are included.
  A very compelling evidence  that the PWA should not be used is presented in the pioneering work \cite{Goldberg} where the experimental angular distribution of the deuterons from the THM reaction ${}^{12}{\rm C}({}^{14}{\rm N},d){}^{24}{\rm Mg}$ at $33$ MeV incident energy of ${}^{14}{\rm N}$. In this experiment the incident energy was higher than in the THM experiment in \cite{Nature}. Besides, the excited bound state of ${}^{24}{\rm Mg}$ was populated rather than the resonance state.  All these facts make the energy of the  outgoing deuterons higher than the Coulomb barrier in the final  $d+{}^{24}{\rm Mg}$ state (note that in the THM experiment the energies of deuterons were below the Coulomb barrier).  Nevertheless, the experimental angular distribution was flat and was perfectly reproduced by the DWBA calculation,  which agrees with the DWBA calculation in \cite{MPK2019}. Moreover, it follows from \cite{Goldberg} that the momentum distribution of the deuterons disagrees with the one extracted in \cite{Nature}. It casts doubt about the mechanism measured in \cite{Nature}. 
   Note that the deuteron angular distribution was not presented  in \cite{Nature}. The only criterium  used in the THM analysis to justify the PWA  was the deuteron momentum distribution.  It worked for lighter nuclei \cite{spitaleri2019}  but not in the case under consideration in which different competing mechanisms do contribute, such as ${}^{10}{\rm B}$ transfer from the ${}^{12}{\rm C}$ target to ${}^{14}{\rm N}$ or  ${}^{8}{\rm Be}$ transfer to ${}^{12}{\rm C}$ leaving ${}^{6}{\rm Li}$ in the resonance state decaying into $d+\alpha$ channel. 
  The angular correlations of the final-state particles, which provide the most crucial information needed to identify   the reaction mechanism \cite{Goldberg}, are missing in \cite{Nature}.
 \item Another evidence casting doubts on the energy dependence of the total $S^{*}$  factor  from \cite{Nature}  is indicated in \cite{EPJALetter1}.
 This $S^{*}$  factor exceeds significantly the theoretical upper limit calculated in \cite{Esbensen}, see Fig. 1 from \cite{EPJALetter1}.
\item The THM doubly DCS  in \cite{Nature}  does not correspond to the one described by the two-step THM mechanism for the THM reaction ${}^{14}{\rm N} + {}^{12}{\rm C} \to d+ {}^{24}{\rm Mg}^{*} \to \alpha (p)+{}^{20}{\rm Ne} ({}^{23}{\rm Na})$. The doubly DCS for the THM mechanism is given by Eq. (39) from \cite{MPK2019}. Specifically, for the reaction under consideration described by the THM mechanism, the energy of the outgoing deuterons corresponding to the  ${}^{24}{\rm Mg}$ resonance energy at $E=2.1$ MeV is $E_{d{}^{24}{\rm Mg}}=1.47$ MeV while for $E=2.6$ MeV $E_{d{}^{24}{\rm Mg}}=0.97$ MeV. Thus the deuterons are well below the Coulomb barrier of $3$ MeV in the system  $d + {}^{24}{\rm Mg}$. 
Hence, the DWBA DCS ${{d\sigma }^{DWZR(prior)}}/{{d{\Omega _{{{\rm {\bf k}}_{sF}}}}}}$ of the transfer reaction ${}^{14}{\rm N} + {}^{12}{\rm C} \to d+ {}^{24}{\rm Mg}^{*}$, which is the first step of the THM reaction, drops by two orders of magnitude on the interval $2.1 - 2.64$ MeV interval what should lead to the decrease of the THM doubly DCS.  Hence, one can expect that as energy $E$ increases the non-THM mechanisms, which are background, should dominate. These mechanisms were not identified (see the discussion above).
 That is why we doubt that the background did not contribute to the THM  $S^{*}$ factor at $E> 2.2$ MeV (see Fig. 2  from \cite{EPJALetter1}).
\item  The comparison of the THM $S$ factors with the experimental ones is another crucial issue. This discussion concerns a general matter of the application of the THM but is especially important for the application of the THM for heavier nuclei like the case under consideration. The main advantage of the THM is the absence of the Coulomb-centrifugal barrier in the entry channel of the 
${}^{12}{\rm C}-{}^{12}{\rm C}$. That is why in the THM one can observe both low and high $l_{xA}$ resonances.
    There is the price one pays for the absence of the barrier: the transferred particle 
    $x={}^{12}{\rm C}$ in the THM reaction is off-the-energy shell. As the result in the doubly DCS appears the off-shell factor $\big|{\cal W}_{l_{xA}}\big|^{2}$, see Eqs. (39) and  (32) from \cite{MPK2019}.
    This off-shell factor can increase the contribution from high spin resonances, which are suppressed in direct measurements.
    Fig. \ref{fig_PlWl1}  shows the off-shell factor $\big|{\cal W}_{l_{xA}}\big|^{2}$ 
    calculated for the ${}^{12}{\rm C}({}^{14}{\rm N},d){}^{24}{\rm Mg}$ reaction at the incident energy of $\,{}^{14}{\rm N}\,$ $\,30\,$ MeV and different $l_{xA}$. 
    \begin{figure}[htbp]
\includegraphics[width=\columnwidth]{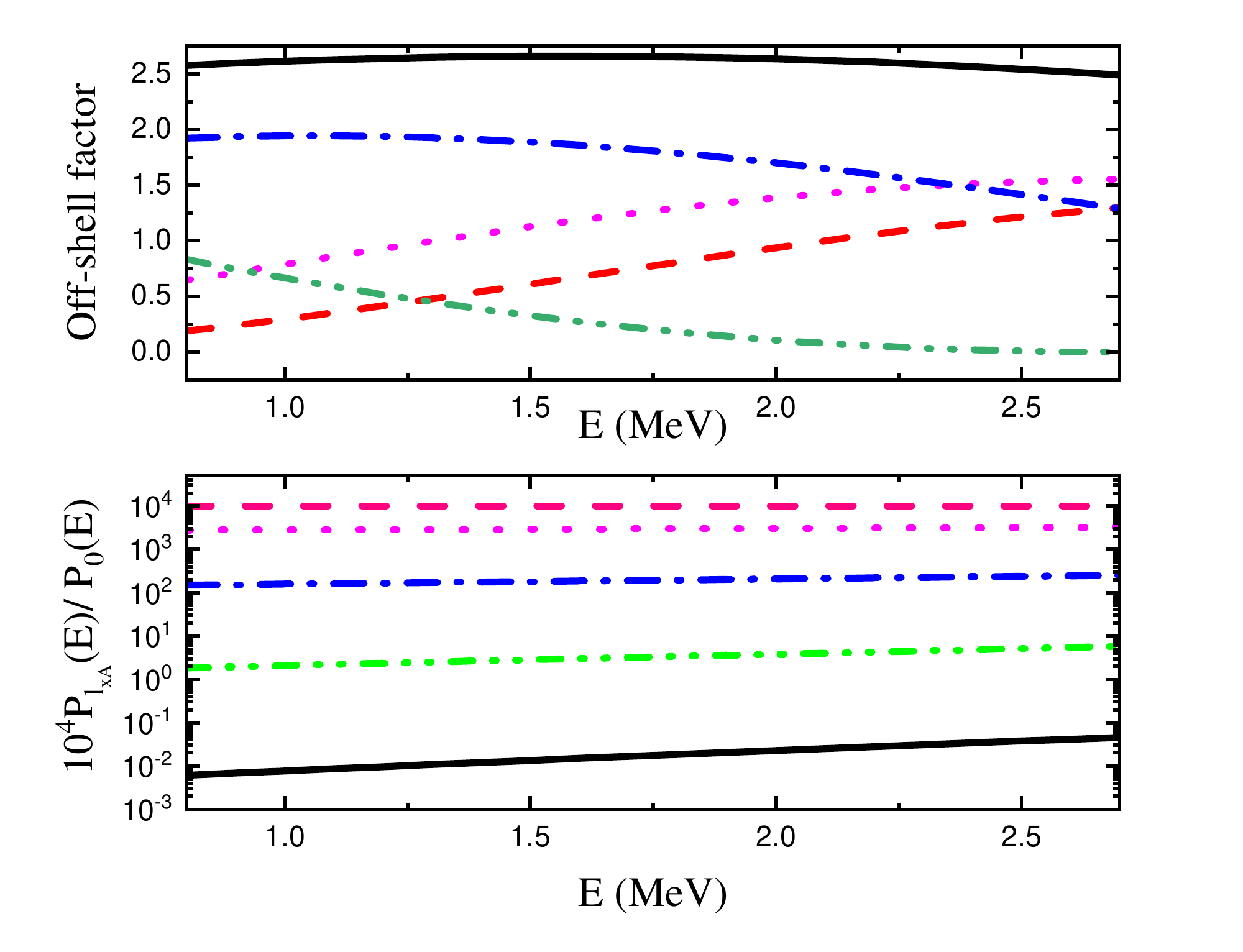}
\caption{(Color online) The upper panel: the off-shell factors $\big|{\cal 
W}_{l_{xA}}\big|^{2}$ for the THM reaction ${}^{12}{\rm C}({}^{14}{\rm N},d){}^{24}{\rm Mg}$ as functions of the ${}^{12}{\rm C}-{}^{12}{\rm C}$
relative energy $E$ at the channel radius $R_{ch}=5$ fm calculated for 5 different ${}^{12}{\rm C}-{}^{12}{\rm C}$ relative orbital angular momenta $l_{xA}$. Red dotted line - $l_{xA}=0$; magenta dotted line- $l_{xA}= 2$; blue dashed-dotted line- $l_{xA}=4$;
green dashed-dotted-dotted line- $l_{xA}=6$ and black solid line - $l_{xA}=8$. The bottom panel: the 
ratio $P_{l_{xA}}/(E)P_{0}(E)$ of the penetrability factors calculated as function of the energy $E$ calculated for the channel radius $R_{ch}= 5$ fm and different $l_{xA}$; $P_{0}$ is the penetrability factor for $l_{xA}=0$. The notations for the lines are the same as in the upper panel. 
First published in\cite{EPJALetter1}.}
\label{fig_PlWl1}
\end{figure}
The THM brings two modifications to the DCS of the binary resonant sub-reactions $x+ A \to F^{*} \to b+B$. It removes the penetrability factor $P_{l_{xA}}$ in the entry channel of the sub-reaction and each partial wave is multiplied by the off-shell factor $\big|{\cal 
W}_{l_{xA}}\big|^{2}$, which may significantly modify the relative weight of the resonances with different $l_{xA}$. The selected channel radius $R_{ch}=5$ fm corresponds to the grazing collision of two carbon nuclei.  From the bottom panel of Fig. \ref{fig_PlWl1}  one can conclude that the dominant contribution to the carbon-carbon fusion in direct measurements at low energies comes from two partial waves: $l_{xA}=0$ and $2$. It is also confirmed by calculations in \cite{ErbBromley}. 
However, in the THM the dominant contribution comes from high spins.  This effect is called kinematical enhancement of higher spin resonances. In particular, in \cite{Nature}  
the resonance spins up to $6$ were assigned at energies below $1.6$ MeV. The  spin $1^{-}$ assigned in \cite{Nature} to the resonance at $0.877$ MeV is a mistake because a resonance with the negative parity cannot be populated in the collision of two identical bosons such as ${}^{12}{\rm C}$ nuclei.  The difference between the low-energy resonance spins contributed to the direct and indirect THM measurements should be taken into account  when comparing the direct and THM $S^{*}$ factors.

\item Despite of the criticism of the analysis in \cite{Nature} we need to acknowledge that 
a compelling evidence of the power of the THM at astrophysically-relevant energies is also clearly demonstrated in \cite{Nature} by discovering   two strong resonances at $0.88$ and $1.5$ MeV. 
This is an undeniable achievement of the THM because such low energies are not yet reachable in direct measurements. 
A strong resonance at $1.5$ MeV, which is inside the Gamow window, is of a crucial importance for the ${}^{12}{\rm C}-{}^{12}{\rm C}$ fusion rates  and will play the same role as the "Hoyle" state \cite{Hoyle54} in the triple-$\alpha$ process of synthesis of ${}^{12}{\rm C}$  \cite{Freer14}. 
Note that the models using the global potentials in \cite{Esbensen,Wiescher,Rowley,Assuncao} can predict resonances only above $E = 2.5$ MeV.
There have been predictions based on phenomenological considerations of explosive stellar events such as superbursts that suggest a  strong $^{12}$C+$^{12}$C cluster resonance around $E = 1.5$ MeV in $^{24}$Mg that would drastically enhance the energy production and may provide a direct nuclear driver for the superburst phenomenon. However, no indication for such a state has not yet been reported in direct measurements in which the minimal measured energy is $E \approx 2.1$ MeV. Hence, the discovery of two strong resonances at $0.88$ and $1.5$ MeV in the THM data is an important contribution of this method in the ${}^{12}{\rm C}-{}^{12}{\rm C}$ fusion research. 
 Improving of the theoretical models, which will be able to predict the low-energy resonances detected in the THM, is another imperative issue.
 
 \item Taking into account the critical importance of the  ${}^{12}{\rm C}+{}^{12}{\rm C}$ fusion in many astrophysical scenarios two major challenges can be outlined: extending direct measurements below $2.1$ MeV with the final goal to reach $1.5$ MeV what would help us to better understand the trend of the $S^{*}$ factors at energies $E < 2.1$ MeV and resolve the question about the validity of the hindrance model. Performing new indirect measurements is another urgent and important issue. One such experiment is in preparation at the Cyclotron Institute, Texas A\&M University~\cite{Rogachev}.

\item 
As a final point, we note that very recent direct measurements \cite{Tan} and \cite{STELLA}  of the experimental $S^{*}$ factors indeed  contradict the ones from \cite{Nature}. This is a practical confirmation of our claim that in the case under consideration the PWA  is not valid.

\end{enumerate}

\section{THM for subthreshold resonances }
\label{subthreshres}

In the previous section we discussed the application of the THM for the resonant rearrangement reaction. Another interesting extension of the THM  is its application for the analysis 
of the rearrangement resonant reactions proceeding through subthreshold resonances. 
A subthreshold bound state (which is close to threshold) reveals itself as a subthreshold resonance in low-energy scattering or reactions. Subthreshold resonances play an important role in low-energy processes, in particular, in astrophysical reactions. 
In this section we consider the application of the THM for analysis of the reactions proceeding through a subthreshold resonance. To describe the subthreshold resonance we use the $R$-matrix approach.
Some notations are modified compared to the ones used in the previous sections.

First, we present new $R$-matrix equations for the reaction amplitudes and astrophysical $S$ factors for analysis of  reactions proceeding through  subthreshold resonances. We consider elastic scattering and a resonant reaction for a subthreshold resonance coupled with open resonance channels for single and two-level cases.  All the equations are expressed in terms of the formal and observable reduced widths.  The observable  reduced width is expressed in terms of the asymptotic normalization coefficient (ANC). We  obtain an equation connecting the ANC with the observable reduced width of the subthreshold resonance, which is coupled with a resonance channel.  Using a generalized $R$-matrix method and the surface-integral method we also derive equations for the  THM reaction amplitude, fully and doubly DCSs in the presence of the subthreshold state. We show that the THM can incorporate equally well the subthreshold and real resonances.  For more details see \cite{muk2017}.


\subsection{Single-channel single-level case}

First we consider  the single-level, single-channel $R$-matrix approach in the presence of the subthreshold bound state (also called subthreshold resonance). 
The resonant elastic-scattering amplitude in the channel $i=x+A$ with the partial wave $l_{xA}$ can be written in the standard $R$-matrix form \cite{lanethomas}:
\begin{eqnarray}
T_{ii} = -2\,i\,e^{-2\,i\,\delta_{i}^{hs}}\,\frac{ P_{i}\,(\gamma_{i}^{(s)})^{2} }{E_{1} - E_{i} -  [S_{i}(E_{i})   - B_{i} + i\,P_{i}]\,(\gamma_{i}^{(s)})^{2} }.\nonumber\\
\label{resTii1}
\end{eqnarray}
Here, $\,\gamma_{i}^{(s)}$ is the reduced width amplitude of the subthreshold bound state $F^{s}=(x\,A)^{(s)}$ with the binding energy $\,\varepsilon_{i}^{(s)} = m_{x} + m_{A} - m_{F^{(s)}}$, $\,m_{j}$ is the mass of the particle $j$, $\,E_{i} \equiv E_{xA}$, $E_{1}$ is the $R$-matrix energy level,  $\,S_{i}(E_{i})  = R_{i} \,Re \Big[{\rm d}\,{\rm ln}O_{l_{xA}}(k_{i},r_{i})/{\rm d}r_{i}\Big|_{r_{i}=R_{i}}\Big]$ is the $R$-matrix shift function in channel $i$,  $r_{i} \equiv r_{xA}$ is the radius connecting centers-off-mass of the particles in the channel $i$ and $k_{i} \equiv k_{xA}$.  $\,B_{i} \equiv B_{l_{xA}}$ is the energy-independent $R$-matrix boundary-condition constant,  $\,P_{i}\equiv P_{l_{xA}}(E_{i},\,R_{i})$ and  $\,R_{i} \equiv R_{ch (i)}$ are the penetrability factor and the channel radius in the channel $i$,  $\,\delta_{i}^{hs} \equiv \delta_{xA\,l_{xA}}^{hs}$   is the hard-sphere scattering phase shift in channel $i$.

If we choose the boundary-condition parameter $\,B_{i}= S_{i}(- \varepsilon_{i}^{(s) } )$,  in the low-energy region  where the linear approximation is valid,
\begin{eqnarray}
S_{i}(E_{i}) - S_{i}(-\varepsilon_{i}^{(s)})  \approx \frac{ {\rm d}\,S_{i}(E_{i}) }{{\rm d}E_{i}}\,\Big|_{E_{i}=-\varepsilon_{i}^{(s)}} (E_i + \varepsilon_{i}^{(s)}),\nonumber\\
\label{linearapproxDelta1}
\end{eqnarray}
then at small $E_{i}$
\begin{eqnarray}
T_{ii} = \frac{2\,i\,e^{-2\,i\,\delta_{i}^{hs}}\, P_{i}({\tilde \gamma}_{i}^{(s)})^{2} }{\varepsilon_{i}^{(s)} + E_{i} +  i\,P_{i}\,({\tilde \gamma}_{i}^{(s)})^{2} },
\label{Tresii2}
\end{eqnarray}
which has a pole at  $\,E_{i}= -\varepsilon_{i}^{(s)}$    because $P_{i}$ vanishes for $E_{i} \leq 0$.   Here,  ${\tilde \gamma}_{i}^{(s)}$ is the observable reduced width of the subthreshold resonance. The observable reduced width $({\tilde \gamma}_{i}^{(s)})^{2}$  is related to the formal $R$-matrix  reduced width $ ( \gamma_{i}^{(s)})^{2}$  as
\begin{eqnarray}
({\tilde \gamma}_{i}^{(s)})^{2}= \frac{  ( \gamma_{i}^{(s)})^{2} }{ 1+(\gamma_{i}^{(s)})^{2} [{\rm d}S_{i}(E_{i})/{\rm d}E_{i}]\big|_{E_{i}=-\varepsilon_{i}^{(s)}}}.
\label{effredwidth1}
\end{eqnarray}
Equation (\ref{Tresii2}) is of fundamental importance because it shows that a subthreshold bound state can be described as a subthreshold resonance, that is, the resonance located at the  negative energy 
$-\varepsilon_{i}^{(s)}$. 

Determining the ANC as the residue in the pole of the scattering amplitude corresponding to the bound-state pole, we get for the ANC of the subthreshold state \cite{muk99}
\begin{eqnarray}  
[C_{i}^{(s)}]^{2}=   \frac{2\,\mu_{i}\,R_{i}\, ({\tilde \gamma}_{i}^{(s)})^{2} }{W_{-\,\eta_{i}^{(s)},\,l_{xA}+1/2}^{2}(  2\,\kappa_{i}^{(s)}\,R_{i})},
\label{ANCredwidth1}
\end{eqnarray}
where   $W_{-i\,\eta_{i}^{(s)},\,l_{xA}+1/2}(\,2\,\kappa_{i}^{(s)}\,R_{i})$ is the Whittaker function, $\eta_{i}^{(s)}= Z_{x}\,Z_{A}/\alpha\mu_{i}/\kappa_{i}^{(s)}$  and $\kappa_{i}^{(s)}$ are the $x-A$ Coulomb parameter and the bound-state wave number of the subthreshold state $F^{(s)}$,  $\,\mu_{i}$ is the reduced mass of $x$ and $A$ in the channel  $i$,  $\,Z_{j}\,e$ is the charge of nucleus $j$.
Equation (\ref{ANCredwidth1}) relates the ANC and the 
the observable reduced width for single-channel case, but in the next section this equation  will be generalized for the two-channel case.

The observable partial resonance width of the subthreshold resonance is given by 
\begin{align}
{\tilde \Gamma}_{i}^{(s)}(E_{i}) &= 2\,P_{i}\,({\tilde \gamma}_{i}^{(s)})^{2} \nonumber\\
&= P_{i}\,\frac{\Big(C_{i}^{(s)}\,W_{-\,\eta_{i}^{(s)}\,l_{xA}+1/2}(2\,\kappa_{i}^{(s)}\,R_{i}) \Big)^{2}}{\mu_{i}\,R_{i}}.
\label{GammaANC1}
\end{align}
Eq. (\ref{GammaANC1}) expresses the resonance width of the subthreshold resonance in terms of the  ANC and the Whittaker function of this subthreshold state taken at the channel radius $R_{i}$. 
We have considered the connection between the ANC and the reduced width amplitude for the bound states.
In the next subsection we consider the connection between the ANC and the observable resonance width
for real resonances.

\subsection{Two-channel single-level case}  

Now  we consider elastic scattering $x+ A \to x+A$ in the presence of the subthreshold bound state $F^{(s)}$ in the channel $i =x+ A$ which is coupled with  the second channel $f=b+B$. The relative kinetic energies in the channel $i$, $E_{i} \equiv E_{xA}$, and channel $f$,  $E_{f} \equiv E_{bB}$,  are related by 
\begin{align}
E_{f} \equiv E_{bB} = E_{i} + Q,  \quad Q= m_{x} + m_{A} - m_{b} - m_{B} > 0.
\label{EfEiQ1}
\end{align}
We assume that $\,Q>0$, that is, the channel $f$ is open for $E_{i} \geq 0$.        

The resonance part of the elastic-scattering amplitude in the channel $i=x+A$ in the single-level two-channel $R$-matrix approach is
\begin{align}
T_{ii} =& -2\,i\,e^{-2\,i\,\delta_{i}^{hs}}\nonumber\\
&\times \frac{ P_{i}\,(\gamma_{i}^{(s)})^{2} }{  E_{1} - E_{i} -  \sum\limits_{n=i,f}\,[S_{n}(E_{n}) - B_{n} + i\,P_{n}]\,\gamma_{n}^{2}},
\label{2chTii1}
\end{align}
where $\gamma_{n}$ is the formal reduced width in the channel $n=i,\,f$. Note that $\gamma_{i}  \equiv \gamma_{i}^{(s)}$. 
 $\,P_{n} \equiv P_{l_{n}} (E_{n},\,R_{n})$ and $\,R_{n}$  are   the penetrability factor and the channel radius in the channel $n$.  There are two fitting parameters, $\gamma_{i}^{(s)}$  and $\gamma_{f}$, in the single-level two-channel $R$-matrix fit at fixed channel radii.

Again, we use the boundary condition 
$B_{n}= S_{n}(-\varepsilon_{i}^{(s)})$ and $E_{1}=-\varepsilon_{i}^{(s)}$. 
The energy in the channel $i$ is $E_{i}=-\varepsilon_{i}^{(s)}$  and that in the channel $f$ is $E_{f}= Q- \varepsilon_{i}^{(s)}$. Assuming a linear energy dependence of $S_{n}(E_{n})$ at small $E_{i}$, we get
\begin{eqnarray}
T_{ii} \approx 2\,i\,e^{-2\,i\,\delta_{i}^{hs}}\,\frac{P_{i}\,\big({\tilde\gamma}_{i}^{(s)}\big)^{2} }{\varepsilon_{i}^{(s)} + E_{i}  +   i\sum\limits_{n=i,f}\,P_{n}\,{\tilde \gamma}_{n}^{2}}, 
\label{2chTii12} 
\end{eqnarray}                 
where the observable reduced width  in the channel $n$ is 
\begin{eqnarray}
{\tilde \gamma}_{n }^{2}= \frac{   \gamma_{n}^{2} }{  1+\sum\limits_{t=i,f}\,\gamma_{t }^{2}[{\rm d}S_{t}(E_{t})/{\rm d}E_{t}]\big|_{E_{t}=E_{t}^{(s) } }},
\label{tildegammapr1a}
\end{eqnarray}
again noticing that $E_{i}^{(s)}= -\varepsilon_{i}^{(s)}$ and $E_{f}^{(s)}=Q - \varepsilon_{i}^{(s)}$.
Correspondingly, the observable partial resonance width in the channel $n$ is
\begin{eqnarray}
{\tilde \Gamma}_{n}(E_{n}) = 2\,P_{n}\,{\tilde \gamma}_{n}^{2},
\label{Gamman1}
\end{eqnarray}
with the total width ${\tilde \Gamma}(E_{i}) = {\tilde \Gamma}_{i}^{(s)}(E_{i}) + {\tilde \Gamma}_{f}(E_{f})$.  

The presence of the open channel coupled to the elastic-scattering channel generates an additional term $\,n=f$ in the denominators of Eqs. (\ref{2chTii1}), (\ref{2chTii12}) and (\ref{tildegammapr1a}). 
Although  the scattering amplitude vanishes at $E_{i}=0$ it can be extrapolated  to the bound-state pole bypassing $E_{i}=0$ using
\begin{align}
T_{ii}  \stackrel{E_{i} \to -\varepsilon_{i}^{(s)}}{\approx}&   2\,\kappa_{i}^{(s)}\,(-1)^{l_{xA}+1}\,e^{i\,\pi\,\eta_{i}^{(s)}}\nonumber\\
&\times \frac{R_{i}}{W_{-\eta_{i}^{(s)},l_{xA}+1/2}^{2}(2\,\kappa_{i}^{(s)}\,R_{i})}\,
\frac{ ( {\tilde \gamma}_{i}^{(s)} )^{2}  }{ \varepsilon_{i}^{(s)} + E_{i} + i\,P_{f}\,{\tilde \gamma}_{f}^{2} }, 
\label{Tiipoleextrap2ch2}
\end{align}
where $\,P_{f} =P_{f}(Q- \varepsilon_{i}^{(s)},R_{f})$.
Again, the ANC, as a residue in the pole of the scattering amplitude, is related to the observable reduced width $( {\tilde \gamma}_{i}^{ (s)} )^{2}$ of the subthreshold state by Eq. (\ref{ANCredwidth1}), in which 
now $({\tilde \gamma}_{i}^{(s)})^{2}$ is determined by 
\begin{eqnarray}
({\tilde \gamma}_{i}^{(s)})^{2}= \frac{ (\gamma_{i}^{(s)})^{2} }{  1+\sum\limits_{t=i,f}\,\gamma_{t }^{2}[{\rm d}S_{t}(E)/{\rm d}E]\big|_{E=E^{(s) } }}.
\label{tildegammapr1}
\end{eqnarray}
The derivation of the connection between the ANC and the reduced width of the subthreshold resonance  in the presence of an open channel $f$ is a generalization of Eq. (\ref{ANCredwidth1}).

It follows from Eq. (\ref{Tiipoleextrap2ch2})  that in the presence of the open channel $f$ coupled with the channel $i$, the 
elastic-scattering amplitude has the bound-state pole  shifted into the $E_{i}$ complex plane, i.e.,
$E_{i}^{(p)} = - \varepsilon_{i}^{(s)}  - i\,P_{f}(Q- \varepsilon_{i}^{(s)},R_{f})\,({\tilde \gamma}_{f})^{2}$.

We have established a connection between the ANC, the observable reduced width and the observable resonance width (at $E_{i} >0$) of the subthreshold resonance state. But, besides the subthreshold bound state in the channel $i$, in Eq. (\ref{2chTii12}) we have also a real resonance in the channel $f$. In Refs. \cite{izvAN,muk99,ANCGamma} the definition of the ANC was also extended to real resonances.
 Here we remind the connection between the resonance width, the ANC, and the reduced width for the real resonance in the channel $f$ whose real part of the complex resonance energy is located at $E_{f}^{(0)}$.  
 The ANC for the resonance state is determined as the amplitude of the outgoing resonance wave \cite{ANCGamma}, which is the generalization of the definition of the ANC for the bound state. For narrow resonances the ANC is expressed in terms of the resonance width as \cite{muk99,izvAN,ANCGamma}:
\begin{align}
C_{f}^{2} = \,\frac{\mu_{f}\,{\tilde \Gamma}_{f}(E_{f}^{(0)})}{k_{f}^{(0)}}\,
e^{i[2\,\delta^{p}_{l_{f}}(k_{f}^{(0)})- i\,\pi\,l_{f}]}.
\label{ANCGammanawrres1}
\end{align}
Here,  $C_{f}$  is the ANC of the resonance state in the channel $f$,  ${\tilde \Gamma}_{f}(E_{f}^{(0)})$  is the observable resonance width in the channel $f$ at the real part of the resonance energy,  $\delta^{p}_{l_{f}}(k_{f}^{(0)})$ is the non-resonant scattering phase shift in the channel $f$ at the real part of the resonance momentum.

\subsection{$S$ factor for reaction proceeding through subthreshold resonance  in $\mathbf{{}^{17}{\rm O}}$}

 In this subsection we present equations for the amplitudes of reactions proceeding through a subthreshold resonance  within the standard $R$-matrix equations generalized for the subthreshold state. Based on these amplitudes we obtain the corresponding  astrophysical S factors which can be used  to analyze experimental data obtained from direct and indirect measurements.  Note that here the expressions for the astrophysical factors are written in a convenient $R$-matrix form and  can be used by experimentalists for the analysis  of  similar reactions proceeding through  subthreshold resonances.

In section \ref{critanalalysisTHM1} we considered the application of the THM for the analysis of the 
binary resonant reaction in which both coupled channels, $x+A$ and $b+B$ were open.
Here we consider the resonant reaction 
\begin{equation}
x + A \to F^{(s)} \to b +B
\label{reaction1}
\end{equation}
with $Q>0$, proceeding through an intermediate resonance,  which is a resonance in the exit channel $f$ and  the subthreshold bound state  $F^{(s)}= (x\,A)^{(s)}$ in the initial channel $i$.  We assume also that $Q- \varepsilon_{i}^{(s)} >0$, that is, the channel $f$ is open at the subthreshold bound-state pole in the channel $i$. 
 
The single-level two-channel $R$-matrix amplitude describing  the resonant reaction in which in the initial state the colliding particles $x$ and $A$ have a subthreshold bound state and a resonance in the final channel $f=b+B$, can be obtained by generalizing the corresponding equations from  Refs. \cite{lanethomas, barker2000}:
\begin{eqnarray}
T_{f\,i}&=&  2\,i\,e^{-\,i(\delta_{i}^{hs} +  \delta_{f}^{hs} ) }\nonumber\\
&&\times \frac{  \sqrt{  P_{f}  }\,\gamma_{f}\, \sqrt{ P_{i} }\,\gamma_{i }^{(s)} }{\varepsilon_{i}^{(s)} + E_{i} +\sum\limits_{n=i,f}\,[S_{n}(E_{n}) - S_{n}(E_{n}^{(s)}) + i\,P_{n}]\,\gamma_{n}^{2}}. \nonumber\\
\label{RmatrRes1}
\end{eqnarray}
Here we remind that $E_{i}^{(s)}= -\varepsilon^{(s)}$ and $E_{f}^{(s)}= Q - \varepsilon^{(s)}$. 
The astrophysical factor $S(E_{i})$  is given by
\begin{widetext}
\begin{eqnarray}
S(E_{i})({\rm keV\,b})= \,\frac{{\hat J}_{F^{(s)}}}{{\hat J}_{x}\,{\hat J}_{A}}\,\lambda_{N}^{2}\,\,E_{N}^{2}\,e^{2\,\pi\,\eta_{i}}\,\frac{20\,\pi}{\mu_{i}}\frac{ P_{f}\,P_{i}\,\gamma_{f}^{2}\,(\gamma_{i}^{(s)})^{2} }{\Big(\varepsilon_{i}^{(s)}  +E_{i} + \sum\limits_{n=i,f}\,[S_{n}(E_{n}) - S_{n}(E_{n}^{(s)})]\,{\gamma}_{n}^{2}\Big)^{2} + \Big[\sum\limits_{n=i,f}\,P_{n}\,{\gamma}_{n}^{2}\Big] ^{2}} ,
\label{Sfactr1l2ch1}
\end{eqnarray}
\end{widetext}
where  $J_{F^{(s)}}$ is the spin of the subthreshold state in the channel $i=x+A$, which is also the spin of the resonance in the channel $f=b+B$, $\,J_{j}$ is the spin of the particle $j$,   ${\hat J}= 2\,J+1$,   $\,\lambda_{N}=0.2118$ fm is the nucleon Compton wavelength,  $E_{N}=931.5$ MeV is the atomic unit mass. All the reduced width amplitudes are expressed in MeV$^{1/2}$.

Assume now that the low-energy binary reaction (\ref{reaction1}) is contributed by a few non-interfering levels. The subthreshold resonance in the channel $\,i=x+A$, which is coupled to the open channel $f=b+B$ is attributed to the first level, $\lambda=1$, while other levels with $\lambda >1$ are attributed to two open coupled channels  $i$ and $f$ of higher energy levels  $E_{\lambda}$ and  spins $J_{F^{\lambda}}$.
The astrophysical factor $S(E_{i})$ is given by
\begin{align}
S(E_{i})({\rm keVb})= \sum\limits_{\lambda=1}^{N}\,S_{\lambda}(E_{i})({\rm keVb}),
\label{Sfctrsumlambda1}
\end{align}
where
\begin{widetext}
\begin{align}
S_{\lambda}(E_{i})({\rm keVb})=
\nu_{N}^{2}\,E_{N}^{2}\,\frac{20\,\pi}{\mu_{i}}\,e^{2\,\pi\,\eta_{i}}\,\frac{ {\hat J}_{F^{(\lambda)}}    }{{\hat J}_{x}\,{\hat J}_{A}}\,  
 \frac{  P_{f_{\lambda}}\,P_{i_{\lambda}}\,\gamma_{f_{\lambda}}^{2}\,\gamma_{i_{\lambda}}^{2} }{ \Big(E_{\lambda} - E_{i}  - \sum\limits_{n=i,f}\,[S_{n\,\lambda}(E_{n}) - B_{n\,\lambda}) ]\,{\gamma}_{n\,\lambda}^{2}\Big)^{2} + \big[\sum\limits_{n=i,f}\,P_{n\,\lambda}\,\gamma_{n\,\lambda}^{2}\big] ^{2} }.  \nonumber\\
\label{Sfactrlambda2}
\end{align}   
\end{widetext}
Here, all the quantities with the subscripts $n,\lambda$  correspond to the channel $n$ and level $\lambda$, 
$\,\gamma_{i\,\lambda}$ and $\gamma_{f\,\lambda}$  are the reduced width amplitudes of the resonance $F^{(\lambda)}$   in the initial and final channels, $\gamma_{i\,1} \equiv \gamma_{i}^{(s)}$, $\,E_{\lambda}$ is the energy level in the channel $\lambda$.									

Now we consider  two interfering levels, $\lambda=1\,$ and $\,2$, and two channels in each level.  All the quantities related to the levels $\lambda=1$ and $2$ have additional subscripts $1$ or $2$, respectively.  We  assume that the level $\lambda=1$ corresponds to the subthreshold state  in the channel $i=x+A$, which decays to a resonant state corresponding to the level $\lambda=1$  in the channel $f=b+B$. The level $2$ describes the resonance in the channel $x+A$, which decays into the resonant state in the channel $f= b+ B$.  The level $\lambda =2$ lies higher than the level $\lambda=1$ but these levels do interfere.  
The reaction amplitude is given by
\begin{eqnarray}
T_{f\,i}=-2\,i\,e^{-i(\delta^{hs}_{f}+ \delta^{hs}_{i} )} \sqrt{P_{f}}\,\sqrt{ P_{i} }\sum\limits_{\lambda\,\tau}\gamma_{f_{\lambda}}\,{\rm {\bf A}}_{\lambda\,\tau}\,\gamma_{i_{\tau}}.
\label{M2lev2ch1}
\end{eqnarray}
Here, ${\rm {\bf A}}$   is the level matrix in the $R$-matrix method, 
\begin{align}
\big({\rm {\bf A}}^{-1}\big)_{\lambda\,\tau}=& (-\varepsilon_{i}^{(s)} - E_{i} )\,\delta_{\lambda\,\tau} - \sum\limits_{n=i,f}\,\gamma_{n\,\lambda}\, \gamma_{n\,\tau}\,
\nonumber \\ &\times
\big[S_{n}(E_{n}) - S_{n}(E_{n}^{(s)}) + i\,P_{n} \big]. \nonumber\\ 
\label{ARmatr1}
\end{align}

The corresponding astrophysical $S(E_{i})$ factor is
\begin{align}
S(E_{xA})({\rm keVb})=& 20\,\pi\,m_{N}^{2}\,E_{N}^{2}\frac{{\hat J}_{F^{(s)}}}{{\hat J}_{x}\,{\hat J}_{A}}\,\frac{1}{\mu_{i}}\,e^{2\,\pi\,\eta_{i}}
 P_{f}\,P_{i}\,  
 \nonumber \\ & \times
 \Big|\sum\limits_{\lambda\,\tau}\,\gamma_{f_{\lambda}}\,{\rm {\bf A}}_{\lambda\,\tau}\,\gamma_{i_{\tau}}\,\Big|^{2}.
\label{Sfactr2l2ch1}
\end{align}

\subsection{$^{13}{\rm C}(\alpha,\,n)^{16}{\rm O}$ reaction}

In this subsection we present the THM analysis of the 
important astrophysical reaction
$^{13}{\rm C}(\alpha,\,n)^{16}{\rm O}$, which is considered to be the main neutron supply to build up heavy elements from iron-peak seed nuclei in AGB stars. At temperature $0.9 \times 10^{8}$ K, the energy range where the $^{13}{\rm C}(\alpha,\,n)^{16}{\rm O}$ reaction is most effective, the so-called Gamow window \cite{illiadis} is within $\approx 140 - 230$ keV with the most effective energy at $\approx 190$ keV.  This reaction was studied using both direct and indirect (TH) methods.  Direct data, owing to the small penetrability factor, were measured with reasonable accuracy down to $E_{\alpha\,{}^{13}{\rm C}} \approx 400$ keV. Data in the interval $300 - 400 $ keV were obtained with much larger uncertainty \cite{davids68,bairhaas,drotleff,brune,Harissopulos}.  In Ref. \cite{Harissopulos}  the unprecedented accuracy of $4\%$  was achieved at energies $E_{\alpha\,{}^{13}{\rm C}}> 600$ keV.  The dominant contribution to the $^{13}{\rm C}(\alpha,\,n)^{16}{\rm O}$  reaction at astrophysical energies comes from the  state ${}^{17}{\rm O}(1/2^{+}, E_{x}=6356 \pm 8 \, {\rm keV})$, where $E_{x}$ is the excitation energy. Taking into account that the $\alpha- {}^{13}{\rm C}$ threshold is located at $6359.2$ keV one finds that this $1/2^{+}$ level is located at $E_{\alpha\,{}^{13}{\rm C}} = -3 \pm 8$ keV, that is, it can be either subthreshold bound state or a resonance \cite{tilley}. This location of the level ${}^{17}{\rm O}(1/2^{+})$ was adopted in the previous analyses of the direct measurements including the latest one in Ref. \cite{heil}. If this level is a subthreshold bound state, then its reduced width is related to ANC of this level.

 However,  in the recent paper \cite{Faestermann} it has been determined that this level is actually a resonance located at $E_{\alpha\,{}^{13}{\rm C}} = 4.7 \pm 3$ keV with the total observable width  of ${\tilde \Gamma} =136 \pm 5 $ keV.  Note that ${\tilde \Gamma}_{\alpha}$ of this resonance with $l_{xA}=1$ is negligibly small because it contains the penetrability factor $P_{1}$.  Hence, ${\tilde \Gamma}= {\tilde \Gamma}_{n}$.
The result obtained in \cite{Faestermann} is a very important achievement in the long history of hunting for this near-threshold level. If this level is actually a resonance located slightly above the threshold then the reduced width is related to the resonance partial $\alpha$ width rather than to  the  ANC. Evidently this resonance is not a Breit-Wigner type and it does not make sense to use the ANC as characteristics of this resonance.

Here we present calculations of the astrophysical $S$ factors for the $^{13}{\rm C}(\alpha,\,n)^{16}{\rm O}$ using the  equations derived above.   We fit the latest TH data \cite{oscar}  using assumptions that the threshold level $1/2^{+}$  is the subthreshold state  located at $-3$ keV and  the resonance state at $4.7$ keV. For the subthreshold state we use parameters from \cite{heil} while for the resonance state we adopt parameters from \cite{Faestermann}. The resonances included in the analysis of this reaction are ($\,1/2^{+},\,l_{xA}=1,\,E_{x}=6.356$ MeV), ($\,5/2^{-},\,l_{xA}=2,\,E_{x}=7.165$ MeV), ($\,3/2^{+},\,l_{xA}=1,\,E_{x}=7.216$ MeV), ($5/2^{+},\,l_{xA}=3,\,E_{x}=7.379$ MeV) and ($5/2^{-},\,l_{xA}=2,\,E_{x}=7.382$ MeV). Only two resonances, the second and the last one have the same quantum numbers and do interfere. Their interference can be taken into account  using  the S factor given by Eq. (\ref{Sfactr2l2ch1}). For non-interfering resonances we  use Eq. (\ref{Sfctrsumlambda1}). 

In Fig. \ref{fig_energdepres1} we present the $S$ factors  contributed by four different resonant states located at $E_{\alpha\,{}^{13}{\rm C}}>0$.  All the parameters of these resonances are taken from \cite{heil}. We only slightly modified the $\alpha$-particle width of the wide resonance at $E_{\alpha\,{}^{13}{\rm C}}=0.857$ MeV taking it to be $0.12$ keV. The adopted channel radii are 
$R_{i}= 7.5$ fm (channel ${\alpha + {}^{13}{\rm C}})$ and $R_{f}= 6.0$ fm (channel $n+{}^{16}{\rm O}$).

\begin{figure}[ht]
\includegraphics[width=\columnwidth]{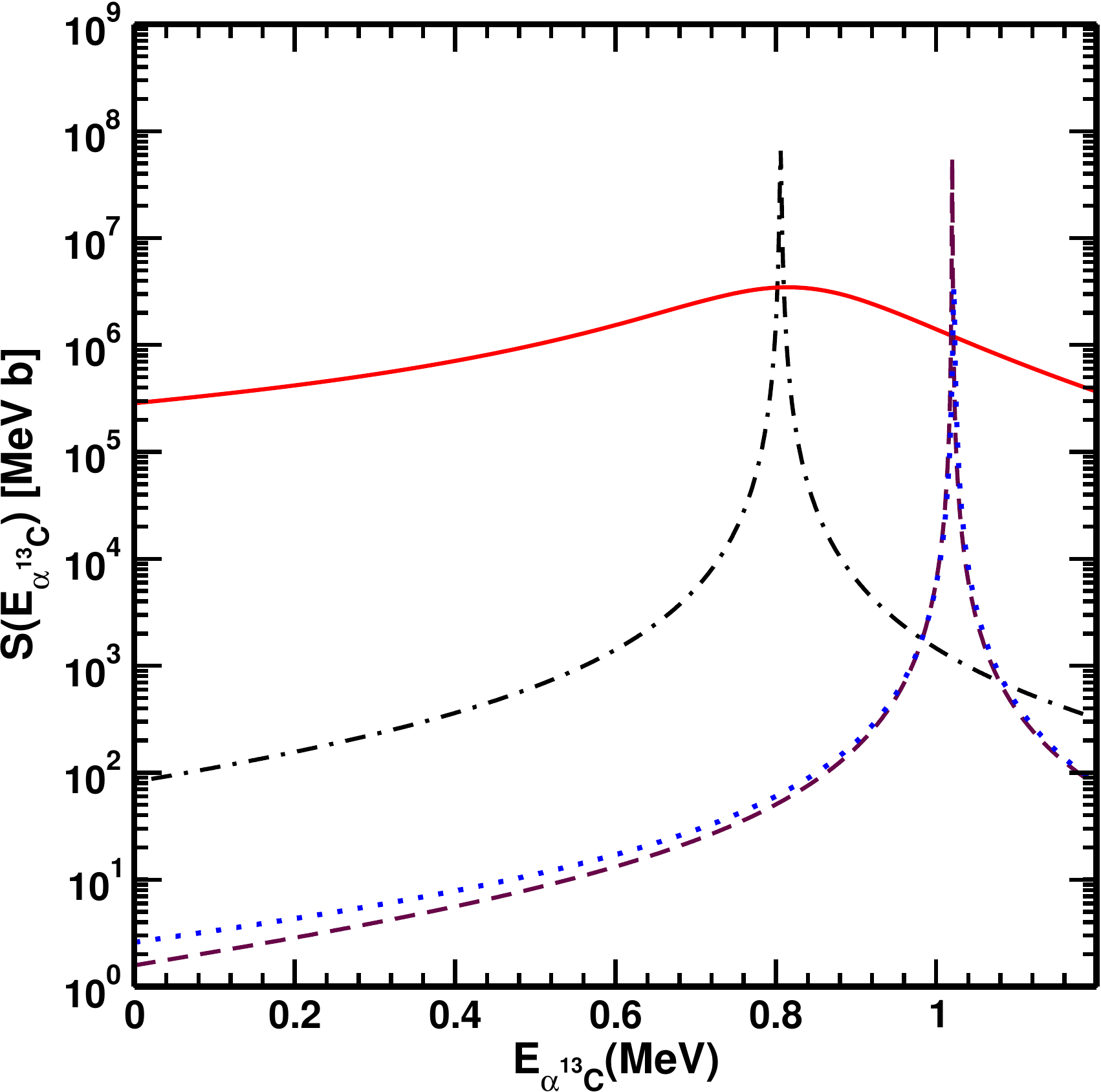}
\caption{(Color online) 
The $S$ factors for the $^{13}{\rm C}(\alpha,\,n)^{16}{\rm O}$ reaction as a function of the $\alpha-{}^{13}{\rm C}$ relative kinetic energy
proceeding through four resonances: black dotted-dashed line -($\,5/2^{-},\,l_{xA}=2,\,E_{x}=7.165$ MeV; solid red line- ($\,3/2^{+},\,l_{xA}=1,\,E_{x}=7.216$ MeV); dashed brown line-($5/2^{+},\,l_{xA}=3,\,E_{x}=7.379$ MeV); dotted blue line- ($5/2^{-},\,l_{xA}=2,\,E_{x}=7.382$ MeV). All the resonant parameters are taken from \cite{heil}.
First published in  \cite{mukshbert2017}.}  
\label{fig_energdepres1}
\end{figure}

As we see from  Fig. \ref{fig_energdepres1}, the contributions of  all the narrow resonances are negligible compared to the wide one (red solid line in Fig. \ref{fig_energdepres1}). That is why we do not take into account the interference between two narrow $5/2^{-}$ resonances. 
Thus eventually we can take into account only the wide resonance ($\,3/2^{+},\,l_{xA}=1,\,E_{x}=7.216$ MeV) and the near-threshold level ($\,1/2^{+},\,l_{xA}=1,\,E_{x}=6.356$ MeV).

\subsection{Threshold level  $\,1/2^{+},\,l_{xA}=1,\,E_{x}=6.356$ MeV}
 
Here we would like to discuss the threshold  level $E_{x}=6.356$ MeV. Until recently 
this level was considered to be the subthreshold resonance located at $E_{\alpha\,{}^{13}{\rm C}} = -3$ keV.
However, as we have mentioned,  in Ref. \cite{Faestermann} this level was shifted to the continuum and is found to be a real resonance located at $E_{\alpha\,{}^{13}{\rm C}}= 4.7$ keV. The astrophysical factor contributed by this $1/2^{+}$ state depends on the  reduced width in the entry channel $\alpha-{}^{13}{\rm C}$ of the $^{13}{\rm C}(\alpha,\,n)^{16}{\rm O}$ reaction and the reduced width in the exit channel $n-{}^{16}{\rm O}$. The latter is determined with an acceptable accuracy, for example, in Refs. \cite{heil,Faestermann}.
If we assume that the level $E_{x}=6.356$ MeV is subthreshold resonance then its reduced width in the $\alpha$-channel is expressed in terms of the ANC for the virtual decay ${}^{17}{\rm O}(1/2^{+},E_{\alpha\,{}^{13}{\rm C}} = -3\,\,{\rm keV} )
\to \alpha + {}^{13}{\rm C}$. This ANC was found in a few papers \cite{pellegriti,guo, marco2}. The latest measurement of this ANC was published in \cite{avila}: ${\tilde C}_{\alpha\,{}^{13}{\rm C}}^{(s)}= 3.6 \pm 0.7$ fm${}^{-1}$, which is the Coulomb renormalized ANC. The problem is that at very small binding energies the ANC of the subthreshold state becomes very large due to the Coulomb-centrifugal barrier. That is why in \cite{blokh84,muk2012}  the Coulomb renormalized ANC was introduced in which the Coulomb-centrifugal factor was removed:
\begin{align}
{\tilde C}= \frac{l_{xA}!}{\Gamma(1+ l_{xA}+ \eta^{(s)})}\,C.
\label{ANCCoulren1}
\end{align} 
Here, $\,\Gamma(x)\,$ is the Gamma function, $\,l_{xA}\,$ is the orbital angular momentum of the bound state. At small binding energies of the bound state, that is, at large $\,\eta^{(s)}$, the factor $\,\Gamma(1+ l_{xA}+ \eta^{(s)})\,$ becomes huge. Usually we are used to see the barrier factor to decrease the cross section, but here we see the opposite effect.  

However, in the $R$-matrix approach the quantity, which we need to calculate the astrophysical $S$ factor, is the reduced width. 
From Eq. (\ref{ANCredwidth1}) one can express the observable reduced width of the bound state  in terms of the ANC: 
\begin{align}  
{\tilde \gamma}^{2}= \frac{C^{2}\,W_{-\,\eta^{(s)},\,l_{xA}+1/2}^{2}(2\,\kappa_{i}^{(s)}\,R_{i}
)}{2\,\mu\,R_{i}}.
\label{redwidthANC2}
\end{align}
 The Coulomb-barrier factor, which significantly enhances the ANC,  has an opposite effect on the Whittaker function $W_{-\,\eta^{(s)},\,l_{xA}+1/2}(2\,\kappa_{i}^{(s)}\,R_{i})$, so that the product
$C\,W_{-\,\eta^{(s)},\,l_{xA}+1/2}(2\,\kappa^{(s)}\,R)$ is unaffected by the Coulomb-centrifugal barrier factor.  It is convenient to rewrite 
Eq. (\ref{redwidthANC2}) as
\begin{align}  
{\tilde \gamma}^{2}= \frac{{\tilde C}^{2}\,{\tilde W}_{-\,\eta^{(s)},\,l_{xA}+1/2}^{2}(2\,\kappa_{i}^{(s)}\,R_{i})}{2\,\mu_{i}\,R_{i}},
\label{redwidthANC3}
\end{align}
where $\mu_{i}=\mu_{\alpha\,{}^{13}{\rm C}}$, $\kappa_{i}^{(s)}= \sqrt{2\,\mu_{i}\varepsilon_{i}^{(s)}}$,
$\varepsilon_{i}^{(s)}=-3$  keV. Also
\begin{align}
{\tilde W}_{-\,\eta^{(s)},\,l_{xA}+1/2}^{2}(2\,\kappa^{(s)}\,R_{i}) =& \frac{\Gamma(1+ l_{xA}+ \eta^{(s)})}{l_{xA}!}
\nonumber\\& \times
W_{-\,\eta^{(s)},\,l_{xA}+1/2}^{2}(2\,\kappa^{(s)}\,R_{i}).
\label{tildeW1}
\end{align}
For example, for the case under consideration, if the subthreshold bound state is located at $-3$ keV then
$\Gamma(1+ l_{xA}+ \eta^{(s)})=2.406 \times 10^{84}$ for $l_{xA}=1$. For the channel radius $R_{i}=7.5$ fm,  
$W_{-\,\eta^{(s)},\,l_{xA}+1/2}(2\,\kappa^{(s)}\,R_{i})=2.44122 \times 10^{-86}$ while ${\tilde W}_{-\,\eta^{(s)},\,l_{xA}+1/2}(2\,\kappa_{i}^{(s)}\,R_{i})= 0.0587$.  Correspondingly, 
\begin{align}
C\,W_{-\,\eta^{(s)},\,l_{xA}+1/2}(2\,\kappa_{i}^{(s)}\,R_{i})&= {\tilde C}\,{\tilde W}_{-\,\eta^{(s)},\,l_{xA}+1/2}(2\,\kappa_{i}^{(s)}\,R_{i})\nonumber\\
&= 0.111 \, {\rm fm}{}^{-1/2}.
\label{CWtildeCW1}
\end{align}

The reduced width changes very little if we assume that the threshold level $1/2^{+}$ is a bound state. We used the single-particle $\alpha-^{13}{\rm C}$
Woods-Saxon potential to generate the bound-state wave function with the binding energy $-3$ keV.  This function has three nodes at $r_{i}>0$. Following the $R$-matrix procedure, we accepted the internal region as $0 \leq r_{i} \leq R_{i}$, where $R_{i}=5.2$ fm is the location of the last peak of the internal wave function, and calculated the wave function, which is normalized over the internal region, at  $R_{i}=5.2$ fm. The obtained value can give estimation of the single-particle reduced width amplitude.  
After that we adopted the binding energy as $-0.1$ keV and repeated the similar procedure and found by decreasing the well-depth that $R_{i}=4.93$ fm. The  value of the single-particle reduced width decreased only by $2.5\%$ compared to the value for the binding energy of $-3$ keV. Because the reduced width of the 
resonance state at $4.7$ keV is unknown and we are not able to reproduce this state using a single-particle Woods-Saxon potential, as we did for the bound states, 
we assume that the reduced width for the resonance state is close to the reduced width for the bound state $-3$ keV, which is $3.3$ keV${}^{1/2}$ for the ANC
${\tilde C}=1.9$ fm${}^{-1/2}$ and $R_{i}=7.5$ fm. To make the fit  to the TH data \cite{oscar} we adopted the reduced width for the resonance state $4.7$ keV  in the interval $\big(2.81-3.6 \big)$ keV${}^{1/2}$. Note that  $R_{i}= 7.5$ fm provided the best fit of the TH data.

\subsection{Low-energy astrophysical factor for ${}^{13}{\rm C}(\alpha,\,n){}^{16}{\rm O}$}

\begin{figure}[ht]
\includegraphics[width=\columnwidth]{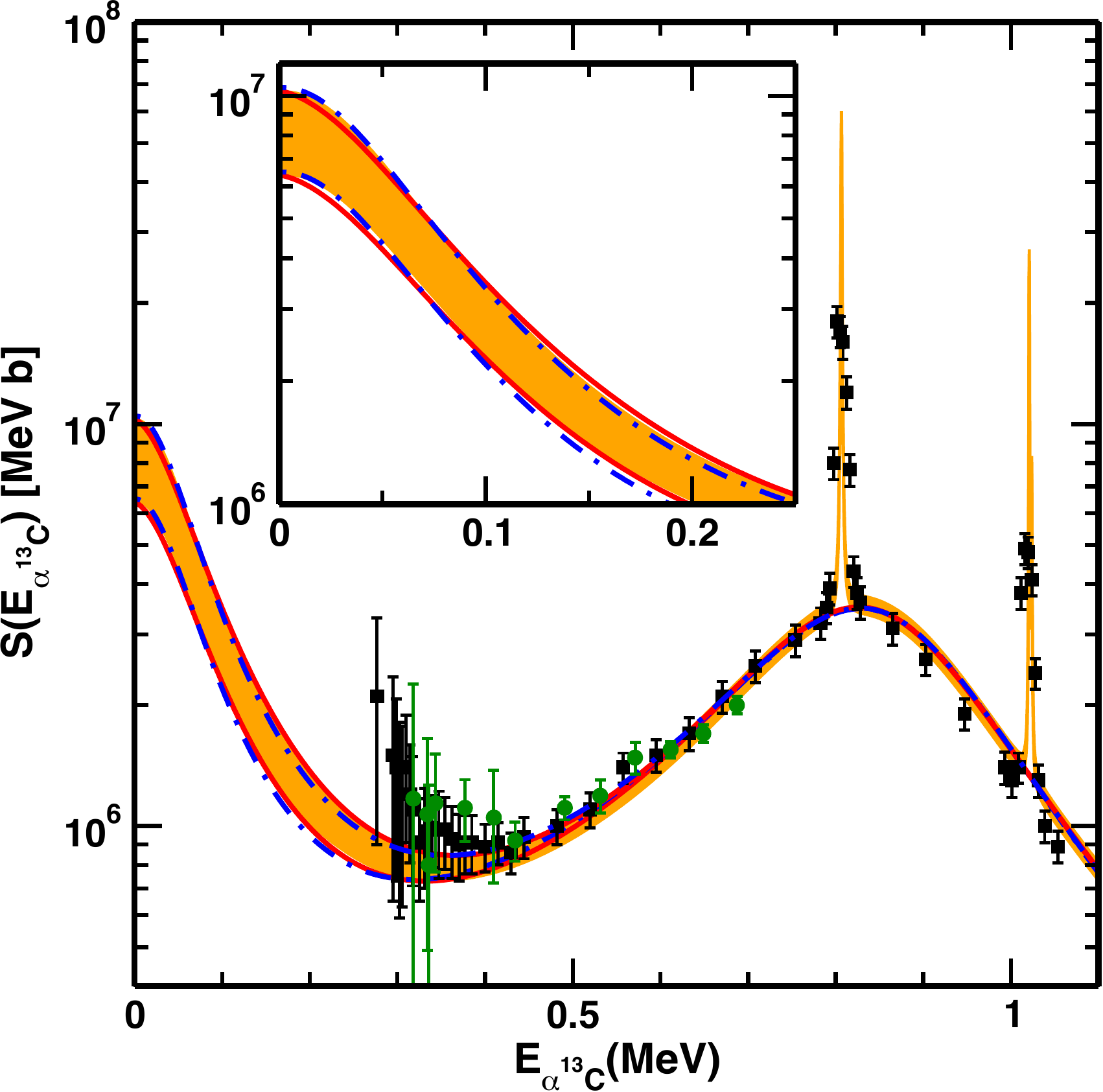}
\caption{(Color online) Astrophysical S factor for the $^{13}{\rm C}(\alpha,\,n)^{16}{\rm O}$ reaction as a function of the $\alpha\,-{}^{13}{\rm C}$ relative kinetic energy. Square black boxes, solid green dots and shaded orange band are data from Refs. \cite{drotleff}, \cite{heil} and \cite{oscar}, respectively. Red solid lines correspond to our calculations for the fit to the lower and upper limits of the TH data  considering  $1/2^{+}$ state as  $-3$ keV subthreshold  resonance with $\Gamma_{n} = 158.1$ keV \cite{heil}. The lower and upper limits of the renormalized ANC square are 2.89 fm$^{-1}$ and 4.7 fm$^{-1}$, respectively. Whereas, the blue dotted-dashed lines correspond to our calculations for the fit to TH data, considering  $1/2^{+}$ state as 4.7 keV threshold resonance with $\Gamma_{n} = 136$ keV \cite{Faestermann} and the corresponding lower and upper values of observable reduced width are 2.81 keV$^{1/2}$ and 3.6 keV$^{1/2}$, respectively.  For our calculations we have used $R_{\alpha\,{}^{13}{\rm C}}= 7.5$ fm, $R_{n\,{}^{16}{\rm O}}= 6.0$ fm. The insert in the figure shows enlarged low energy $S$ factor.  
 First published in  \cite{mukshbert2017}.} 
\label{fig_Sfactor}
\end{figure}
From Fig. \ref{fig_energdepres1} it is clear that only the experimental $S$ factor generated by the broad resonance   $3/2^{+},\,E_{\alpha\,{}^{13}{\rm C}}=0.857$ MeV can be used for normalization of the TH doubly DCS at $E_{\alpha\,{}^{13}{\rm C}} > 0.5$ MeV. The problem of the normalization of the TH data for this specific reaction was discussed in details in \cite{marco2,oscar,reviewpaper}.
We use the results from \cite{oscar} as fitting data but need to renormalize them because in \cite{marco2,oscar} the factor ${\cal W}_{1}$, see Eq. (\ref{calWr1}), was calculated without the integral term in Eq. (\ref{calWr1}). Recalculating  ${\cal W}_{1}$  taking into account the integral term we find that the TH results in \cite{oscar} should be renormalized by $0.948$. After renormalization of the TH data from \cite{oscar} we did a new fit. In Fig. \ref{fig_Sfactor} we present our final results for the $S$ factor for the reaction  ${}^{13}{\rm C}(\alpha,\,n){}^{16}{\rm O}$.

Our numerical values of  the $S(0)$ factors are:\\
(1) for  $1/2^{+}$,  $-3$ keV and $\Gamma_{n}= 158.1$ keV  \cite{heil},  $S(0)=7.62 _{-1.23}^{+2.65}\times 10^{6}$ MeV b;\\
(2) for  $1/2^{+}$,   4.7 keV   and $\Gamma_{n}= 136$ keV \cite{Faestermann}, $S(0)=7.51 _{-1.1}^{+2.96}\times 10^{6}$ MeV b. 

Thus, even the TH data, which  provides the astrophysical factor at significantly lower energies than direct measurements \cite{heil}, cannot  answer the question whether the threshold level  is  a subthreshold bound state or resonance.

In the analysis of the TH data, in the previous TH papers (see Refs. \cite{oscar,marco2}),  only the two-stage mechanism proceeding through the intermediate threshold state $1/2^+$ has been taken into account.  However, the single-step direct reaction $^{13}$C($\alpha, n$)$^{16}$O also can contribute to the low-energy cross section. Although the S factor of the direct mechanism is flat and can be small, its interference with the two-stage resonant mechanism can change the total S factor. However, the accuracy of the existing data does not allow us to determine the contribution of the direct mechanism.

\begin{figure}[tp] 
  \centering
 \includegraphics[
 width=0.7\columnwidth]{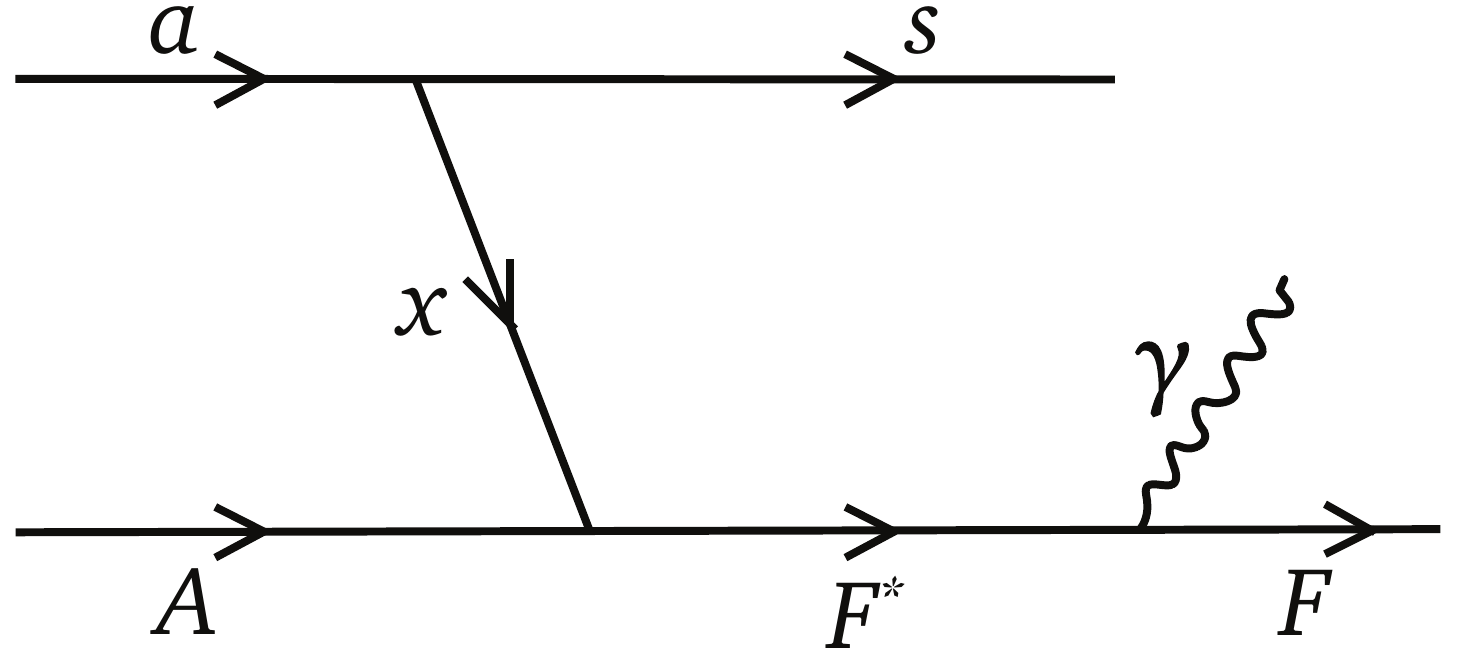}
  \caption{Pole diagram describing the indirect radiative capture reaction proceeding through
the intermediate excited state $F^{*}$. }
  \label{fig:fig_polediagram1}
\end{figure}

\section{THM for radiative capture reactions}
\label{THMradcapt}

\subsection{Introduction}
\label{RadTHM1}
The THM  can be used to determine different reactions proceeding through the resonance. In the previous sections we considered the THM for resonant rearrangement reactions. In this section we focus on the application of  the THM  to measure resonant radiative-capture reactions
at energies so low that direct measurements  can hardly be performed due to the negligibly small penetrability factor in the entry channel of the reaction. We present here the theory of the indirect THM to treat the resonant radiative-capture processes when only a few subthreshold bound states and resonances are involved, and statistical methods cannot be applied. As in the previous sections, the developed formalism is based on the generalized multi-level $R$-matrix approach and surface-integral formulation of the transfer reactions, which are the first stage of the indirect reaction mechanism described by  the diagram depicted in Fig.  \ref{fig:fig_polediagram1} \cite{mukrog}.  The developed formalism allows one to study the photon's angular distribution correlated with the scattering angle of one of the final nuclei formed in the transfer reaction.

There are many papers devoted to the angular correlation of the photons emitted in nuclear transfer reactions with final nuclei, see, for example, \cite{Austern} and references therein. The approach, which is outlined below, can be applied  for the ${}^{12}{\rm C}({}^{6}{\rm Li},d\,\gamma){}^{16}{\rm O}$ reaction, which can provide important information about the astrophysical $\,{}^{12}{\rm C}(\alpha,\,\gamma){}^{16}{\rm O}$ process. This astrophysical reaction is contributed by two interfering subthreshold resonances (\cite{RolfsRodney}, sect. 4.5).   The subthreshold bound state may reveal itself as a resonance in the case of the radiative capture, which can occur to the wing of the subthreshold state at positive energy forming the intermediate excited state. The excited bound state subsequently decays to lower lying states by emitting a photon. In this case, the subthreshold bound state is characterized by a resonance width in complete analogy with the real resonance \cite{muk99}. 

Numerous attempts to obtain the astrophysical factor of the $\,{}^{12}{\rm C}(\alpha,\,\gamma){}^{16}{\rm O}$ reaction, both experimental and theoretical, have been made for almost 50 years \cite{fowler,arnett,angulo,abia,lugaro,redder,Ouellet,caughlan,barker78,barker87,barkerkajino,azuma,azuma1997,brune1999,brune2001,kunz,hammer,hammer05,assuncao,plug,schurmann2005,schurmann2006,Schurmann,gai2015,sayre,gai2013,brune2013,gai2014, gai2010,avila1,Jas70, dyer74,ket82,duke,eli,tischhauser,bertulani2016}. This reaction is contributed by interfering $E1$ and $E2$ transitions. 
The $E1$ transition is complicated by the interference of the capture through the wing of the subthreshold $1^{-}$ resonance at $-0.045$ MeV with the low-energy tail  of the resonance  $1^{-},\,E_{\alpha\,{}^{12}{\rm C}}=2.423$ MeV, where $E_{\alpha\,{}^{12}{\rm C}}$ is the $\alpha-{}^{12}{\rm C}$ relative kinetic energy. The $E2$ transition is dominated by the capture to the ground state of ${}^{16}{\rm O}$ through the wing of the subthreshold  bound state $2^{+},E_{\alpha\,{}^{12}{\rm C}}=-0.245$ MeV.  In addition, to fit the experimental data, usually
a few artificial levels are added to fit $E1$ and $E2$ data \cite{barkerkajino,brune1999}. The difficulty of the direct measurements of the $E1$ transition can be easily understood if even in the peak of the resonance at
$1^{-},\,E_{\alpha\,{}^{12}{\rm C}}=2.423$ MeV the cross section is only about $40-50$ nb \cite{Jas70,dyer74,ket82} . Moreover , the $E1$ transition from $1^{-}$ states to the ground state of ${}^{16}{\rm O}$ is isospin forbidden for the $T=0$ component and is possible only due to a small admixture of the $T=1$ components.

Extremely small penetrability factor at  $E_{\alpha\,{}^{12}{\rm C}} \leq 1$ MeV makes it impossible or very difficult to measure the astrophysical factor for the $\,{}^{12}{\rm C}(\alpha,\,\gamma){}^{16}{\rm O}$ reaction at energies $E_{\alpha\,{}^{12}{\rm C}} \leq  1$ MeV  with reasonable accuracy. For the sensitivity of the extracted  astrophysical factor from the existing data, see  works \cite{gai2013,gai2014,gai2015}. Note that from the astrophysical point of view  the required uncertainty  of this astrophysical factor at $E_{\alpha\,{}^{12}{\rm C}} \sim 0.3$ MeV should be $\leq 10\%$. New gamma-ray facilities, an upgraded gamma-source (HIGS2) \cite{duke} in USA and Compton gamma-ray source of Eli-NP \cite{eli}
in Romania,  are supposed to measure the astrophysical factor for the $\,{}^{12}{\rm C}(\alpha,\,\gamma){}^{16}{\rm O}$ reaction down to $1$ MeV.

The THM allows one not only to determine the astrophysical $S$ factor down to energies $E_{\alpha\,{}^{12}{\rm C}}  \sim 0.3$ MeV but also investigate the interference pattern between the subthreshold bound state and higher resonance for the $E1$ transition. The method, which we describe here, can be used for a broader type of radiative-capture experiments  $A(a,\,s\,\gamma)F$ proceeding through the subthreshold and real resonances.

To measure the cross section of the binary process
\begin{equation}
x + A \to F^{*} \to \gamma + F
\label{radcapt1}
\end{equation}
proceeding through the intermediate resonance $F^{*}$  at astrophysical energies we suggest to measure the surrogate reaction [two-body to three-body process ($2 \to 3$ particles)]:
\begin{align}
a + A \to s + F^{*} \to s + \gamma + F
\label{THMreactiongamma1}
\end{align}
in the vicinity of the quasi-free 
kinematics region \cite{reviewpaper}.
It is assumed that the incident particle $a=(sx)$ is accelerated at energies above the Coulomb barrier. The first stage of the THM reaction is the transfer reaction $a + A \to s+ F^{*}$ populating the wing of the subthreshold bound state at $E_{xA} >0$ or leading to a real resonance. On the second stage, the excited state $F^{*}$ decays to the ground state $F\,$ by emitting a photon. 

Physics of the THM has already been discussed in the previous sections in which the expression for the amplitude of the transfer reaction (\ref{THMreaction1})  in the surface-integral approach and the DWBA was derived. 
We assume that the energies of the particles in the initial and final states in reaction (\ref{THMreactiongamma1})  are above the Coulomb barrier. 
In this case, in contrast to the ${}^{14}{\rm N} + {}^{12}{\rm C}$ one in which the deuterons were below the Coulomb barrier,   it makes sense to use the PWA, see Section \ref{THMPWA1} and \cite{reviewpaper}.   The comparison of the PWA and  DWBA has been done in many THM papers 
\cite{reviewpaper,lacognata2010,sergi2010,gulino2013}. In these papers, the momentum distribution of the spectator was calculated in the PWA and DWBA. Both calculations agreed with each other and experimental data within the range of the QF peak. The most detailed comparison of the PWA and DWBA was done in \cite{gulino2013}. It was confirmed  that the
angular distributions of the spectator calculated in the DWBA and the plane-wave impulse approximation agree quite well within the QF peak.   The differences
in the ratios of the integrated resonance cross sections calculated in the PWA and DWBA  are less than $19\%$,
compared with the experimental uncertainties. Therefore, when no absolute values of the cross sections are extracted, the PWA description is more preferable  than DWBA because the former does not depend on the optical potentials which for light nuclei are not known accurately at low energies.

\subsection{Amplitude of indirect resonant radiative-capture reaction}
\label{Amplitradcapt1}
Let us consider the radiative capture reaction  (\ref{radcapt1}) proceeding through the wing (at $E_{xA} >0$) of the subthreshold bound state (aka subthreshold resonance) $F^{*} = F^{(s)}$, where  $F^{(s)} =(x\,A)^{(s)}$, or a real resonance at $E_{xA} >0$. We assume that both can decay to the ground state $F=(xA)$. To measure the cross section of this reaction at astrophysically relevant energies where subthreshold resonances can be important (for the reasons explained above),  we  use the indirect reaction (\ref{THMreactiongamma1}).
First, we derive the reaction amplitude of the indirect radiative-capture process and then the fully DCS of reaction (\ref{THMreactiongamma1}). After that, by integrating over the angles of the emitted photons, we get the doubly DCS.
The interference of  the subthreshold bound state and the resonance, which both decay to the ground state
$F=(xA)$, is taken into account. Cleary, this case can be applied for the $E1$ and $E2$ transitions of the reaction ${}^{12}{\rm C}(\alpha,\gamma){}^{16}{\rm O}$.

To describe the radiative capture to the ground state through two interfering states  we use the single-channel two-level generalized $R$-matrix equations developed for the three-body reactions $2\,\,{\it particles} \to 3\,\,{\it particles}$ \cite{reviewpaper,muk2011}.  
We also take into account the interference of transitions with different multipolarities $L$. Thus, we take into account the interference of radiative decays from different levels with the same multipolarity and interference of transitions from various levels with different multipolarities.  

The indirect reaction described by the diagram shown in Fig. \ref{fig:fig_polediagram1}  proceeds as a two stage process. The first stage is transfer of particle $x$ (stripping process) to the excited state $F_{\tau},\,\,\tau=1,2$, where $F_{1} =F^{(s)}$ is the subthreshold resonance and $F_{2}$ is the resonance state at $E_{xA} >0$.  No gamma is emitted during the first stage. On the second stage the excited state $\,F_{\tau}$ decays to the ground state $\,F=(xA)$ by emitting a photon.
Then the indirect reaction amplitude followed by the photon emission from the intermediate subthreshold resonance and resonance takes the form:
\begin{align}
M_{M_{a}\,M_{A}}^{M_{s}\,M\,M_{F}} = \sum\limits_{\tau, \nu =1}^{2}\,\sum\limits_{M_{F_{\nu}^{(s)}}\,M_{F_{\tau}^{(s)}}}\, V_{M_{F_{\nu}^{(s)}}\,\nu}^{M_{F}\,M}\,{\rm {\bf A}}_{\nu\,\tau}\,M_{M_{a}\,M_{A}\,\tau}^{M_{s}\,M_{  F_{\tau}^{(s)}}}.
\label{THradamplfull1}
\end{align}
Here 
$V_{\nu} \equiv V_{M_{F_{\nu}^{(s)}}\,\nu}^{M_{F}\,M}$
is the amplitude of the radiative decay of the excited state  $\,F_{\nu} \,\,(\nu=1,2)$ to the ground state $\,F=(xA)$, $\,{\rm {\bf A}}_{\nu\,\tau}\,$ is the matrix element of the level matrix in the $\,R$-matrix method, $M_{i}$ is the projection of the spin $J_{i}$ of particle $i$,   $M_{F_{\tau}^{(s)}}$ is the projection of the spin $J_{F^{(s)}}$ of the subthreshold resonance ($\tau =1$) and resonance ($\tau=2$), $M$ is the projection of the angular momentum of the emitted photon. Also number $2$ for the upper limit of the first sum represents the number of the levels included. We assume that the spins of the subthreshold resonance and real resonance are equal $F_{1} = F_{2}= F^{(s)}$ and these resonances do interfere. At the moment we confine ourselves by transition with one multipolarity $L$. That is why the index $L$ is omitted. Later on we take into account transitions with different $L$.
$M_{M_{a}\,M_{A}\,\tau }^{M_{s}\,M_{F_{\tau}^{(s)}}}$ is the amplitude of the direct transfer reaction 
\begin{align}
a+ A \to s+ F_{\tau},
\label{transfreact1}
\end{align} 
populating the intermediate excited state $F_{\tau}$. The reaction (\ref{transfreact1}) is the first stage of the indirect reaction (\ref{THMreaction1}).

Below we present the general equations for the fully and doubly differential cross sections of the indirect radiative-capture reactions  proceeding through subthreshold and isolated resonances.

\subsection{Triply DCS}

Let us consider the indirect resonant reaction (\ref{THMreactiongamma1}) contributed by different interfering multipoles $L$. For each $L$ we assume a two-level contribution.  The derivation of the amplitude of this reaction is given Appendix \ref{THMRCampl1}.
Then the fully DCS of the resonant indirect radiative-capture reaction for unpolarized initial and final particles (including the photon) in the center-of-mass of the reaction (\ref{THMreactiongamma1}) is  given by
\begin{widetext}
\begin{align}
\frac{ {\rm d}\sigma}{  {\rm d}{\Omega}_{ {s} }\,{\rm d}{\Omega}_{{\gamma}}\, {\rm d}{E_{sF}} }=&
 \frac{\mu_{aA}\,\mu_{sF}}{{\hat J}_{a}\,{\hat J}_{A}(2\,\pi)^{5}}\,\frac{k_{sF}\,k_{\gamma}^{2}}{k_{aA}}
\sum\limits_{M_{a}\,M_{A}\,M_{s}\,M_{F}\,M\,\lambda}\Big|M_{M_{a}\,M_{A}}^{M_{s}\,M_{F}\,M\,\lambda}\Big|^{2}
 \nonumber\\
=& - \frac{1}{(2\,\pi)^{7} }\,\frac{\mu_{aA}\,\mu_{sF}}{{\hat J}_{x}\,{\hat J}_{A}} \,\frac{\varphi_{a}^{2}(p_{sx})\,R_{xA}}{4\,\mu_{xA}}\,\frac{ k_{sF}}{k_{aA}}\,(-1)^{J_{F}  - j_{xA} }\,
  \sum\limits_{L'\,L} (-1)^{L'+L}\, k_{\gamma}^{L'+L+1}\,\,{\hat J}_{F^{(s)}}^{L'}\,{\hat J}_{F^{(s)}}^{L}\,\sqrt{ {\hat L}'\,{\hat L}}\,  
    \nonumber\\
& \times \sum\limits_{l_{xA}'\,l_{xA}\,l}i^{L'-l_{xA}'- L+l_{xA}}
\sqrt{ {\hat l}_{xA}'\,{\hat l}_{xA} }  {\cal {\tilde W}}_{l_{xA}'}^{*}\,{\cal {\tilde W}}_{l_{xA}}\, \Bigg\lbrace \begin{array}{ccc} j_{xA}\,l_{xA}'\, J_{F^{(s)}}^{L'} \\  l\,J_{F^{(s)}}^{L}\,l_{xA} \end{array} \Bigg\rbrace\,\Bigg\lbrace \begin{array}{ccc} J_{F^{(s)}}^{L'}\, J_{F}\,L' \\
L\,l\,J_{F^{(s)}}^{L} \end{array} \Bigg\rbrace   
\nonumber\\
& \times
\gamma_{\tau'\,j_{xA}l_{xA}'J_{F^{(s)}}^{L'} }\,\gamma_{\tau\,j_{xA}l_{xA}J_{F^{(s)}}^{L}} \,
\sum\limits_{\nu', \,\nu,\,\tau',\, \tau=1}^{2}\,\big[\gamma_{ (\gamma)\,\nu'\,J_{F}\,L'}^{J_{F^{(s)}}^{L'}}  \big]^{*} [\gamma_{(\gamma)\,\nu\,J_{F}\,L}^{J_{F^{(s)}}^{L}} ]\,\big[{\bf A}_{v'\,\tau'}^{L'} \big]^{*}\,\big[{\bf A}_{\nu\,\tau}^{L} \big]\     \nonumber\\
& \times   \blb l_{xA}'\,0\,\,l_{xA}\,0\big|l\,0 \blk \, \blb L'\,1\,\,L\,-1 \big| l\,0 \blk \,[1+ (-1)^{L'+L+l}]\,P_{l}(cos{\theta}).
\label{tripplediffcrsect1}
\end{align}
\end{widetext}
To obtain Eq. (\ref{tripplediffcrsect1}) we adopted $z\, ||\, {\rm {\bf {\hat  p}}}_{xA} $,  that is,  
$Y_{l\,m_{l}}({\rm {\bf {\hat  p}}}_{xA}) = \frac{\hat l}{\sqrt{4\,\pi}}\,\delta_{m_{l}\,0}$.
Thus, in the PWA, the direction ${\rm {\bf {\hat  p}}}_{xA}$ becomes the axis of the symmetry.
Note that if we replace the plane waves by the distorted waves, the vestige of this symmetry will still survive \cite{Austern}.

We remind that the radiative transition  $J_{F^{(s)}}^{L} \to J_{F}$ is the electric $EL$ where $J_{F^{(s)}}^{L}$ is the spin of the intermediate state (subthreshold resonance or resonance).

For a more simple case when only one multipole $L$ contributes into the radiative transition, the fully DCS
takes the form:
\begin{widetext}
\begin{align}
\frac{ {\rm d}\sigma}{  {\rm d}{\Omega}_{ {s} }\,{\rm d}{\Omega}_{{\gamma}}\, {\rm d}{E_{sF}} }=& - \frac{1}{(2\,\pi)^{7} }\,\frac{\mu_{aA}\,\mu_{sF}}{{\hat J}_{x}\,{\hat J}_{A}} \,\frac{\varphi_{a}^{2}(p_{sx})\,R_{xA}}{2\,\mu_{xA}}\,\frac{ k_{sF}}{k_{aA}}\,
 k_{\gamma}^{{\hat L}}\,(-1)^{J_{F}  - j_{xA} }\,{\hat L}\,({\hat J}_{F^{(s)}}^{L})^{2}\,
\sum\limits_{l_{xA}\,l}\,{\hat l}_{xA}\,                                   \nonumber\\
& \times \big|{\cal {\tilde W}}_{l_{xA}}\,\big|^{2}\,\Bigg\lbrace \begin{array}{ccc} j_{xA}\,l_{xA}\, J_{F^{(s)}}^{L} \\  l\,J_{F^{(s)}}^{L}\,l_{xA} \end{array} \Bigg\rbrace\,\Bigg\lbrace \begin{array}{ccc} J_{F^{(s)}}^{L}\, J_{F}\,L \\
L\,l\,J_{F^{(s)}}^{L} \end{array} \Bigg\rbrace   \sum\limits_{\nu', \,\nu,\,\tau',\, \tau=1}^{2}\,\big[\gamma_{ (\gamma)\,\nu'\,J_{F}\,L}^{J_{F^{(s)}}^{L}}  \big]^{*} [\gamma_{(\gamma)\,\nu\,J_{F}\,L}^{J_{F^{(s)}}^{L}} ]\,\big[{\bf A}_{v'\,\tau'}^{L} \big]^{*}\,\big[{\bf A}_{\nu\,\tau}^{L} \big]\     \nonumber\\
& \times ,\gamma_{\tau'\,j_{xA}l_{xA}J_{F^{(s)}}^{L}}\,\gamma_{\tau\,j_{xA}l_{xA}J_{F^{(s)}}^{L}} \,  \blb l_{xA}\,0\,\,l_{xA}\,0\big|l\,0 \blk \, \blb L\,1\,\,L\,-1 \big| l\,0 \blk \,P_{l}(cos{\theta}).
\label{tripplediffcrsect2}
\end{align}
\end{widetext}
The off-shell factor ${\cal {\tilde W}}_{l_{xA}}$ is defined in Appendix B, see Eq. (\ref{caltildeWr1})
Though we formally keep the summation over $l_{xA}$, in the long-wavelength approximation for given $L$ at astrophysically relevant energies only minimal allowed $j_{xA}$ contributes. 

The fully DCS depends on ${\rm {\bf k}}_{s}$ and $\,{\rm {\bf k}}_{\gamma}$. Because we neglected the recoil of the final nucleus $F$, $\,k_{s}$ and $k_{\gamma}$ are related by Eq. (\ref{energconserv2}). We remind that we selected axis $z||{\rm{\bf p}}_{xA}$.  Hence the photon's scattering angle is counted from ${\rm{\bf p}}_{xA}$, which itself is determined by ${\rm {\bf k}}_{s}$.
 Thus the angular dependence of the fully DCS determines the angular correlation between the emitted photons from the intermediate excited state $F^{*}$ and the spectator $s$. Since we consider the three-body reaction (\ref{THMreaction1}),
the angular correlation function also depends on the spins $J_{F^{(s)}}^{L}$ of the intermediate nucleus $F^{*}$ which decays to $F$.

We note that \\
(i) The most important feature of the indirect reaction fully DCS  is that it does not contain 
the penetrability factor $P_{l_{xA}}(E_{xA},R_{xA})$ in the entry channel of the sub-reaction (\ref{reaction1}). This factor is the main obstacle to measure the astrophysical factor of this reaction if one uses direct measurements. 
The absence of this penetrability factor in the entry channel of the sub-reaction allows one to use the indirect method to get the information about the astrophysical factor of the sub-reaction.\\
(ii) The indirect reaction fully DCS is parameterized in terms of the formal $R$-matrix  width amplitudes, which are connected to the observable resonance widths.\\
(iii) The final expression for the indirect reaction triply DCS does not depend on the $R$-matrix hard-sphere scattering phase shift. 

By choosing QF kinematics, $p_{sx}=0$, one can provide the maximum of the fully DCS due to the maximum of $\varphi_{a}^{2}(p_{sx})$. At fixed ${\rm {\bf k}}_{s}$ the fully DCS determines the emitted photon's angular distribution, which is contributed by different interfering multipoles $L$.
By measuring the photon's angular distributions at different photon's energies (that is, at different $k_{s}$ or $E_{xA}$) one can determine the energy dependence of the photon's angular distribution. However, a wide variation of ${\rm {\bf k}}_{s}$ 
away from the QF kinematics ${\rm {\bf p}}_{sx} =0$ will decrease the DCS due to the drop of $\varphi_{a}^{2}(p_{sx})$. Usually, in indirect methods $\,{\rm {\bf k}}_{s}$ is varied in the interval in which $p_{sx} \leq \kappa_{sx}$ \cite{reviewpaper}. \\

\subsection{Doubly DCS}

Integrating the fully DCS over the the photon's solid angle ${\Omega}_{{\gamma}}$ we get the non-coherent sum of the doubly DCSs with different multipoles $L$:
\begin{widetext}
\begin{align}
\frac{ {\rm d}\sigma}{  {\rm d}{\Omega}_{{s} }\, {\rm d}{E_{sF}} }=&
 \frac{1}{(2\,\pi)^{6} }\,\frac{\mu_{aA}\,\mu_{sF}}{{\hat J}_{x}\,{\hat J}_{A}} \,\frac{\varphi_{a}^{2}(p_{sx})\,R_{xA}}{\mu_{xA}}\,\frac{ k_{sF}}{k_{aA}}\,
 \sum\limits_{L}\,\sqrt{{\hat L}\,{\hat J}_{F^{(s)}}^{L}}\,k_{\gamma}^{{\hat L}}\,\sum\limits_{l_{xA}}\,\big| {\cal {\tilde W}}_{l_{xA}}\big|^{2}                                  \nonumber\\
& \times\,\sum\limits_{\nu', \,\nu,\,\tau',\, \tau=1}^{2}\,\big[\gamma_{ (\gamma)\,\nu'\,J_{F}\,L}^{J_{F^{(s)}}^{L}}  \big]^{*} [\gamma_{(\gamma)\,\nu\,J_{F}\,L}^{J_{F^{(s)}}^{L}} ]\,\big[{\bf A}_{v'\,\tau'}^{L} \big]^{*}\,\big[{\bf A}_{\nu\,\tau}^{L} \big]\ \gamma_{\tau'\,j_{xA}\,l_{xA}\,J_{F^{(s)}}^{L}}\,\gamma_{\tau\,j_{xA}\,l_{xA}\,J_{F^{(s)}}^{L}}.
\label{doublediffcrsect11}
\end{align}

Despite of the virtual transferred particle $x$ in the diagram of Fig. \ref{fig:fig_polediagram1}, using the surface-integral approach and the generalized $R$-matrix we can rewrite the doubly DCS in terms of the 
OES astrophysical factor for the resonant radiative capture $A(x,\,\gamma)F$ for the electric transition of the multipolarity $L$ and the relative orbital angular momentum $l_{xA}$ of particles $x$ and $A$ in the entry channel of the $A(x,\,\gamma)F$ radiative capture. In the $R$-matrix formalism this astrophysical factor is given by \cite{muk2011a}
\begin{align}
S_{ELl_{xA}}(E) (MeVb) = 2\,\pi\,\lambda_{N}^{2}\,\frac{{\hat J}_{F^{(s)}}^{L}}{{\hat J}_{x}\,{\hat J}_{A}}\,\frac{1}{\mu_{xA}}\,m_{N}^{2}\,
e^{2\,\pi\,\eta_{xA}}\,P_{l_{xA}}(E,\,R_{xA})\,10^{-2}\,k_{\gamma}^{{\hat L}}\,\Big|\sum\limits_{\nu,\tau=1}^{2}[\gamma_{(\gamma)\,\nu\,J_{F}\,L}^{J_{F^{(s)}}^{L}} ]\,\big[{\bf A}_{\nu\,\tau}^{L} \big]\ \gamma_{\tau\,j_{xA}\,l_{xA}\,J_{F^{(s)}}^{L}}\Big|^{2}.
 \label{Sfactor1}
\end{align}   
Here, 
 $\mu_{xA}$ is the $x-A$ reduced mass expressed in MeV, $\eta_{xA}$ is the $x-A$ Coulomb parameter at their relative energy $E \equiv E_{xA}$.  
Then the indirect doubly DCS takes the form:
\begin{align}
\frac{ {\rm d}\sigma}{  {\rm d}{\Omega}_{{s} }\, {\rm d}{E_{sF}} }=& K(E) N_{F}\,\varphi_{a}^{2}(p_{sx})\,R_{xA}
\,\sum\limits_{L} \,\sqrt{\frac{{\hat L}}{{\hat J}_{F^{(s)}}^{L}}}
\sum\limits_{l_{xA}}e^{-2\,\pi\,\eta_{xA}}
\frac{\big| {\cal {\tilde W}}_{l_{xA}} \big|^{2}}{P_{l_{xA}}(E,\,R_{xA})}
\,S_{ELl_{xA}}(E),                                 
\label{doublediffcrsectSfctr1}
\end{align}
\end{widetext}
where 
\begin{align}
K(E) =  \frac{10^{2}}{(2\,\pi)^{7} }\,\frac{\mu_{aA}\,\mu_{sF}}{m_{N}^{2}\,\lambda_{N}^{2}}\,\frac{ k_{sF}}{k_{aA}}
\label{KF1} 
\end{align}
is the kinematical factor,   $N_{F}$ is an energy-independent THM normalization factor.  To determine the astrophysical factor from the indirect doubly 
DCS we need to identify the region where accurate direct data are available and only one resonance dominates with given $L$ and $l_{xA}$.  By normalizing in this region the astrophysical factor obtained from the indirect measurement to the experimental one we get 
\begin{align}
 S_{ELl_{xA}}(E)=& N_{F}\,\frac{ {\rm d}\sigma}{  {\rm d}{\Omega}_{  {\rm{\bf {\hat k}}}_{s} }\, {\rm d}{E_{sF}} }\,\sqrt{\frac{ {\hat J}_{F^{(s)}}^{L} }{ {\hat L}}}  \,\frac{1}{KF\,\varphi_{a}^{2}(p_{sx})\,R_{xA}}\,
 \nonumber \\ &\times
 e^{2\,\pi\,\eta_{xA}}\,P_{l_{xA}}(E,\,R_{xA})\,
\big| {\cal {\tilde W}}_{l_{xA}} \big|^{-2}.
\label{Sfactrdblcrsect11}
\end{align}
%
Using this normalization factor we can determine the astrophysical factors at energies $E_{xA} \to 0$ with accuracy, that is not achievable in any direct approach. This is the main advantage  of the indirect approach.
We remind that in our formalism we use the PWA rather than the distorted wave assuming that both methods give similar energy dependence of the transfer reaction DCS. If they predict different energy dependences then the distorted wave method should be used.   
 
We summarize the methodology of the indirect method to obtain the astrophysical factor for the radiative capture reactions proceeding trough subthreshold and/or real resoances:\\
(1) Measurements of the photon's angular distribution (photon-spectator angular correlation) at different $x-A$ energies covering the interval from low energies relevant to nuclear astrophysics up to higher energies at which direct data are available. To cover a broad energy range at fixed energy of the projectile, the energy, and scattering angle of the spectator should be varied near the QF kinematics ($p_{sx}=0$). \\
(2) Obtaining the indirect doubly DCS by integrating the fully DCS over the photon's scattering angle.\\
(3) Expressing the astrophysical factor in terms of the indirect doubly DCS.\\
(4) Normalization of astrophysical factor to the available experimental data at higher energy.\\
(5) Determination of the astrophysical factor at astrophysical energies.

\section{Radiative capture ${}^{12}{\rm C}(\alpha,\,\gamma){}^{16}{\rm O}$ via  indirect reaction ${}^{12}{\rm C}({}^{6}{\rm Li},d\,\gamma){}^{16}{\rm O}$}
\label{Radcaptapplication1}

\subsection{Introduction}
\label{RadcaptIntroduction1}

In this section we discuss the practical application of the developed formalism for the analysis
of the indirect reaction ${}^{12}{\rm C}({}^{6}{\rm Li},d\,\gamma){}^{16}{\rm O}$ to obtain the information about the astrophysical factor for the ${}^{12}{\rm C}(\alpha,\,\gamma){}^{16}{\rm O}$ at energies $< 1$ MeV. For our analysis we use the energy levels from \cite{tiley}.

At low energies the astrophysical reaction under consideration is contributed by the $L=1$ and $L=2$ electric transitions \cite{barkerkajino,brune1999,kunz,assuncao,gai2015}. $E1$ transition to the ground state of ${}^{16}{\rm O}$ with $J_{F}=0$ and zero $\alpha-{}^{12}{\rm C}$  orbital angular momentum proceeds as resonant capture through the wing at $E_{\alpha\,{}^{12}{\rm C}}>0 $ of the subthreshold bound state $1^{-}$ at $E_{\alpha\,{}^{12}{\rm C}} = -0.045$ MeV, which works as the subthreshold resonance. Besides, the $E1$ transition to the ground state  is contributed by the resonant capture through the low-energy tail of the $1^{-}$ resonance located at $E_{R}= 2.423$ MeV. The $E2$ transition is contributed by the subthreshold $2^{+}$ state at $E_{\alpha\,{}^{12}{\rm C}} = -0.2449$ MeV and low-energy tail of $2^{+}$ resonance at $2.68$ MeV.  

These four states are observable physical states contributing to the low-energy radiative capture under consideration. In addition to  these 
states, when fitting the data an artificial level was added for $E1$ transition (see, for example, 
\cite{barkerkajino,brune1999,kunz,assuncao,sayre} and references therein).
 In the present paper we calculate the  photon's angular distribution (the angular photon-deuteron correlation) at low energies down to the most effective astrophysical energy of $E_{\alpha\,{}^{12}{\rm C}}=0.3$ MeV.

We take into account the mentioned four physical states and added one artificial state for the $E1$ transition. It can be explained qualitatively why the background level is necessary to include into the fit of the $E1$ transition. The problem is that the subthreshold state $J=1,\,E_{\alpha\,{}^{12}{\rm C}} = -0.045$ MeV and the resonance $J=1,\,E_{R}= 2.423$ MeV cannot decay by the $E1$ transition to the ground state of ${}^{16}{\rm O}$ because all of them have isospin $T=0$. Note that the observed weak $E1$  transition from the first two $J=1,\,T=0$ states is possible only due to the small admixture of the higher lying $J=1,\,T=1$ states \cite{redder}. 

Let us estimate the recoil effect of the nucleus $F$ for the THM reaction ${}^{12}{\rm C}({}^{6}{\rm Li},d\,\gamma){}^{16}{\rm O}$ at the most effective astrophysical energy $E_{xA}=E_{\alpha\,{}^{12}{\rm C}}= 0.3$ MeV; the energy of the emitted photon is $k_{\gamma} \approx 7$ MeV and $E_{aA}= 7$ MeV. 
As we will see below (Fig. \ref{fig_Angdistr1}) at $0.3$ MeV the maximum of the photon's angular distribution is at 
$\theta =52^{\circ}$, where $\theta$ is the angle between ${\rm {\bf p}}_{\alpha\,{}^{12}{\rm C}}$ and ${\rm {\bf k}}_{\gamma}$. In the QF kinematics
${\rm {\bf p}}_{\alpha\,{}^{12}{\rm C}} || {\rm {\bf k}}_{d}$, where ${\rm {\bf k}}_{s} = {\rm {\bf k}}_{d}$, that is, $\theta'=\theta$.  At $\theta =52^{\circ}$, which is the maximum of the photon's angular distribution and is close to the maximum of the angular distribution for the $E2$ transition, the recoil effect is $\sim 6.5\%$. 
Note that for the $E1$ transition the photon's angular distribution has a peak at $90^{\circ}$ where the recoil effect vanishes.

The reduced widths of the subthreshold resonances are known from the experimental ANCs \cite{brune1999,avila} and the reduced width of the $1^{-},\,2.423$ MeV resonance is determined from the resonance width. We disregard the cascade transitions to the ground state of ${}^{16}{\rm O}$ through subthreshold states. According to \cite{redder}, the sum of all cascade transitions contributes only $7-10\%$. Because we don't pursue here a perfect fit, we neglect all the cascade transitions. Following  \cite{redder}, in our fit we also disregard the $E2$ direct radiative capture to the ground state ${}^{16}{\rm O}$. 

For the case under consideration $J_{x}=0,\,J_{A}=0,\,j_{xA}=0,\,l_{xA}=L=J_{F^{(s)}}^{L},\,J_{F}=0$. Hence, the expression for the fully DCS for the case under consideration
simplifies to
\begin{widetext}
\begin{align}
\frac{ {\rm d}\sigma}{  {\rm d}{\Omega}_{{s}}\,{\rm d}{\Omega}_{{\gamma}}\, {\rm d}{E_{sF}} }
=& - \frac{\mu_{aA}\,\mu_{sF}}{(2\,\pi)^{7} }\,\frac{\varphi_{a}^{2}(p_{sx})\,R_{xA}}{2\,\mu_{xA}}\,\frac{ k_{sF}}{k_{aA}}\,\sum\limits_{L'\,L} (-1)^{L'+L}\, k_{\gamma}^{L'+L+1}\,
\sqrt{ {\hat L}'\,{\hat L}}                                   \nonumber\\
& \times\, {\cal {\tilde W}}_{L'}^{*}\,{\cal {\tilde W}}_{L}\,  
 \gamma_{\tau'\,0\,L'\,L'} \,\gamma_{\tau\,0\,L\,L} \,\sum\limits_{l}\,  \blb L'\,0\,\,L\,0\big|l\,0 \blk \, \blb L'\,1\,\,L\,-1 \big| l\,0 \blk \,P_{l}(\cos{\theta})    \nonumber\\
& \times\sum\limits_{\nu', \,\nu,\,\tau',\, \tau=1}^{2}\,\big[\gamma_{ (\gamma)\,\nu'\,0\,L'}^{L'}\,\big]^{*} [\gamma_{(\gamma)\,\nu\,0\,L}^{L} ]\,\big[{\bf A}_{v'\,\tau'}^{L'} \big]^{*}\,\big[{\bf A}_{\nu\,\tau}^{L} \big]\  
\label{tripplediffcrsect11}
\end{align}
\end{widetext}
Here, $a={}^{6}{\rm Li},\,A={}^{12}{\rm C},\,s=d,\,x=\alpha,\,F={}^{16}{\rm O}$.
This expression is used for the analysis of the indirect reaction ${}^{12}{\rm C}({}^{6}{\rm Li},d\,\gamma){}^{16}{\rm O}$ at low energies. We outline below some details of the calculations. 

After integration over the photon's solid angle we get the indirect doubly DCS (\ref{doublediffcrsectSfctr1})
in which $l_{xA}=L$. 
Then at energies near the $1^{-}$ resonance at $2.423$ MeV where, as we will see below, the $E1$ transition completely dominates,
\begin{align}
 S_{E1}(E_{xA})=& N_F\,\frac{ {\rm d}\sigma}{  {\rm d}{\Omega}_{  {\rm{\bf {\hat k}}}_{s} }\, {\rm d}{E_{sF}} }\,\frac{1}{K(E_{xA})\,\varphi_{a}^{2}(p_{sx})\,R_{xA}}\,
 \nonumber \\ & \times
 e^{2\,\pi\,\eta_{xA}}\,P_{1}(E_{xA},\,R_{xA})\,
\big|{\cal {\tilde W}}_{1} \big|^{-2}.
\label{Sfactrdblcrsect412}
\end{align}
The $S_{E1}$ astrophysical factor was measured at energies near $2.423$ MeV with a very good accuracy   \cite{redder,assuncao,Schurmann}. Should we have the experimental indirect doubly DCS expressed in arbitrary units, we can use Eq. (\ref{Sfactrdblcrsect412}) to normalize the $S_{E1}(E_{xA})$ to the experimental one at higher energies. After that, having 
indirectly measured the doubly DCS at $0.3$ MeV, we can determine the $S_{E1}(0.3 \ {\rm MeV}) + S_{E2}(0.3 \ {\rm MeV})$. 

We can calculate  the photon's angular distribution at different $E_{\alpha\,{}^{12}{\rm C}}$ energies for the 
${}^{12}{\rm C}(\alpha,d\,\gamma){}^{16}{\rm O}$ reaction and study how it is affected by the interference character (constructive or destructive) of the $1^{-}$ subthreshold bound state and $1^{-}$ resonance.
In the $R$-matrix approach the fitting parameters are the formal reduced widths $\gamma_{\tau\,j_{xA}l_{xA}J_{F^{(s)}}}^{2}$ which are related to the 
observable ones  ${\tilde \gamma}_{\tau\,j_{xA}l_{xA}J_{F^{(s)}}}^{2}$  by 
\begin{align}
{\tilde \gamma}_{\tau\,j_{xA}l_{xA}J_{F^{(s)}}}^{2}= \frac{\gamma_{\tau\,j_{xA}l_{xA}J_{F^{(s)}}}^{2}}{ 1+ \gamma_{\tau\,j_{xA}l_{xA}J_{F^{(s)}}}^{2} [{\rm d}S_{l_{xA}}(E_{xA})/{\rm d}E_{xA}] \big|_{E_{xA}= E_{\tau} }}. 
\label{obsformwidth1}
\end{align}
$R$-matrix energy levels, $E_{\tau}$, are $E_{1} = -\varepsilon_{xA}^{(s)}$ and $E_{2} = E_{R}$, where $\varepsilon_{xA}^{(s)}$ is the binding energy of the subthreshold bound state 
$(xA)^{(s)}$ and  $\,E_{R}\,$ is the resonance energy corresponding to the level $\tau=2$.
 The observable reduced widths $({\tilde \gamma}_{1\,0\,1\,1})^{2}$ and $({\tilde \gamma}_{1\,0\,2\,2})^{2}$ are expressed in terms of the corresponding ANCs of the subthreshold bound states by Eq. (\ref{ANCredwidth1}). 
For the ANCs of the $1^{-}$ and $2^{+}$ subthreshold states we adopted $[C_{(\alpha\,{}^{12}{\rm C})1}^{(s)}]^{2}=4.39 \times 10^{28}$ fm${}^{-1}$ and $\,[C_{(\alpha\,{}^{12}{\rm C})2}^{(s)}]^{2}=1.48 \times 10^{10}$ fm${}^{-1}$ \cite{avila}, respectively.
In all the calculations, following \cite{brune1999}, we use the channel radius $R_{ch(\alpha\,{}^{12}{\rm C})}= 6.5$ fm. 
The observable reduced width of the resonance $1^{-}$ is expressed in terms of the observable resonance width of this resonance (see Eq. (\ref{Gamman1})). For this resonance we adopt ${\tilde \Gamma}_{2\,011}=0.48$ MeV \cite{tiley}. 
In the case under consideration for the $E1$ transition we take into account three states and select the boundary condition at the energy of the first level, which is the $1^{-}$ subthreshold bound state, that is, $E_{1}=-\varepsilon_{xA(1)}^{(s)}$. 
For the $E2$ transition we take into account two levels and select the boundary condition at the energy of the $2^{+}$ subthreshold bound state $E_{2}=-0.245$ MeV.

Now we discuss the radiative width amplitudes.
We use Eq. (\ref{observformradwidth1}) to express the formal radiative widths amplitudes
$\gamma_{(\gamma)\,1\,0\,1}^{1},\,\gamma_{(\gamma)\,2\,0\,1}^{1}$ and 
$\gamma_{(\gamma)\,1\,0\,2}^{2}$ in terms of corresponding 
observable reduced widths, which are related to the observable radiative resonance widths by Eq. (\ref{Gg2}) \cite{lanethomas}.

Another important point to discuss is the kinematics of the indirect reaction. The fully DCS is proportional to $\varphi_{d\,\alpha}^{2}(p_{d\,\alpha})$,  which is shown in Fig. \ref{fig_6Libstwfpspace}. Here  $\varphi_{{}^{6}{\rm Li}   }(p_{d\,\alpha})   \equiv    \varphi_{d\,\alpha}(p_{d\,\alpha})$. The maximum of $\varphi_{d\,\alpha}^{2}(p_{d\,\alpha})$ at $p_{d\,\alpha}=0$ (QF kinematics) also provides the maximum of the fully DCS. Here $p_{d\,\alpha}$ is the $d-\alpha$ relative momentum  in the three-ray vertex ${}^{6}{\rm Li} \to d + \alpha$ of the diagram in Fig. \ref{fig:fig_polediagram1}.
\begin{figure}
[tbp] 
  \includegraphics[width=\columnwidth]{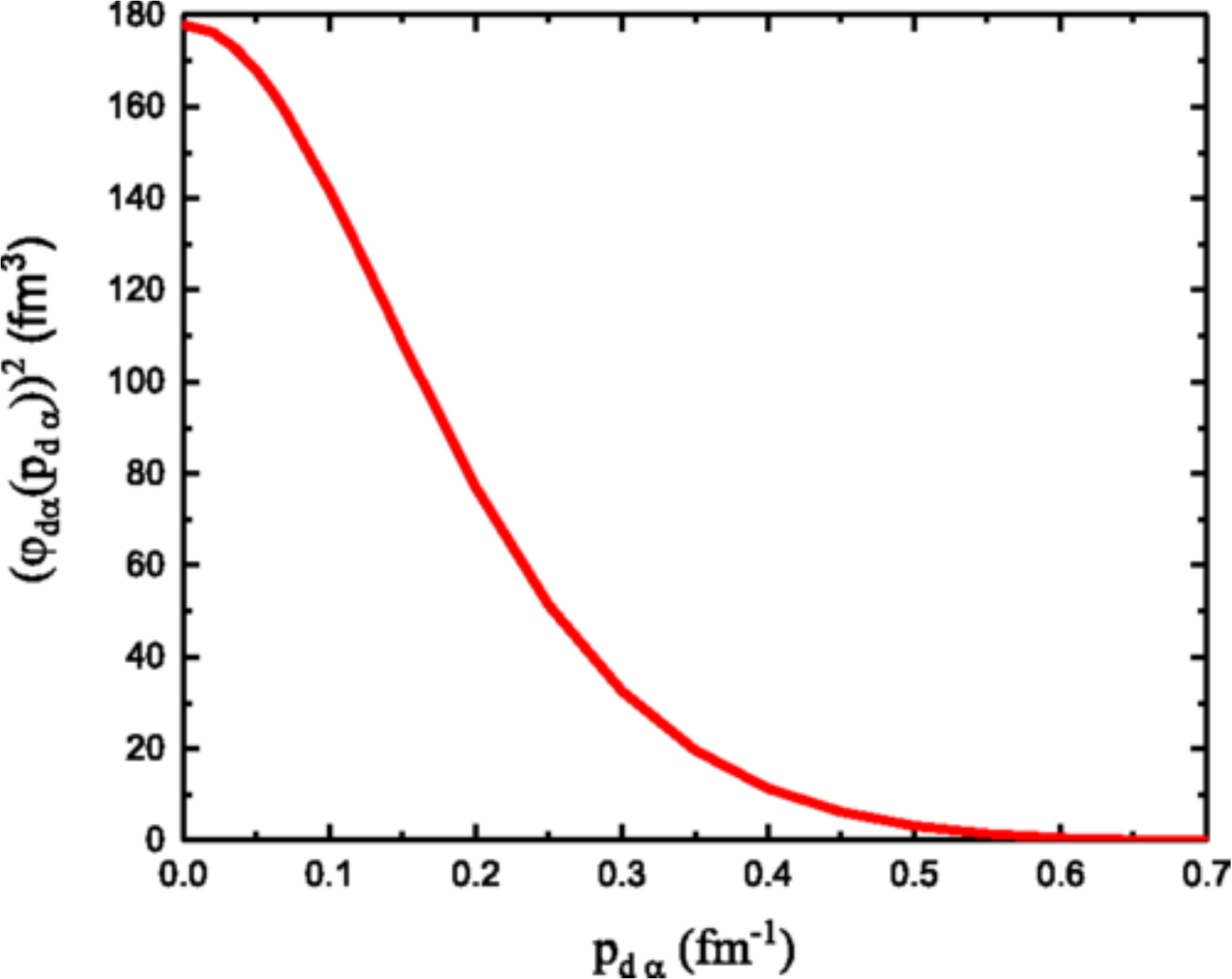}
  \caption{Square of the $d-\alpha$ bound state wave function in the momentum space.
  First published in \cite{mukrog}.} 
 \label{fig_6Libstwfpspace}
\end{figure} 
 
To calculate the Fourier transform of the $\,{}^{6}{\rm Li}=(d\,\alpha)$ bound-state wave function we use the Woods-Saxon potential with the depth $\,V_{0}= 60.0$ MeV, radial parameter $r_{0}=r_{C}=1.25$ fm and diffuseness $a=0.65$ fm. This potential provides the $d-\alpha$ bound state with the binding energy $\,\varepsilon_{d\,\alpha}= 1.474$ MeV \cite{tiley} . The corresponding bound-state wave number of the $\,d\,\alpha$ bound state is $\kappa_{d\,\alpha} = \sqrt{2\,\mu_{d\,\alpha}\,\varepsilon_{d\,\alpha}} = 0.31$ fm$^{-1}$.
The square of ANC for the virtual decay ${}^{6}{\rm Li} \to d + \alpha$ is $[C_{(d\,\alpha)0}]^{2}= 7.28$ fm${}^{-1}$. This value is higher than the realistic value of this square of $[C_{(d\,\alpha)0}]^{2}= 5.29$ fm${}^{-1}$ \cite{blokh1993}. To get the correct ANC from the one obtained in the Woods-Saxon potential we need to introduce the spectroscopic factor.  However, since we are not interested in the absolute cross section, we keep using the ANC generated by the Woods-Saxon potential.

Usually the indirect experiments are performed at fixed incident energy of the projectiles \cite{reviewpaper}. In the case under consideration the projectile is ${}^{6}{\rm Li}$ or ${}^{12}{\rm C}$ (in the inverse kinematics). To cover the $E_{\alpha\,{}^{12}{\rm C}}$ energy interval $\sim 2$ MeV at fixed relative kinetic energy $E_{^{6}{\rm Li}\,{}^{12}{\rm C}}$, one needs to change $p_{d\,\alpha}$. Since ${\rm {\bf k}}_{{}^{6}{\rm Li}}$ is fixed to change $p_{d\,\alpha}$ 
we have to change ${\rm {\bf k}}_{d}$ so that $p_{d\,\alpha}
\leq \kappa_{d\,\alpha}$. It can be achieved by changing $k_{d}$ or its direction ${\rm {\bf {\hat k}}}_{d}$ or both.
Experimentally one can select all the events falling into the region $p_{d\,\alpha} \leq \kappa_{\alpha\,d}$. Here, 
to simplify calculations, we assume that ${\rm {\bf k}}_{d} || {\rm {\bf k}}_{{}^{6}{\rm Li}}$. It means that the variation of $p_{d\,\alpha}$ is achieved by changing of $k_{d}$. 
Owing to the energy conservation by changing $k_{d}$ we can vary $E_{\alpha\,{}^{12}{\rm C}}$ but simultaneously we change the $d-\alpha$ relative momentum $p_{d\,\alpha}$. The fully
DCS given by Eq. (\ref{tripplediffcrsect11}) is proportional to the $d-\alpha$ bound-state wave function in the momentum space $\varphi_{d\,\alpha}^{2}(p_{d\,\alpha})$, which 
decreases with increase of $p_{d\,\alpha}$, see Fig. \ref{fig_6Libstwfpspace}.

 To avoid significant decrease of the fully DCS when covering the $E_{\alpha\,{}^{12}{\rm C}}$ energy interval $\approx 2$ MeV it is better to take a lower $E_{^{6}{\rm Li}\,{}^{12}{\rm C}}$ 
but not too close to the Coulomb barrier in the initial channel of the indirect reaction (\ref{THMreaction1}). Taking into account the fact that this Coulomb barrier is $\approx 5$ MeV we consider as an example the relative kinetic energy $E_{^{6}{\rm Li}\,{}^{12}{\rm C}}=7$ MeV. 
In this case for $E_{\alpha\,{}^{12}{\rm C}}=2.28$ MeV, which is close to the resonance energy of the $1^{-}$ resonance, $p_{d\,\alpha}= 0.141$ fm${}^{-1}$ while at $E_{\alpha\,{}^{12}{\rm C}}=0.3$ MeV  $p_{d\,\alpha}= 0.281$ fm${}^{-1}$. Hence, when covering the $E_{\alpha\,{}^{12}{\rm C}}$ energy interval from   $E_{\alpha\,{}^{12}{\rm C}}=2.28$ MeV to the most effective astrophysical energy for the process 
${}^{12}{\rm C}(\alpha,\,\gamma){}^{16}{\rm O}$  the square of the Fourier transform $\varphi_{d\,\alpha}^{2}(p_{d\,\alpha})$ drops by a factor of $2.97$. Note that the drop of $\varphi_{d\,\alpha}^{2}(p_{d\,\alpha})$, when moving from $E_{\alpha\,{}^{12}{\rm C}}=2.28$ MeV to $0.9$ MeV, is $2.1$. Note that $\,\varphi_{d\,\alpha}^{2}(p_{d\,\alpha})$ appears because we consider the indirect three-body reaction. There is another energy-dependent factor ${\cal {\tilde W}}_{L}$, which is also result of the consideration of the three-body indirect reaction. This factor will be considered below.

Our goal is to calculate the photon's angular distributions at different $E_{\alpha\,{}^{12}{\rm C}}$
energies. It allows us to compare the indirect cross sections at higher energy $E_{\alpha\,{}^{12}{\rm C}} = 2.28$ MeV and the most effective astrophysical energy $E_{\alpha\,{}^{12}{\rm C}} = 0.3$ MeV. Since the indirect fully DCS does not contain the penetrability factor in the channel $\alpha-{}^{12}{\rm C}$ of the binary sub-reaction (\ref{reaction1}), the indirect method allows one to measure the fully DCS at $E_{\alpha\,{}^{12}{\rm C}} = 0.3$ MeV what is impossible by any direct method. In particular: 
\begin{enumerate}
\item By comparing the fully DCSs at higher energies and at $0.3$ MeV we can determine how much the indirect cross section will drop when we reach $E_{\alpha\,{}^{12}{\rm C}} = 0.3$ MeV. It will help understand whether it is feasible to measure the fully DCS  at such a low energy.
\item The second goal is to determine whether the interference of the $1^{-}$ subthreshold resonance and $1^{-}$ resonance at $2.423$ MeV is constructive or distractive because the pattern of this interference may affect the photon's angular distribution.
\item The third goal is to compare the relative contribution of the $E1$ and $E2$ transitions. 
\end{enumerate}

\subsection{Astrophysical factors for ${}^{12}{\rm C}(\alpha,\,\gamma){}^{16}{\rm O}$}

First, to determine the parameters, which we use to calculate the fully DCSs, we fit the experimental  astrophysical factors $S_{E1}$ for the $E1$ transition and 
$S_{E2}$ for the $E2$ transition for the ${}^{12}{\rm C}(\alpha,\,\gamma){}^{16}{\rm O}$ reaction from \cite{redder}. We do not pursue a perfect fit and  are mostly interested in fitting energies below the $1^{-}$ resonance at $2.423$ MeV,  and at low energies $E_{\alpha\,{}^{12}{\rm C}} \leq 1$ MeV. To get an acceptable fit for the $E1$ transition we needed  to include three levels, two physical states, subthreshold $1^{-}$ state and the $1^{-}$ resonance, and one background state. 
Note that in the $R$-matrix fits one  must take into account all the upper-lying $1^{-}$ levels
which is practically impossible. Sometimes taking into account closest levels, subthreshold and resonance ones 
 in the case under consideration, is enough. But in all the previous  publications when fitting the $E1$ transition an artificial  background is added. We also introduce one additional state as a background.

For the $E2$ transition it was enough to include only two physical states, $2^{+}$ subthreshold resonance and $2^{+}$ resonance at $2.683$ MeV.

Our goal is to demonstrate the pattern of the fully DCS using reasonable parameters. A more accurate fit can be done when indirect data will be available.
In our fit, we kept fixed only the parameters of the subthreshold resonances $\,1^{-}\,$ and $\,2^{+}\,$ while the parameters of the higher lying resonances $\,1^{-}\,$ and $\,2^{+}\,$ were varying.
The fixed parameters are shown in Table \ref{table_parameters} in parentheses. This table shows
the set of the parameters used to fit the astrophysical factors $\,S_{E1}\,$ and $\,S_{E2}\,$. These parameters are also used to calculate the fully DCS. 
\tabcolsep=3pt
\begin{table}
\begin{center}
\caption{Parameters used in calculations of the astrophysical factors of the ${}^{12}{\rm C}(\alpha,\,\gamma){}^{16}{\rm O}$ radiative capture and the photon's angular distributions from the indirect ${}^{12}{\rm C}({}^{6}{\rm Li}, d\,\gamma){}^{16}{\rm O}$ reaction.}
\vspace{0.4cm}
\begin{tabular}[t]{c c c}
\hline \hline \\
 &  $L=1$ &  $L=2$  \\[2mm]
\hline\\[1mm]
$E_{1}$ [MeV] & $(-0.45)$ &  $(-0.245)$   \\[1ex]
 $\gamma_{1\,0\,L\,L}$ [MeV$^{1/2}$]\,\,\, & $(0.0867)$ & $(0.1500)$  \\[1ex] 
  $\gamma_{(\gamma)\,1\,0\,L\,L}^{L}$ [MeV${}^{1/2}$fm${}^{L+1/2}$] & $(0.0241)$ & $(0.9415)$  \\[1ex]
    $E_{2}$ [MeV] & $3.0$ & $2.8$  \\[1ex]
   $\gamma_{2\,0\,L\,L}$ [MeV$^{1/2}$] & $0.3254$ & $0.75$  \\[1ex]
$\gamma_{(\gamma)\,2\,0\,L\,L}^{L}$ [MeV${}^{1/2}$fm${}^{L+1/2}$]  & $-0.00963$ & $-0.09257$  \\[1ex]
$E_{3}$ [MeV] & $33.8$ &   \\ [1ex]  
$\gamma_{3\,0\,L\,L}$ [MeV$^{1/2}$]  & $1.1$ &  \\[1ex]
$\gamma_{(\gamma)\,3\,0\,L\,L}^{L}$ [MeV${}^{1/2}$fm${}^{L+1/2}$] & $-0.00239$ &  \\[2mm]
\hline \hline \\
\label{table_parameters}
\end{tabular}
\end{center}
\end{table}
$E_{n}$ is the energy of the $n$-th level. 

Note that in the $R$-matrix approach, which includes a few interfering levels, it is convenient to choose one of the energy levels to coincide with the location of the observable physical state \cite{barker2008,muk2016} while energies of other levels become fitting parameters.

We adopted $E_{1}=-\varepsilon_{\alpha\,{}^{12}{\rm C}(1)}^{(s)}=-0.045$ MeV for $L=1$ and $E_{2}=-\varepsilon_{\alpha\,{}^{12}{\rm C}(2)}^{(s)}= -0.245$ MeV for $L=2$ transitions. 
Then the boundary condition for the second and third levels of the $E1$ transition is taken at $E_{\alpha\,{}^{12}{\rm C}}=-0.045$ MeV while for $L=2$ the boundary condition is taken at $E_{\alpha\,{}^{12}{\rm C}}= -0.245$ MeV. Moreover, because in our choice the locations of the subthreshold bound states for $L=1$ and $L=2$ are fixed, the energies of other levels are  fitting parameters and deviate from the real resonance energies. For example, the $1^{-}$ resonance at $2.423$ MeV in the fit is shifted to $E_{\alpha\,{}^{12}{\rm C}}= 3.0
$ MeV and the $2^{+}$ resonance at $2.683$ MeV is shifted to $2.8$ MeV. Hence, the statement that we take into account the radiative capture through the wing of the subthreshold $1^{-}$ resonance at $E_{\alpha\,{}^{12}{\rm C}}=-0.045$ MeV and the $1^{-}$ resonance at $E_{\alpha\,{}^{12}{\rm C}}= 2.423$ MeV does not contradict  the fact that in the fit the resonance at $2.423$ MeV is shifted to $3.0$ MeV. To fit the $E1$ transition we needed to add the background state at $33.8$ MeV with parameters given in Table \ref{table_parameters}. 

The parameters given in  this table provide the constructive interference of the subthreshold $1^{-}$ resonance and resonance at $2.423$ MeV at low energies. Changing the sign of 
$\gamma_{(\gamma)\,2\,0\,1\,L}^{1}=-0.00963$ MeV${}^{1/2}$fm${}^{3/2}$ to positive  provides the destructive interference between the first two $1^{-}$ levels. In what follows by the $E1$ 
constructive (destructive) interference we mean the constructive (destructive) interference between the first two $1^{-}$ levels.

\begin{figure*}
[tbp] 
    \includegraphics[width=1.2\columnwidth]{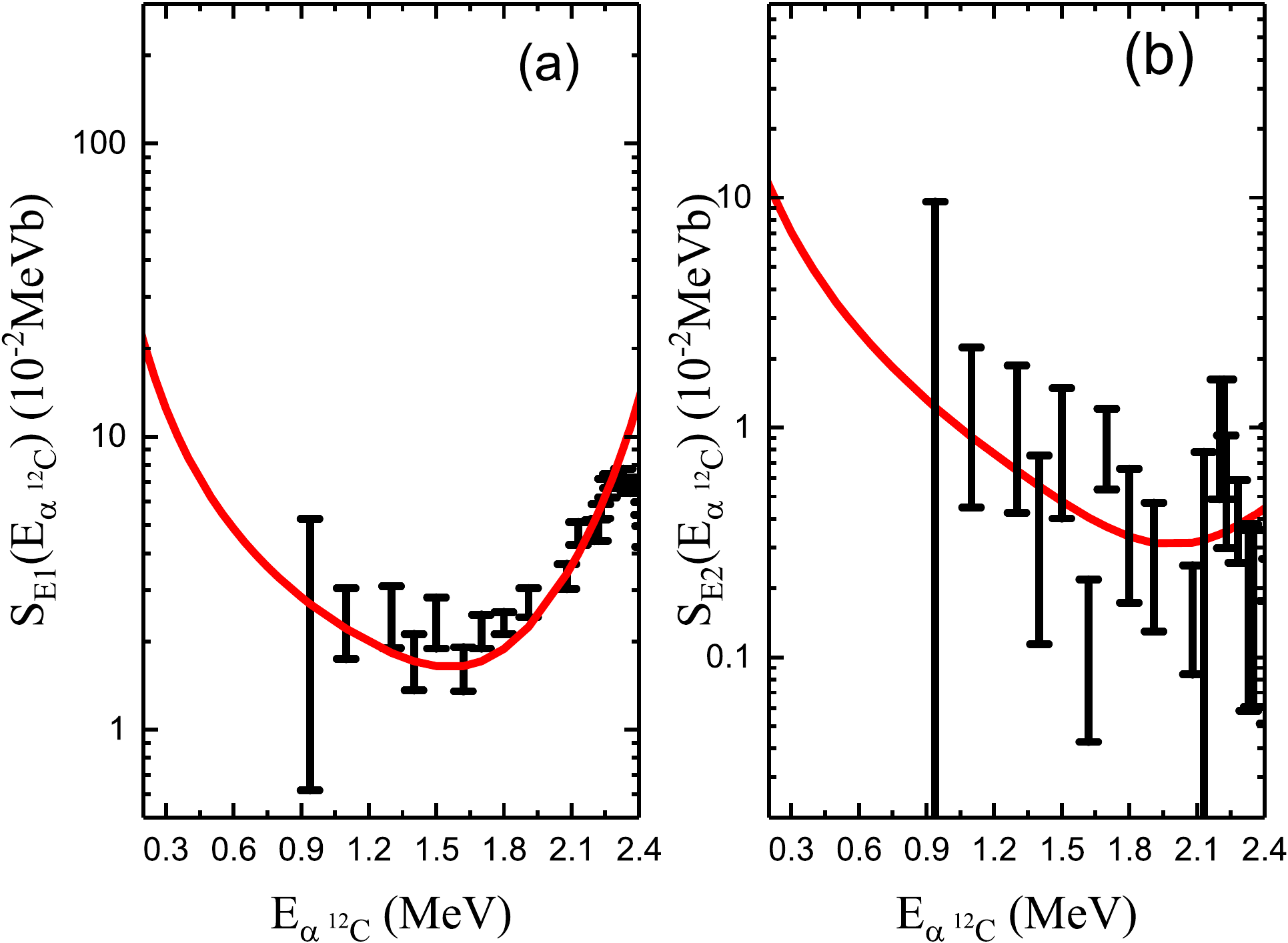}
  \caption{Low-energy astrophysical $S_{E1}(E_{\alpha\,{}^{12}{\rm C}})$ and $S_{E2}(E_{\alpha\,{}^{12}{\rm C}})$ factors for $E1$ and $E2$ transitions for the  ${}^{12}{\rm C}(\alpha,\,\gamma){}^{16}{\rm O}$ radiative capture. Black dots are astrophysical factors from \cite{redder}, solid red line is present paper fit.
Panel (a): $S_{E1}(E_{\alpha\,{}^{12}{\rm C}})$ astrophysical factor; panel (b): $S_{E2}(E_{\alpha\,{}^{12}{\rm C}})$ astrophysical factor.
First published in \cite{mukrog}.}  
\label{fig_Sfactors}
\end{figure*} 

In Fig. \ref{fig_Sfactors} the calculated $S_{E1}$ and $S_{E2}$ astrophysical factors for the $E1$ and $E2$ transitions, respectively, are compared with the experimental ones from \cite{redder}. 
Our fitted astrophysical factors are: $S_{E1}(0.3\,{\rm MeV})= 124.6$ keVb for the $E1$ transition and $S_{E2}(0.3\,{\rm MeV})= 71.1$ keVb for the $E2$ transition. One can see that our 
value for the $E1$ transition is higher than the contemporary accepted value of $80$ keVb for constructive interference but the value for the $E2$ transition is close to the low value $60$ keVb \cite{gai2015}. But, as we have underscored, our values should not be taken quantatively.
In the absence of indirect data we use the parameters obtained from fitting the data from \cite{redder} to generate the photon's angular distributions to make some qualitative predictions.
We also show how the photon's angular distributions are affected by lowering $S_{E1}(0.3\,{\rm MeV})$.

\subsection{Photon's angular distributions}
\label{Photangdistr1}

\begin{figure*}
[tbp]   
  \includegraphics[width=0.7\textwidth
  ]{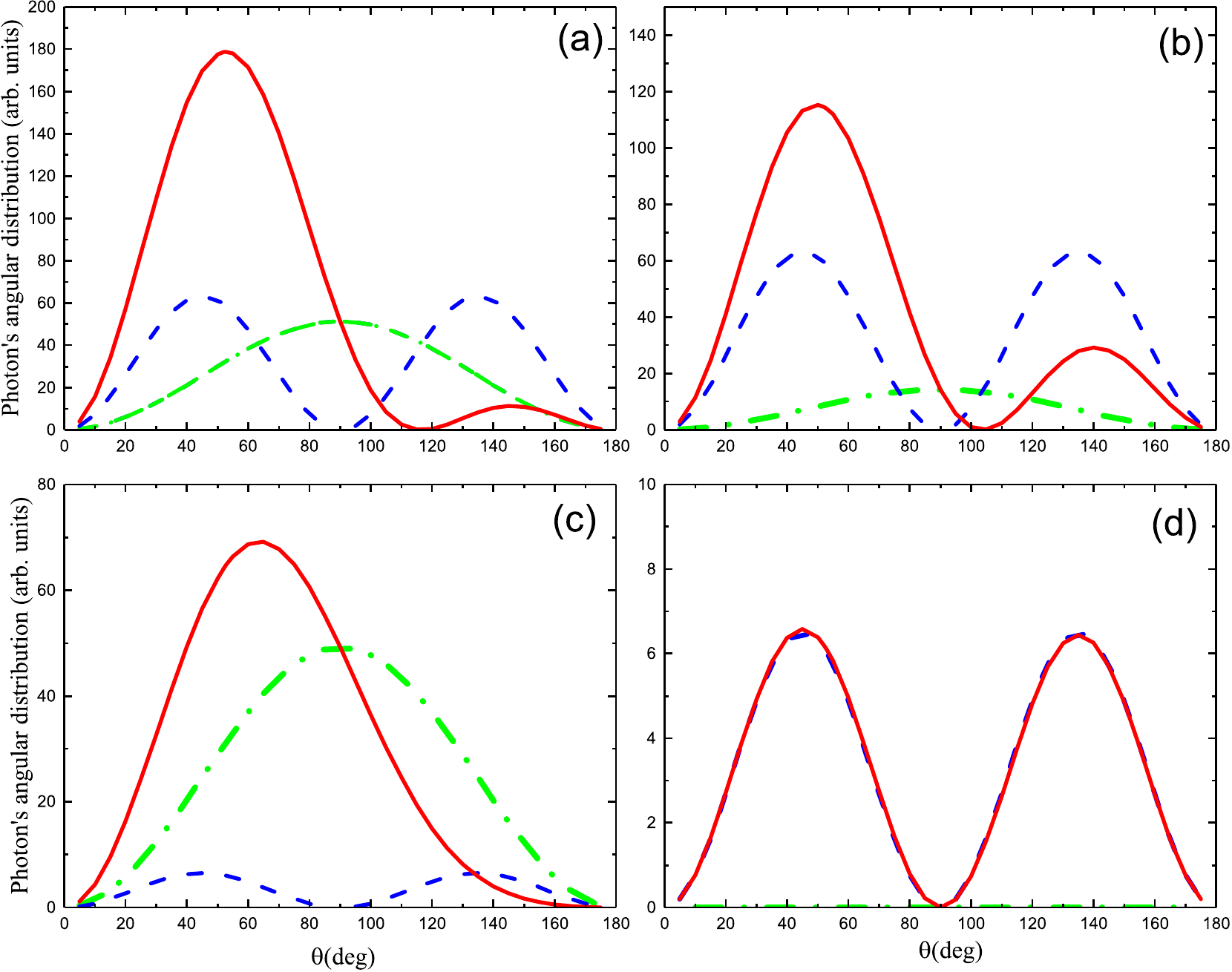}
  \caption{(Color online) Angular distribution of the photons emitted from the reaction ${}^{12}{\rm C}({}^{6}{\rm Li},d\,\gamma){}^{16}{\rm O}$ proceeding through the wings of two subthreshold resonances $1^{-},\,E_{\alpha\,{}^{12}{\rm C}} = -0.045$ MeV, $2^{+},\,E_{\alpha\,{}^{12}{\rm C}}= -0.245$ MeV, and the resonances at $E_{\alpha\,{}^{12}{\rm C}} >0$. The green dashed-dotted line is the angular distribution for the electric dipole transition $E1$, the blue dashed line is the angular distribution generated by the electric quadrupole $E2$ transition, and the red solid line is the total angular distribution resulted from the interference of the $E1$ and $E2$ radiative captures.
Panel (a):  $\,E_{\alpha\,{}^{12}{\rm C}}= 0.3$ MeV, constructive interference of the $\,E1$ transitions through the wing of $\,1^{-},\,E_{\alpha\,{}^{12}{\rm C}} = -0.045$ MeV and the resonance $\,1^{-},E_{R}= 2.423$ MeV; panel (b): $\,E_{\alpha\,{}^{12}{\rm C}}= 0.3$ MeV, destructive interference of the $E1$ transitions through the wing of $\,1^{-},\,E_{\alpha\,{}^{12}{\rm C}} = -0.045$ MeV and the resonance $\,1^{-},E_{R}= 2.423$ MeV;  panel (c): the same as panel (a) but for $\,E_{\alpha\,{}^{12}{\rm C}}= 0.9$ MeV; panel (d): the same as panel (b) but for $\,E_{\alpha\,{}^{12}{\rm C}}= 0.9$ MeV.
 First published in \cite{mukrog}.}  
 \label{fig_Angdistr1}
\end{figure*}

\begin{figure*}
[tbp] 
 \includegraphics[width=0.7\textwidth
 ]{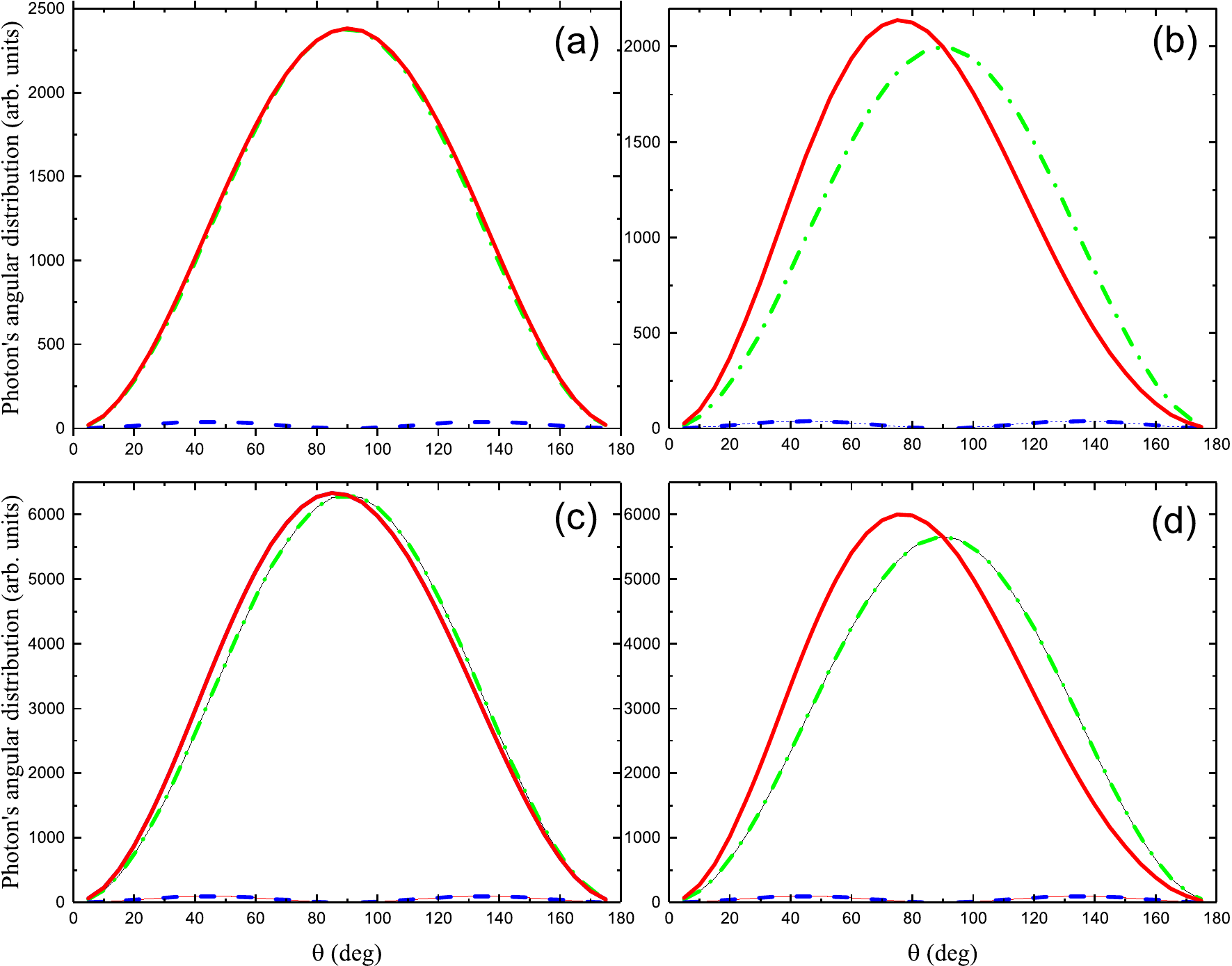}
 \caption{(Color online) Angular distribution of the photons emitted from the reaction ${}^{12}{\rm C}({}^{6}{\rm Li},d\,\gamma){}^{16}{\rm O}$ proceeding through the wings of the two subthreshold resonances, $1^{-},\,E_{\alpha\,{}^{12}{\rm C}} = -0.045$ MeV, $2^{+},\,E_{\alpha\,{}^{12}{\rm C}}= -0.245$ MeV,  and the resonances at $E_{\alpha\,{}^{12}{\rm C}} >0$. Notaions of the lines are the same as in Fig. \ref{fig_Angdistr1}. 
Panel (a): the same as panel (a) in Fig. \ref{fig_Angdistr1} but for $E_{\alpha\,{}^{12}{\rm C}}= 2.1$ MeV; panel (b): the same as panel (b) in Fig. \ref{fig_Angdistr1} but for $E_{\alpha\,{}^{12}{\rm C}}= 2.1$ MeV;
panel (c): the same as panel (c) in Fig. \ref{fig_Angdistr1} but for $E_{\alpha\,{}^{12}{\rm C}}= 2.28$ MeV;
panel (d): the same as panel (d) in Fig. \ref{fig_Angdistr1} but for $E_{\alpha\,{}^{12}{\rm C}}= 2.1$ MeV.
First published in \cite{mukrog}.}
\label{fig_Angdistr2}
\end{figure*}

In Figs. \ref{fig_Angdistr1}   and  \ref{fig_Angdistr2} the photon's angular distributions are shown at four
different $\,E_{\alpha\,{}^{12}{\rm C}}\,$ energies: $\,0.3,\,0.9,2.1$ and $\,2.28$ MeV. 
We do not show the angular distributions at an intermediate energy of $\,1.5$ MeV because it is very similar to the angular distributions at higher energies and is completely dominated by the $E1$ transition. The calculations are performed at $E_{^6{\rm Li}\, ^{12}{\rm C}}=7$ MeV ($10.5$ MeV in the laboratory frame of reference with $\,{}^{6}{\rm Li}\,$ as a projectile), which is higher than the Coulomb barrier $\,V_{CB} \approx 5\,$ MeV in the entry channel $\,{}^{6}{\rm Li} + {}^{12}{\rm C}\,$ of the indirect reaction.  The angular distributions are calculated in the c.m. 
of the reaction ${}^{6}{\rm Li}({}^{12}{\rm C},d\,\gamma){}^{16}{\rm O}$. We selected axis $z||{\rm {\bf p}}_{xA}$.
Hence the photon scattering angle is counted from ${\rm {\bf p}}_{xA}$, which itself is determined by ${\rm {\bf k}}_{s}$. The calculated photon angular distribution determines  the angular correlation between the
emitted photons from the intermediate excited state $F^{*}$ and the spectator $s$.

Figures \ref{fig_Angdistr1} and \ref{fig_Angdistr2} are very instructive. 
First, we note that the $E1$ angular distributions of the photons at all energies are peaked at $90^{\circ}$ while the $E2$ angular distributions are double-humped and peaked at $45^{\circ}$ and $135^{\circ}$. However, the interference of the $E1$ and $E2$ transitions leads to different total angular distributions.
The angular distributions at $0.3$ MeV are quite similar for the $E1$ transitions with constructive and destructive interferences, panels $(a)$ and $(b)$ in Fig. \ref{fig_Angdistr1}, with pronounced peaks at $52^{\circ}$ and $50^{\circ}$, respectively. The character of the total angular distribution at $0.3$ MeV depends on the relative weight of the $E1$ and $E2$ transitions.

The photon's angular distributions at $0.9$ MeV, panels $(c)$ and $(d)$, are the most instructive.  The patterns of the photon's angular distributions are different for the constructive and destructive $E1$ transitions. This allows one to distinguish between the two types of $E1$ interferences. However, the cross sections for the destructive $E1$ interference is too small compared to the cross section at $0.3$ MeV.

Finally, we proceed to the angular distributions at higher energies. These are shown in Fig. \ref{fig_Angdistr2}. At higher energies, the $E1$ transition dominates and we see profound $E1$ type angular distributions both for the $E1$ constructive and destructive interferences of the two first $1^{-}$ levels. Hence, the angular distributions at higher energies cannot distinguish between constructive and destructive $E1$ interferences.

Comparing the relative values of the fully DCSs of Fig. \ref{fig_Angdistr2}, panel (c) and Fig. \ref{fig_Angdistr1}, panel (a) we can make, presumably, the most important conclusion: 
the fully DCS near the $1^{-}$ resonance at $2.28$ MeV exceeds the one at $0.3$ MeV by approximately an order of magnitude. We recall that in the case of the direct measurements when moving from $2.28$ MeV to $0.3$ MeV, the cross section drops by a factor of  $10^{9}$. Our estimation 
shows that measurements of the indirect fully DCS at $0.3$ MeV are feasible. Thus, for the first time, we provide a possibility  to measure the ${}^{12}{\rm C}(\alpha,\,\gamma){}^{16}{\rm O}$ right at the most effective astrophysical energy $0.3$ MeV. 

\section{Summary}
\label{Summary1}
In this review, we focused on the theory of the Trojan horse method.
The THM  is a powerful indirect technique allowing to obtain the astrophysical factors for resonant rearrangement  reactions at astrophysically-relevant energies which often cannot be reached in direct experiments.
 The  THM  was suggested by Baur  \cite{Baur1986}. However, in its suggested form the method was not practical because  the low astrophysical energies of the binary subreaction $x+A \to b+B$  can be achieved  at  the high $s-x$ relative $p_{sx}$  momentum.  High  $p_{sx}$  correspond to small $r_{sx}$,  meaning  $s$ is close to $x$ and cannot be treated  as a spectator.  It was Spitaleri \cite{Spitaleri1999} who first described the correct  kinematics making the THM method suitable for practical applications. This kinematics follows from Eqs.  
(\ref{ExATHM1})--(\ref{ExAkazero1}).  In particular, from Eq.  (\ref{ExATHMQF1}) it   
follows  that at the quasi-free kinematics ($p_{sx}=0$),  low $E_{xA}$  are achieved due to the compensation by the binding
energy of the TH particle $a=(sx)$.  

Another important point regarding the application of the THM concerns types 
of the binary reactions which can be studied  using the THM. There are two main types of binary rearrangement subreactions playing an important role in nuclear astrophysics: direct and resonant reactions. Due to the presence of the Coulomb-centrifugal penetrability 
factors in each partial wave, the direct rearrangement reactions at astrophysically-relevant energies  are contributed by the lowest partial waves $l_{xA}=0,1$.    Meanwhile,  in the THM
the penetrability  factor in the  channel $x+A$  is absent. Hence, a significant number of the partial waves can contribute to the THM amplitude of the subreaction $x+A \to b+B$ and, a priori, the THM cannot be used  to extract
$S$  factors for the direct reactions except for  some exceptional cases for which  $Q_{ 2 } =Q_{2} =m_{x} + m_{A} - m_{b}- m_{B}$ is so large that  condition $k_{bB} >> k_{xA}$ takes place.  Then the angular distribution  of the  DCS of the binary subreaction $x+A \to b+B$  will be  isotropic or almost isotropic, so that only $l_{xA}=0,1$ partial waves contribute.

The primary goal of the THM is to analyze the resonant rearrangement reactions when the density of the resonance levels is low and the statistical model cannot be applied. 
The main  difficulty of the analysis is related with the facts that in the final state of the TM reaction  involves three particles and that 
the intermediate particle $x$, which is transferred from the TH particle $a$ to nucleus $A$ to form a resonance state $F^{*}$, is virtual. Another difficulty is associated with the Coulomb interaction between the particles, especially, taking into account that  the goal of the THM is to study resonant rearrangement reactions at very low energies important for nuclear astrophysics. The exact theory of such reactions with three charged particles is very complicated and is not  available. That is why different approximations are being used to analyze THM reactions. The simplest one the plane-wave approximation in which  all the rescaterring  effects are neglected.  This approach has significant shortcomings and may provide incorrect results, especially  for heavier particles.  The shortcoming of the PWA  has been discussed in this paper. 

In this review paper  we described a new approach based on a few-body approach, which provides a solid basis for deriving the THM reaction amplitude taking into account rescattering of the particles in the initial, intermediate and final states of the THM reaction.  Since the THM uses a two-step reaction in which the first step is the transfer reaction populating a resonance state, we addressed the theory of the transfer reactions. The theory is based on the surface-integral method and $R$-matrix formalism. The surface-integral approach allows one to take into account explicitly the off-shell character of the transferred particle in the transfer reaction, which represents the first step of the THM reaction. That is why the quasi-free character of the resonant sub-reaction is not needed. Moreover, the initial and final-state Coulomb-nuclear distortions make the assumption of the quasi-free character of the resonant  sub-reaction invalid.   To single out the amplitude of the resonance sub-reaction $x+A \to b+B$, which represents the second step of the THM reaction, we extrapolate the amplitude to the second energy sheet over $E_{xA}$.
We discussed application of the THM for a resonant reaction populating both resonances located on the second energy sheet and subthreshold resonances, which are subthreshold bound states located at negative energies close to thresholds. We also discussed  the application of  the THM  to determine the astrophysical factors of  resonant radiative-capture reactions at energies so low that direct measurements  can hardly be performed due to the negligibly small penetrability factor in the entry channel of the reaction.  We elucidated the main ideas of the THM and outlined necessary conditions to perform the THM experiments.  The shortcoming of the surface-integral based THM amplitude is that it neglects the contribution of the internal post-form DWBA part.  Finally, it is assumed that the THM mechanism dominates, however, no estimation of the background is given. It would be interesting to investigate the accuracy of this assumption.

Numerous examples are presented throughout the review to demonstrate practical applications of the THM. One of them is the analysis of the neutron generator reaction in AGB stars  ${}^{13}{\rm C }(\alpha, n){}^{16}{\rm O}$, which is contributed by the  subthreshold bound state. We also critically analyzed the application of the THM method for the analysis of the one of  the important astrophysical  reactions,  ${}^{12}{\rm C}+ {}^{12}{\rm C}$ fusion.  We showed that inclusion of the Coulomb-nuclear distortions in the initial and final states of the THM reaction used to determine the astrophysical $S^{*}$ factor for the carbon-carbon fusion drops $S^{*}$ by three orders of the magnitude bringing its behavior with the agreement with the theoretical prediction of the upper limit of the $S^{*}$ factor.  We considered also a new extension of the THM, namely, its application to radiative capture reactions.  It is demonstrated how this method can be applied to investigate a crucially-important astrophysical radiative capture reaction of  ${}^{12}{\rm C}(\alpha,\,\gamma){}^{16}{\rm O}$. We explained what measurements should be done to correctly identify the THM mechanism. 

\section*{ACKNOWLEDGMENTS} 

A.M.M. acknowledges support from the U.S. DOE Grant No. DE-FG02-93ER40773 and NNSA Grant No. DENA0003841. A.S.K. acknowledges support from the Australian Research Council and thanks the staff of the Cyclotron Institute, Texas A\&M University for hospitality during his visit. D.Y.P. acknowledges support from the NSFC Grant No. 11775013.

\newpage
\begin{appendices}
\section{Spectral decomposition of  the two-channel Green's function}
 \label{spectraldecGrfunct1}

To single out  the resonance in the subsystem $\,F= x+A\,$ from the Green's function $G$ we follow Ref. \cite{mukfewbody2019} and rewrite it as 
\begin{align}
 G= {G}_{s}\,\big(1 +  {\overline V}_{xA}^{N}\,G \big),
\label{GGs1}
\end{align}
where ${\overline V}_{xA}^{N}= V_{sx}^{N} + V_{sA}^{N}$ and 
\begin{align}
{G}_{s}(z) = \frac{1}{z - {\hat T}_{sF}  -  {\hat H}_{F} -  {\overline V}_{xA}^{C}}.
\label{tildeGs1}
\end{align}
Note that ${\overline V}_{xA}^{C}=V_{sx}^{C}+V_{sA}^{C}$.
Substituting Eq. (\ref{GGs1}) into Eq. (\ref{Mf11})  one gets
 \begin{align}  
 { M}'=  \blb{{\overline \Phi}_{bB}^{C(-)}}\bl\blb{X_f}\bl{V_{bB}\, G_{s}\,{ {\tilde U}}_{sA}}\bl X_{i}\blk \bl I_{x}^{a}\,
{\Psi_{{\rm {\bf k}}_{aA}}^{C(+)} }\blk,
\label{Mf2}
\end{align} 
where the transition operator ${ {\tilde U}}_{sA}$  is
\begin{align}
{ {\tilde U}}_{sA}  = {\overline V}_{sx}^{N} + {\overline V}_{xA}^{N}\,G\,{\overline V}_{sx}^{N}.
\label{tildeUSA1}
\end{align}

Now we use approximation by replacing ${\overline V}_{xA}^{C}$  in Eq. \eqref{tildeGs1} with $U_{sF}^{C}$ to get
\begin{align} 
{G}_{s} (z)&
\approx    \frac{1}{z - {\hat T}_{sF}  -  {\hat H}_{F} - U_{sF}^{C} },
\label{GstildeGs1}
\end{align}
where $U_{sF}^{C}$ is the channel Coulomb potential describing the  interaction between the c.m. of nuclei $s$ and $F$. 
Then the reaction amplitude ${ M}$ takes the form 
 \begin{align}                                     
{ M}'=  
\blb {{\overline \Phi}_{bB}^{C(-)}} \bl
\blb {X_f} \bl V_{bB}\,G_{s}\, {\tilde U}_{sA}\bl {X_i} \blk \bl I_{x}^{a}
{\Psi_{{\rm {\bf k}}_{aA}}^{C(+)}} \blk.                              
\label{Mf3}
\end{align} 
Singling out the first step of the THM reaction  $\,a+ A \to s + F^{*}\,$  we get 
\begin{align}                                     
{ M}' =& 
\blb {{\overline \Phi}_{bB}^{C(-)}} \bl \blb {X_{f}} \bl {V_{bB}} \bl {X_{f}} \blk
  \blb   {X_{f}} \bl { G_{s}} \bl {X_{i}}   \blk \nonumber\\
& \times
   \blb {X_{i}} \bl {{ {\tilde U}}_{sA}} \bl {X_i} \blk  \bl
  I_{x}^{a}\,{\Psi_{{\rm {\bf k}}_{aA}}^{C(+)}} \blk             \nonumber\\
= &
\blb{{\overline \Phi}_{bB}^{C(-)}} \bl
{{\tilde V}_{bB} \bl
  \blb  {X_{f}} \bl  { G_{s}} \bl {X_{i}} \blk \bl  {\cal U}_{sA}} \bl I_{x}^{a}\,
{\Psi_{{\rm {\bf k}}_{aA}}^{C(+)}} \blk,         
\label{Mfpgrst2}
\end{align}
where the short-hand notations $\,{\tilde V}_{bB}\,= 
\blb {X_f}\big|{V_{bB}}\bl {X_f}\blk $ and ${\cal U}_{sA}= \blb{X_{i}}\big|{{ {\tilde U}}_{sA}}\big|{X_i}\blk$ are introduced. We assume that the potential $V_{bB}$  is spin-independent.

To single out a resonance state in the intermediate subsystem $\,F\,$ one can introduce the spectral decomposition  of  $\,G_{s} (z)$:
\begin{widetext}
\begin{align}
\blb X_{f} \big| G_{s}  \bl  \,X_{i}\big>\, =&\,  \sum_{n}\,\int\,\frac{{\rm d} {\rm {\bf k}}_{sF}}{(2\,\pi)^{3}}\,\frac{  \bl  I_{f}^{F_{n}}\,\Psi_{{\rm {\bf k}}_{sF} } ^{(-)} \blk \blb \Psi_{{\rm {\bf k}}_{sF} }^{(-)} I_{i}^{F_{n}} \bl }
{E_{aA} + Q_{n} - k_{sF}^{2}/(2\,\mu_{sF})   + i0 }                                                     \nonumber\\
&+ \,\int\, \frac{{\rm d} {\rm {\bf k}}_{bB}}{(2\,\pi)^{3}}\,\frac{{\rm d} {\rm {\bf k}}_{sF}}{(2\,\pi)^{3}}\,  \frac{ \bl \Psi_{{\rm {\bf k}}_{bB};f}^{(-)}
\, \Psi_{{\rm {\bf k}}_{sF} }^{C(-)} \blk  \blb \Psi_{ {\rm {\bf k}}_{sF} }^{C(-)}\,\Psi_{{\rm{\bf k}}_{bB};i}^{(-)}\bl }
{  E_{aA} - \varepsilon_{a} +Q_{if} -  k_{bB}^{2}/(2\,\mu_{bB}) - k_{sF}^{2}/(2\,\mu_{sF})  + i0 }.
\label{spectrtildeGs1}
\end{align} 
\end{widetext}
Here we use the notion of the overlap function $I_{f}^{F_{n}}(r_{bB}) = \blb \varphi_{B}(\xi_{B})\,\varphi_{b}(\xi_{b}) \bl \varphi_{n}(\xi_{B},\xi_{b}; r_{bB}) \blk $, which is the projection of the $n$-th bound-state of the many-body wave function $\varphi_{n}$  of $\,F$  on $X_{f}$. Integration in the 
matrix element 
is carried out over all the internal coordinates $\xi_{b}$ and $\xi_{B}$ of nuclei $b$ and $B$.
Hence, $I_{f}^{F_{n}}$ depends only on $r_{bB}$ (for non-zero orbital angular momenta it depends on ${\rm {\bf r}}_{bB}$).
Similar meaning has the second overlap function $I_{i}^{F_{n}}(r_{xA})= \blb \varphi_{n}(\xi_{A},\xi_{x}; r_{xA}) \bl \varphi_{x}(\xi_{x})\,\varphi_{A}(\xi_{A}) \blk $ introduced in Eq. \eqref{spectrtildeGs1}.

Usually the overlap functions are determined for bound states. But here we also introduce the overlap functions for 
the continuum states. In particular, we define
\begin{align}
\Psi_{ {\rm {\bf k}}_{bB};f}^{(-)}({\rm {\bf r}}_{bB})  &=  \blb \varphi_{B}(\xi_{B})\,\varphi_{b}(\xi_{b}) \bl  \Psi_{{\rm {\bf k}}_{bB}}^{(-)}(\xi_{B},\xi_{b}; {\rm {\bf r}}_{bB}) \blk        \\
\Psi_{ {\rm {\bf k}}_{bB};i}^{(-)*}({\rm {\bf r}}_{xA}) & =  \blb  \Psi_{{\rm {\bf k}}_{bB}}^{(-)}(\xi_{A},\xi_{x}; {\rm {\bf r}}_{xA}) \bl \varphi_{A}(\xi_{A})\,\varphi_{x}(\xi_{x}) \blk 
\label{ovfPsif1}
\end{align}
to be the projections of the  wave function $\Psi_{{\rm {\bf k}}_{bB}}^{(-)}$ 
 of the system $F$ in the continuum on $X_{f}$ and $X_{i}$, respectively. 
 We  assume that the continuum wave function $\Psi_{{\rm {\bf k}}_{bB}}^{(-)}$ has the incident wave in the channel $f=b+B$  with ${\rm {\bf k}}_{bB}$ being the $b+B$ relative momentum.

Also in Eq. \eqref{spectrtildeGs1},  $\;E_{aA}$ is the $a-A$ relative kinetic energy,  $Q_{n}= m_{a} + m_{A} - m_{s} - m_{F_{n}}= \varepsilon_{F_{n}} - \varepsilon_{a}$,  $\,\varepsilon_{F_{n}}\, = m_{x} + m_{A} - m_{F_{n}}$  is   the binding energy of the bound state $\,F_{n}\,$ for the virtual decay $\,F_{n} \to x+A$,  $\,\Psi_{{\rm {\bf k}}_{sF} }^{C(-)}\,$ is the Coulomb scattering wave function of particles $s$ and $F$ with the relative momentum  $\,{\rm {\bf k}}_{sF}$, $\,\mu_{sF}\,$  is the reduced mass of particles $s$ and $F$, $\;\varepsilon_{sx} = m_{s} + m_{x} - m_{a}$ is the binding energy  of $a$, $\,m_{i}$  is the mass of particle $i$,  $\,E_{aA}-\varepsilon_{a} + Q_{if}$  is the total kinetic energy of the three-body system   $s+b +B$,   $\,Q_{if}= m_{x} + m_{A} - m_{b} - m_{B}$,  $m_{i}$ is the mass of particles $i$,  $i=x+A$ and $f=b+B$  are the initial and final channels of the binary subreaction $x+ A \to b+B$. 

  Here, for simplicity, we consider  the wave function $\, \Psi_{{\rm {\bf k}}_{bB}}^{(-)}$ only in the external region. The internal region can be taken similarly using the $R$-matrix approach (see Appendix A in \cite{muk2011}).
 In the external region   the wave function  $\, \Psi_{{\rm {\bf k}}_{bB}}^{(-)}$ with the incident wave  in the channel $f=b+B$   becomes an external multichannel scattering wave function  \cite{muk2011,lanethomas}:
\begin{align}
\Psi _{{\rm {\bf k}}_{bB}}^{(-)}({\rm {\bf r}}_{bB}) =& -i\frac{1}{2\,k_{bB}}\sum\limits_c {\sqrt {\frac{{{v_f}}}{{{v_c}}}} } \frac{1}{{{r_c}}}{{ X}_c} 
\nonumber\\ & \times
\big [I^*(k_{bB} ,\,r_{bB})\,\delta _{cf} - \,\,S^*_{cf}\,O^*({k_c},{r_c}) \big] .
\label{multuchPsi1} 
\end{align}
We recall that $f$ stands for the channel $b+B$.  The sum over $c$ is taken over all open final channels $\,c\,$ coupled with the initial channel $f$.  $\,X_{c}$ stands for the product  of the bound-state wave functions of the fragments in the channel $c$,  $\,v_{c}\,$ is the relative velocity of the nuclei in the channel $c$, $S_{c\,f}$ is the scattering $S$ matrix for the transition $\, f \to  c$, $\;O(k_{c},r_{c})$ is the Coulomb Jost singular solution of the Schr\"odinger equation  with the outgoing-wave asymptotic behavior.

In the case under consideration we consider only two coupled channels, $\,i= x+A$ and  $\,f=b+B$.  
In the external region the channels are decoupled  and the overlap function  $\Psi_{{\rm {\bf k}}_{bB}; i}^{(-)}$ 
    is written as
  \begin{align}
  \Psi_{{\rm {\bf k}}_{bB};i}^{(-)*}({\rm {\bf r}}_{xA}) = \,i\frac{1}{{{2\,k_{bB}}{\,r_{xA}}}} {\sqrt {\frac{{{\mu_{xA}\,k_{bB}}}}{{{\mu_{bB}\,k_{xA}}}}} }\,S_{  f\, i}\,O({k_{xA}},{r_{xA}}). 
   \label{overlapfunctionsxA1}
 \end{align} 
 Equation (\ref{overlapfunctionsxA1})  determines the projection of the external two-channel wave function $\Psi_{F}^{(-)}$ , which has an incident wave in the channel $\,f=b+B$, onto the channel $\,i=x+A$.  
 
The second overlap function takes the form
 \begin{align}
\Psi_{{\rm {\bf k}}_{bB};f}^{(-)}({\rm {\bf r}}_{bB}) =&  \, -i\frac{{1 }}{{{2\,k_{bB}}{r_{bB}}}}\,
\nonumber \\ & \times
\big[ I^{*}({k_{bB}},{r_{bB}})  - \,\,S_{f\,f}^{*}\,O^{*}({k_{bB}},{r_{bB}}) \big],
\label{Psiff1}
\end{align}
where $I^{*}({k_{bB}},{r_{bB}})=  O({k_{bB}},{r_{bB}})$.    
 
We denote 
the second (continuum) term in Eq. (\ref{spectrtildeGs1}) as $ \blb  X_{f} \big|{ G}_{s}^{\rm cont} \bl X_{i} \blk$:
\begin{widetext}
\begin{align}
  \blb  X_{f} \big|{ G}_{s}^{\rm cont} \bl X_{i} \blk = &  \int   \frac{{\rm d} {\rm {\bf k}}_{bB}}{(2 \pi)^{3}} \frac{{\rm d} {\rm {\bf k}}_{sF}}{(2 \pi)^{3}}   \frac{ \bl \Psi_{ {\rm {\bf k}}_{bB};f}^{(-)} 
  \Psi_{{\rm {\bf k}}_{sF} }^{C(-)} \blk \blb \Psi_{{\rm {\bf k}}_{sF} }^{C(-)}  \Psi_{{\rm {\bf k}}_{bB};i}^{(-)} \bl  }
{ E_{aA} - \varepsilon_{a} + Q_{if} - k_{bB}^{2}/(2 \mu_{bB})  - k_{sF}^{2}/(2 \mu_{sF})  + i0 }. 
\label{spectrtildeGscont1}
\end{align}
\end{widetext}
 The resonant term corresponding to the subsystem  $F$  can be singled out from Eq. (\ref{spectrtildeGscont1}). 
To this end
  \begin{enumerate}
  \item  We first perform the integration over the solid angle $\Omega_{{\rm {\bf { k}}}_{bB}}$. 
  \item The $S$-matrix element  $S_{i\,f}$  has a resonance pole on the second Riemann sheet at the $b-B$ relative energy $E_{{\rm R}(bB)} =E_{0(bB)} - i\Gamma/2$,  where $\Gamma$ is the total resonance width.  In the momentum plane this resonance pole  occurs at the $b-B$ relative momentum  $k_{{\rm R}(bB)}= k_{0(bB)} - i\,k_{I(bB)}$. We assume that  the resonance is narrow:  $\Gamma  << E_{0(bB)}$  or $k_{I(0)} << k_{0(bB)}$. 
   \item  When $k_{bB}  \to k_{{\rm R}(bB)}$  the integration contour over $k_{bB}$  moves down to  the fourth quadrant  pinching the contour to the pole at $k_{{\rm R}(bB)}$. Taking the  residue at the pole  $\,E_{bB}= E_{{\rm R}(bB)}\,$ one can single out the contribution to  $ \blb X_{f} \big|{ G}_{s}^{\rm cont} (z) \bl X_{i} \blk\,$ from the resonance term in the subsystem $\,F$. 
  \end{enumerate}
  
Following all these steps  we get the desired spectral decomposition of $G_{s}$ for two coupled channels \cite{mukfewbody2019}:
\begin{widetext}
\begin{align}
& \blb  X_{f} \big|  G_{s}^{{\rm R}} \bl X_{i} \blk =- \frac{i}{4\,\pi}\,\int\,\frac{{\rm d} {\rm {\bf k}}_{sF}}{(2\,\pi)^{3}}\,                                                
 \frac{ {\, \bl  \phi_{{\rm R}(bB)}(r_{bB})\, \Psi_{{\rm {\bf k}}_{sF} }^{C(-)} \blk \blb \Psi_{{\rm {\bf k}}_{sF} }^{C(-)} \,  {\tilde \phi}_{{\rm R}(xA)}(r_{xA})} \bl  }
{ E_{aA} - \varepsilon_{a} + Q_{if} - E_{{\rm R}(bB)}  - k_{sF}^{2}/(2\,\mu_{sF}) },
\label{spectrtildeGsRtf1}
\end{align} 
\end{widetext}
where  
\begin{align}
& {\tilde \phi}_{{\rm R}(xA)}(r_{xA}) = e^{-i\,\delta^{p}(k_{0(xA)})}\,\sqrt{    \frac{\mu_{xA}}{k_{{\rm R}(xA)} }\,\Gamma_{xA} }\,\frac{O^{*}({k_{{\rm R}(xA)}},{r_{xA}})}{r_{xA}},  \nonumber\\
& \phi_{{\rm R}(bB)}(r_{bB}) =e^{i\,\delta^{p}(k_{0(bB)})}\, \sqrt{    \frac{\mu_{bB}}{k_{{\rm R}(bB)} }\,\Gamma_{bB} }\,\frac{O({k_{{\rm R}(bB)}},{r_{bB}})}{r_{bB}} 
\label{Gamowwfi1}
\end{align}
are the Gamow resonant wave functions in channels $i$ and $f$. Note that $\, {\tilde \phi}_{{\rm R}(xA)}(r_{xA})\,$  is the Gamow wave function from the dual basis, $\Gamma_{xA}$ and $\Gamma_{bB}$ are the partial resonance widths in the initial channel $i$ and final channel $f$, respectively,
$\delta^{p}(k_{0(xA)}$ and $\delta^{p}(k_{0(bB)})$ are the potential (non-resonant) scattering phase shifts in the initial and final channels.

After deriving the expression for  the resonance term  in the  spectral decomposition of   $\blb X_{f} \big| G_{s}^{{\rm R}} (z) \big| X_{i}\blk$  we can substitute it into Eq. (\ref{Mfpgrst2})  and  derive an equation for the amplitude of the reaction  $a+ A  \to s+b+B$  with three charged particles in the final state, proceeding through an intermediate resonance in the subsystem  $F=x+ A=b+B$.  
As mentioned above, the TH reaction amplitude is described by the two-step process: the first step is the transfer reaction populating the resonance state 
$a+ A \to s +F^{*}$ and the second step is the decay of the resonance into two-fragment channel $F^{*} \to b+B$   leading to the formation of the three-body final state, $s+ b+ B$.  We derive below the expression for the TH reaction amplitude taking into account the Coulomb interactions in the intermediate and final states. 

Substituting Eq. (\ref{spectrtildeGsRtf1})  into Eq. (\ref{Mfpgrst2}) and writing it in the momentum representation  one gets                                 
\begin{align}                                     
{ M}'=& - \frac{i\,\mu_{sF}}{2\,\pi} \,\int\,
\frac{{\rm d} {\rm {\bf p}}_{B}}{(2\,\pi)^{3}}\,\frac{{\rm d} {\rm {\bf p}}_{b}}{(2\,\pi)^{3}}                                               
 {\overline \Phi}_{  (bB){\rm {\bf k}}_{B}\, {\rm {\bf k}}_{b} }^{(+)}\big({\rm {\bf p}}_{B},\,{\rm {\bf p}}_{b}\big)        \nonumber\\
& \times W_{bB}\,\big({\bf p}_{bB} \big)\, 
 {\cal J}({\rm {\bf p}}_{sF},\, {\rm {\bf k}}_{aA}),
\label{THMres2}
\end{align} 
where 
\begin{align}
{\cal J}({\rm {\bf p}}_{sF},\, {\rm {\bf k}}_{aA})   = 
\int \,
\frac{{\rm d} {\rm {\bf k}}_{sF}}{(2\,\pi)^{3}}\,
\frac{ \Psi_{ {\rm {\bf k}}_{sF} }^{C(-)}\big({\rm {\bf p}}_{sF}\big)\,{ M}_{(tr)}\big({\rm {\bf k}}_{sF} ,\,{\rm {\bf k}}_{aA} \big) }
{  k_{\rm R}^{2}  - k_{sF}^{2} },
\label{J11int} 
\end{align}
and
\begin{align}
 { M}_{(tr) } \big({\rm {\bf  k}}_{sF},\,{\rm {\bf k}}_{aA}\big)= \blb \Psi_{  {\rm {\bf k}}_{sF} }^{C(-)}\, {\tilde \phi}_{{\rm R}(xA)} \bl {\cal U}_{sA} \bl I_{x}^{a}\,\Psi_{  {\rm {\bf k}}_{aA}    }^{C(+)} \blk
 \label{M11}
 \end{align}
is the amplitude of the transfer reaction $a + A \to s + F^{*}$  populating the resonance $F^{*}$.  
This amplitude can be approximated by $M_{{M_F}{M_s};{M_A}{M_a}}^{(prior)}({k_{0(sF)}}{{\bf{\hat k}}_{sF}},{{\bf{k}}_{aA}})$ defined in Eq. (\ref{MDSextpr1}).

${\cal M}_{(tr)}= M_{{M_F}{M_s};{M_A}{M_a}}^{(prior)}({k_{0(sF)}}{{\bf{\hat k}}_{sF}},{{\bf{k}}_{aA}})$

The expression for the form factor $W_{bB}\big({\rm {\bf p}}_{bB}\big)$ can be obtained 
 using Eq. (36) from  \cite{blokh84}:  
\begin{align}
W_{bB}\big({\bf p}_{bB} )  =& \int {d{{\bf r}_{bB}}} {e^{-i{{\bf p}_{bB}} \cdot {{\bf r}_{bB}}}}{\widetilde V_{bB}}({ r}_{bB}){ \varphi _{{\rm R}(bB)}}({r}_{pB})                            \nonumber\\
=&{e^{i{\delta ^p}({k_{0(bB)}})}}\,{e^{\frac{{ - \pi {\eta _{{\rm R}(bB)}}}}{2}}}\,{\left( {\frac{{p_{bB}^2 - k_{{\rm R}(bB)}^2}}{{4k_{{\rm R}}^2}}} \right)^{i{\eta _{R(bB)}}}}\,
\nonumber\\
& \times \Gamma (1 - i{\eta _{{\rm R}(bB)}})\,\sqrt{\frac{\mu_{bB}\,\Gamma_{bB}}{k_{{\rm R}(bB)}  }}\,g(p_{bB}^{2}),
 \label{formfactor1}
\end{align}
where $g(p_{bB}^{2})$   is the so-called reduced form factor, which satisfies $g(k_{{\rm R}(bB)}^{2})=1$.  
Also
\begin{align}
k_{\rm R}^{2}/(2\,\mu_{sF}) = E_{aA} - \varepsilon_{a} + Q_{if} - E_{{\rm R}(bB)}.
\label{k0sF1}
\end{align}
Note that the imaginary part  ${\rm Im} (k_{\rm R}) >0$.  

The above equations allow us to obtain the expression for the resonant THM reaction (\ref{THMreaction1}).
\\[0.1in]

\section{Derivation of the THM radiative capture amplitude}
\label{THMRCampl1}
We analyze the THM radiative capture reaction (\ref{THMreactiongamma1})  using the PWA.  The derivation of the PWA for the transfer reaction is presented in \cite{reviewpaper} and in Section \ref{THMPWA1}.  For the case under consideration, we use only the surface term, which 
takes the form:
\begin{widetext}
\begin{align}
M_{M_{a}\,M_{A}\,\tau}^{M_{s}\,M_{F_{\tau}^{(s)}}}({\rm {\bf k}}_{sF_{\tau}},{\rm {\bf k}}_{aA}) =&\frac{\sqrt{\pi}}{\mu_{xA}}\,i^{l_{xA}}\,\varphi_{a}(p_{sx})\,\sqrt{2\,\mu_{xA}\,R_{xA}}\,\gamma_{\tau\,j_{xA}l_{xA}J_{F^{(s)}}}
{\cal {\tilde W}}_{l_{xA}}\,\sum\limits_{M_{x}\,m_{j_{xA}}\,m_{l_{xA}}}\,  \blb J_{s}\,M_{s}\,\,J_{x}\,M_{x} \big|J_{a}\,M_{a} \blk          \nonumber\\
& \times \blb J_{x}\,M_{x}\,\,J_{A}\,M_{A} \big|j_{xA}\,m_{j_{xA}} \blk  \, \blb j_{xA}\,m_{j_{xA}}\,\,l_{xA}\,m_{l_{xA}} \big| J_{F^{(s)}}\,M_{F_{\tau}^{(s)}} \blk \,Y_{l_{xA}\,m_{l_{xA}}}^{*}({\rm {\bf {\hat p}}}_{xA}),
\label{THradamplSurf11} 
\end{align}
\end{widetext}
where ${\cal {\tilde W}}_{l_{xA}}$  is given by 
\begin{align}
{\cal {\tilde W}}_{l_{xA}}  =& \Big[
 j_{{l_{xA}}}({p_{xA}}{r_{xA}})\Big[R_{ch}\,\frac{ {\partial {\ln[O_{{l_{xA}}}}({k_{xA}}{r_{xA}})]}}{{\partial {r_{xA}}}}
-1\Big] 
\nonumber\\
& - R_{ch}\,\frac{\partial{{\ln\,[j_{{l_{xA}}}}({p_{xA}}{r_{xA}})]}}{{\partial {r_{xA}}}}\Big]{\Big|_{{r_{xA}} = {R_{ch}}}}                    \nonumber\\
&+  2\,\mu_{xA}\,\frac{Z_{x}\,Z_{A}}{\alpha}\,\int\,{\rm d}r_{xA}\,j_{l_{xA}}(p_{xA}r_{xA})    \nonumber\\
& \times \frac{O_{l_{xA}}(k_{xA}r_{xA})}{O_{l_{xA}}(k_{xA}R_{ch})}.
\label{caltildeWr1}
\end{align}
This is the generalization of Eq. (\ref{calWr1}) by including the integral giving the external term contribution. Momenta ${\rm {\bf p}}_{xA}$ and ${\rm {\bf p}}_{sx}$  are defined by Eqs. (\ref{pxApsx1}).

Note that $M_{M_{a}\,M_{A}\,\tau}^{M_{s}\,M_{F_{\tau}^{(s)}}}({\rm {\bf k}}_{sF_{\tau}},{\rm {\bf k}}_{aA}) $ does not contain the hard-sphere scattering phase shift $\delta_{j_{xA}l_{xA}J_{F^{(s)}}    }^{hs}$.
Also, $\varphi_{a}(p_{sx})$ is the Fourier transform of the radial part of the $\,s$-wave bound-state wave function $\varphi_{a}(p_{sx})$ of the $a=(s\,x)$.  Also, $\kappa_{ sx}=\sqrt{ 2\,\mu_{sx}\,\varepsilon_{sx}}\,$ is the wave number of the bound-state $\,a=(s\,x)$, $\,\varepsilon_{sx}\,$ is its binding energy for the virtual decay  $a \to s+ x$. Since  particles $s$ and $x$  are structureless,  the spectroscopic factor of the bound state $a=(s\,x)$ is unity and we can use just the bound-state wave function $\varphi_{sx}$. 
Also $\,k_{s}$ and $E_{xA}$ are related by the energy conservation \cite{reviewpaper}:
\begin{align}
E_{aA} - \varepsilon_{sx} = E_{xA} + k_{s}^{2}/(2\,\mu_{sF}), 
\label{energconserv1}
\end{align}
where ${\rm {\bf k}}_{s}$ is the momentum of the spectator $s$  in the c.m. of the THM reaction, $\mu_{sF}$ is the reduced mass of particles $s$ and $F$.

Now we consider the amplitude  $V_{\nu},\,\,\nu=1,2,$ describing the radiative decay of the intermediate resonance $F_{\nu} \to F + \gamma$ \cite{muk2016}:
\begin{align}
V_{M_{F_{\nu}}^{(s)}\,\nu}^{M_{F}\,M\,\lambda}=& -\int\,{\rm d}{\rm {\bf r}}_{xA}
\nonumber \\ & \times
 \blb I_{xA}^{F}({\rm {\bf r}}_{xA})\,\big|{\rm {\bf {\hat J}}}({\rm {\bf r}}) \big|\Upsilon_{\nu}({\rm {\bf r}}_{xA}) \blk \cdot {\rm {\bf A}}^{*}_{\lambda\,{\rm {\bf k}}_{\gamma}}({\rm {\bf r}}),
\label{Vnu1}
\end{align}
where  $I_{xA}^{F}({\rm {\bf r}}_{xA})$ is the overlap function of the bound-state wave functions of $\,x,$ $\,A$ and the ground state of $\,F=(x\,A)$. Again, for the point-like nuclei $x$ and $A$  the overlap function $I_{xA}^{F}({\rm {\bf r}}_{xA})$  can be replaced by the single-particle  bound-state wave function of $(xA)$ in the ground state. Also ${\rm {\bf A}}^{*}_{\lambda\,{\rm {\bf k}}_{\gamma}}({\rm {\bf r}}) $ is the electromagnetic  vector potential of the photon with helicity $\lambda=\pm 1$ and momentum ${\rm {\bf k}}_{\gamma}$ at coordinate ${\rm {\bf r}}_{xA}$,
$\,{\rm {\bf {\hat J}}}({\rm {\bf r}})$  is the charge current density operator. Matrix element in Eq. (\ref{Vnu1}) is written assuming that on the first stage of the reaction the excited state $F_{\nu},\,\nu=1,2,$ is populated, which subsequently decays to the ground state $F$.

Using the multipole expansion of  the vector potential, leaving only
the electric components with the lowest allowed multipolarities $L$ and using the long wavelength approximation for ${\rm {\bf {\hat J}}}({\rm {\bf r}}) $ we get (see Ref. \cite{muk2016} for details)
\begin{align}
 V_{M_{F_{\nu}^{(s)}}\,\nu}^{M_{F}\,M\,\lambda}  =& \frac{\sqrt{2}}{4\,\pi}\,\sum\limits_{L}i^{-L}\,(-1)^{L+1}\sqrt{\hat L} \,k_{\gamma}^{L-1/2} \,[\gamma_{(\gamma)\,\nu\,J_{F}\,L}^{J_{F^{(s)}}^{L}} ]\, \nonumber \\ & \times
\big[D^{L}_{M\,\lambda}(\phi,\,\theta,0)\big]^{*}\, \blb J_{F}\,M_{F}\,\,L\,M \big|J_{F^{(s)}}^{L}\,M_{F_{\nu}^{(s)}}^{L} \blk ,
 \label{VMFgamma2}
\end{align}
where $\,\gamma_{(\gamma)\,\nu\,J_{F}\,L}^{J_{F^{(s)}}^{L}}$ is the formal $R$-matrix radiative width amplitude for the electric  ($EL$) transition $\,J_{F^{(s)}}^{L} \to J_{F}$  given by the sum of the internal and external radiative width amplitudes, see Eqs (32) and (33) from \cite{muk2017},  in  which we singled out $\sqrt{2}\,k_{\gamma}^{L+1/2}$. Since now we take into account a few multipolarities $L$, we replace the previously introduced spin of the intermediate resonance $J_{F^{(s)}}$ by $J_{F^{(s)}}^{L}$, where the superscript $L$ denotes the multipolarity of the $EL$ transition to the ground state $F$. Replacement of $J_{F^{(s)}}$ by $J_{F^{(s)}}^{L}$ takes into account that the spins of the intermediate excited states are different for different multipolarities. Since we added the superscript $L$ to the spin of the intermediate resonance we added the same superscript to its projection $M_{F_{\nu}^{(s)}}^{L}$.
Also in Eq. (\ref{VMFgamma2})  $\,M$  is  the projection  of the angular momentum $L$ of the emitted photon (multipolarity of the electromagnetic transition).
We remind that $V_{M_{F_{\nu}^{(s)}}\,\nu}^{M_{F}\,M\,\lambda}$ does not depend on the 
hard-sphere scattering phase shift.

 The determined  radiative width amplitude is related to the formal resonance radiative width by the standard equation
\begin{align}
\Gamma_{(\gamma)\,\nu\,J_{F}\,L}^{J_{F^{(s)}}^{L}}= 2\,k_{\gamma}^{L+1/2}\,(\gamma_{(\gamma)\,\nu\,J_{F}\,L}^{J_{F^{(s)}}^{L}})^{2}.
\label{Gg1}
\end{align}
Note that the observable radiative width is related to the formal one by [see also Eq. (\ref{tildegammapr1})]
\begin{align}
\left({\tilde \gamma}_{(\gamma)\,\nu\,J_{F}\,L}^{J_{F^{(s)}}^{L}}\right)^{2} = \frac{(\gamma_{(\gamma)\,\nu\,J_{F}\,L}^{J_{F^{(s)}}^{L}})^{2}}{ 1+ \gamma_{\nu\,j_{xA}l_{xA}J_{F^{(s)}}}^{2} 
\left[\frac{{\rm d}S_{l_{xA}}(E_{xA})}{{\rm d}E_{xA}}\right]
_{E_{xA}= E_{\nu}}}.
\label{observformradwidth1}
\end{align}

We consider the two-level approach with $\nu=1$ ($\nu=2$) corresponding to the subthreshold resonance  (the resonance at $E_{xA}>0$). Then $E_{\nu}=-\varepsilon_{xA}^{(s)}$ for $\nu=1$  and  $E_{\nu}= E_{R}$ for $\nu=2$ with $E_{R}$ being the resonance energy corresponding to the level
$\nu=2$.
This observable radiative width is related to the observable resonance radiative width as
\begin{align}
{\tilde \Gamma}_{(\gamma)\,\nu\,J_{F}\,L}^{J_{F^{(s)}}^{L}}= 2\,k_{\gamma}^{L+1/2}\,({\tilde \gamma}_{(\gamma)\,\nu\,J_{F}\,L}^{J_{F^{(s)}}^{L}})^{2}.
\label{Gg2}
\end{align}

Substituting Eqs. (\ref{THradamplSurf11}) and (\ref{VMFgamma2})  into Eq.  (\ref{THradamplfull1}) we get  the expression for the indirect  reaction amplitude
\begin{widetext}
\begin{align}
M_{M_{a}\,M_{A}}^{M_{s}\,\,M_{F}\,M\,\lambda} =&\frac{\varphi_{a}(p_{sx})}{2}\,\sqrt{\frac{R_{xA}}{\pi\,\mu_{xA}}}\,\sum\limits_{L}\,(-1)^{L+1}\,{\hat L}^{1/2}\,k_{\gamma}^{L-1/2}\,\big[D^{L}_{M\,\lambda}(\phi,\,\theta,0)\big]^{*}\,\sum\limits_{l_{xA}}\,i^{l_{xA} -L}\,{\cal {\tilde W}}_{l_{xA}} \nonumber\\
& \times  \sum\limits_{\nu,\,\tau=1}^{2}\,\gamma_{(\gamma)\,\nu\,J_{F}\,L}^{J_{F^{(s)}}^{L}}\,{\rm {\bf A}}_{\nu\,\tau}^{L} \,\gamma_{\tau\,j_{xA}l_{xA}J_{F^{(s)}}^{L}}\,\sum\limits_{ M_{F_{\nu}^{(s)}}^{L} }\,  \blb J_{F}\,M_{F}\,\,L\,M \big|J_{F^{(s)}}^{L}\,M_{F_{\nu}^{(s)}}^{L} \blk                                                            \nonumber\\
& \times  \sum\limits_{m_{j_{xA}}\,m_{l_{xA}}\,M_{x} } \blb j_{xA}\,m_{j_{xA}}\,\,l_{xA}\,m_{l_{xA}}\big| J_{F^{(s)}}^{L}\,M_{F_{\tau}^{(s)}}^{L} \blk \,  \blb J_{x}\,M_{x}\,\,J_{s}\,M_{s} \big|J_{a}\,M_{a} \blk 
 \nonumber \\ & \times  
 \blb J_{x}\,M_{x}\,\,J_{A}\,M_{A} \big| j_{xA}\,m_{j_{xA}} \blk \,Y_{l_{xA}\,m_{l_{xA}}}^{*} ({\rm {\bf {\hat p}}}_{xA}).
\label{MTHtot1}
\end{align}
\end{widetext}

The amplitude $M_{M_{a}\,M_{A}}^{M_{s}\,\,M_{F}\,M\,\lambda}$  describes the indirect reaction proceeding through the intermediate resonances, which decay to the ground state $F=(x\,A)$ by emitting  photons. Equation (\ref{MTHtot1}) is generalization of Eq. (\ref{THradamplfull1})
by including the sum over multipolarities $L$ corresponding to the radiative electric transitions from the intermediate resonances with the spins $J_{F^{(s)}}^{L}$ to the ground state $F$ with the spin $J_{F}$. Note also that we assume that two levels contribute to each transition of multipole $L$. It requires the two-level generalized $R$-matrix approach. 
The generalization of Eq. (\ref{MTHtot1}) for three- or more-level cases is straightforward. 
In Eq. (\ref{MTHtot1}) the reaction part and radiative parts are interconnected by the $R$-matrix level matrix elements ${\rm {\bf A}}_{\nu\,\tau}^{L}$.  

The sum over $\nu$ and $\tau$ in Eq. \eqref{MTHtot1}
is the standard $R$-matrix term for the binary resonant radiative-capture reaction. However, we analyze the three-body reaction  $a(x\,s) + A \to s +F +\gamma$ with the spectator $s$ in the final state rather than the standard two-body radiative-capture reaction $x +A \to F+ \gamma$. This difference leads to the 
generalization of the standard $R$-matrix approach for the three-body reactions resulting
in the appearance of the additional factor $\varphi_{a}(p_{sx})\,{\cal W}_{l_{xA}}$ which should be familiar to the reader from Eq. (\ref{LSPWAfinal1}). That is why we call the developed approach the generalized $R$-matrix method for the indirect resonant radiative-capture reactions. 

We take the indirect reaction amplitude at fixed projections of the spins of the initial and final particles including the fixed projection $M$ of the orbital momentum $L$ of the emitted photon and fixed its chirality $\lambda$.  For example, for the ${}^{12}{\rm C}(\alpha,\,\gamma){}^{16}{\rm O}$ reaction the electric dipole $\,E1$ ($L=1$) and quadrupole $\,E2$ ($L=2$) transitions do contribute and they interfere. In the long-wavelength approximation only minimal allowed $l_{xA}$ for given $L$ does contribute. For example, for the case considered below $\,l_{f}=0$ $\,\,l_{xA}=L=1$ for the dipole  and $\,l_{xA}=L=2$ for the quadrupole electric transitions do contribute. The dimension of the $R$-matrix level matrix ${\rm {\bf A}}^{L} $ depends on the number of the levels taken into account for each $L$. 

The indirect reaction amplitude  depends on the off-shell momenta ${\rm {\bf p}}_{sx}$ and 
${\rm {\bf p}}_{xA}$. Both off-shell momenta are expressed in terms of ${\rm {\bf k}}_{a}$ and ${\rm {\bf k}}_{s}$, see Eq. (\ref{pxApsx1}). Also the the indirect reaction amplitude depends on the momentum of the emitted photon ${\rm {\bf k}}_{\gamma}$  whose direction is determined by the angles in the Wigner $D$-function. 
 
In the center-of-mass of the  reaction 
(\ref{THMreaction1})  neglecting the recoil effect of the nucleus $F$ during the photon emission from the energy conservation we get
\begin{align}
&E_{aA} + Q = E_{sF} + k_{\gamma}, 
\label{energconserv2}\\
&k_{\gamma} = E_{xA} + \varepsilon_{xA},
\label{kgExA1}
\end{align}
where $\,E_{sF}= k_{s}^{2}/(2\,\mu_{sF}),$ $\,Q=\varepsilon_{xA} - \varepsilon_{sx}$ and $\varepsilon_{xA}$ is the binding energy of the ground state of the nucleus $F$.

To estimate the recoil effect we take into account that in the center-of-mass of the reaction (\ref{THMreaction1}) the momentum conservation in the final state gives
\begin{align}
 {\rm {\bf k}}^{'}_{F} = -{\rm {\bf k}}_{\gamma} - {\rm {\bf k}}_{s},
\label{momconservfinalst1}
\end{align}
where ${\rm {\bf k}}_{F}^{'}$ is the momentum of the final nucleus $F$ after emitting the photon. 
Then the energy conservation leads to
\begin{align}
E_{aA}- \varepsilon_{sx}&= \frac{k_{s}^{2}}{2\,\mu_{sF}} + E_{xA}= \frac{k_{s}^{2}}{2\,m_{s}} + \frac{ (k_{F}')^{2} }{2\,m_{F} }  + k_{\gamma} 
\label{1stenergconerv1} \\
&= \frac{k_{s}^{2}}{2\,\mu_{sF}} +  2\frac{k_{s}\,k_{\gamma}}{2\,m_{F}}\cos\theta' +
\frac{k_{\gamma}^{2}}{2\,m_{F}} + k_{\gamma}.
\label{energconTHreac1}
\end{align}
We remind that we use the system of units in which $\hbar=c=1$, hence, $E_{\gamma}=k_{\gamma}$.
Clearly, the term $\frac{k_{\gamma}^{2}}{2\,m_{F}} = E_{\gamma}\,\frac{E_{\gamma}}{2\,m_{F}}$ can be neglected because $E_{\gamma} << m_{F}$.
The contribution of the term $2\frac{k_{s}\,k_{\gamma}}{2\,m_{F}}\cos\theta'$ depends on $\cos\theta' = {\rm {\bf {\hat k}}}_{s} \cdot {\rm {\bf {\hat k}}}_{\gamma}$. 

Neglecting the recoil effect of the nucleus $F$ in Eq. (\ref{1stenergconerv1}) we can replace $k_{F}^{'}$ by $k_{s}$. Then $k_{\gamma}$ and $k_{s}$ are related by Eq. (\ref{energconserv2})  while $k_{\gamma}$ and $E_{xA}$ are related by Eq. (\ref{kgExA1}).
If we would take into account the recoil effect then the relationship between $k_{\gamma}$ and $E_{xA}$ is more complicated than Eq. (\ref{kgExA1}) and is given by
\begin{align}
k_{\gamma} = \frac{E_{xA}}{\frac{k_{s}\,\cos\theta'}{m_{F}} +1},
\label{recoilkgExA1}
\end{align}
where we neglected the extremely small term ${k_{\gamma}^{2}}/{(2\,m_{F})}$. 

The expression for $p_{xA}$ is needed to calculate ${\tilde M}_{l_{xA}}$. From the energy-momentum conservation law in the three-ray vertices $a \to s+x$ and $ x+ A \to F^{(s)}$ of the diagram in Fig. \ref{fig:fig_polediagram1} we get \cite{reviewpaper}
\begin{align}
E_{xA}= \frac{p_{xA}^{2}}{2\,\mu_{xA}} - \frac{p_{sx}^{2}}{2\,\mu_{sx}} -\varepsilon_{s\,x}.
\label{ExApxA1}
\end{align}
In the QF kinematics $p_{sx}=0$ and 
\begin{align}
E_{xA}= \frac{p_{xA}^{2}}{2\,\mu_{xA}} - \varepsilon_{s\,x}.
\label{ExApxAQF1}
\end{align}
Thus always ${p_{xA}^{2}}/{(2\,\mu_{xA})} > E_{xA}$.

\end{appendices}


\end{document}